\documentclass[12pt,letterpaper]{article}
\usepackage{epsfig}
\usepackage{amsmath}
\usepackage{amssymb}
\usepackage{amsthm}
\usepackage{indentfirst}
\usepackage{xspace}
\usepackage{multirow}
\usepackage{hyperref}
\usepackage{xcolor}
\definecolor{darkred}{rgb}{0.5,0.15,0.15}
\hypersetup{colorlinks=true,urlcolor=darkred,linkcolor=darkred,citecolor=darkred}
\usepackage{verbatim}
\usepackage[letterpaper,margin=1in,headheight=15pt]{geometry}
\usepackage{mathpazo}
\usepackage{authblk}
\usepackage{tikz-cd}
\usepackage{capt-of}
\usepackage{silence}
\usepackage{enumerate}
\usepackage{youngtab}
\usepackage[missing=]{gitinfo2}

\usepackage{tikz}
\usetikzlibrary{calc}   
\usetikzlibrary{topaths}
\usetikzlibrary{decorations}
\usetikzlibrary{decorations.pathmorphing}

\numberwithin{equation}{section}

\newtheorem{theorem*}{Theorem}

\newcommand{\cC}{\ensuremath{\mathcal C}}

\newcommand{\cL}{\ensuremath{\mathcal L}}
\newcommand{\cS}{\ensuremath{\mathcal S}}

\newcommand{\cM}{\ensuremath{\mathcal M}}
\newcommand{\cO}{\ensuremath{\mathcal O}}

\newcommand{\cX}{\ensuremath{\mathcal X}}

\newcommand{\cW}{\ensuremath{\mathcal W}}
\newcommand{\cG}{\ensuremath{\mathcal G}}

\newcommand{\R}{\ensuremath{\mathbb R}}
\newcommand{\C}{\ensuremath{\mathbb C}}
\newcommand{\PP}{\ensuremath{\mathbb P}}

\newcommand{\half}{\ensuremath{\frac{1}{2}}}

\newcommand{\N}{{\mathcal N}}

\newcommand{\ab}{\mathrm{ab}}

\newcommand{\eps}{\epsilon}

\newcommand{\ti}[1]{\textit{#1}}

\DeclareMathOperator{\Tr}{Tr}

\DeclareMathOperator{\End}{End}

\DeclareMathOperator{\diag}{diag}

\newcommand{\insfigscaled}[3]{

\medskip
\noindent
\begin{minipage}{\linewidth}

\makebox[\linewidth]{\includegraphics[keepaspectratio=true,scale=#2]{figures/#1.pdf}}

\captionof{figure}{#3}

\label{fig:#1}
\end{minipage}
\medskip

}

\newenvironment{definition}[1][Definition.]{\begin{trivlist}
\item[\hskip \labelsep {\bfseries #1}]}{\end{trivlist}}
\newenvironment{example}[1][Example.]{\begin{trivlist}
\item[\hskip \labelsep {\bfseries #1}]}{\end{trivlist}}

\begin{document}

\title{Higher length-twist coordinates, generalized Heun's opers, and twisted superpotentials}
\date{{{\tiny \color{gray} \tt \gitAuthorIsoDate}}
{{\tiny \color{gray} \tt \gitAbbrevHash}}}
\author[1]{Lotte Hollands}
\author[2]{Omar Kidwai}
\affil[1]{Maxwell Institute for Mathematical Sciences and 
 Department of Mathematics, Heriot-Watt University}
\affil[2]{Mathematical Institute, University of Oxford}

\maketitle

\begin{abstract}
In this paper we study a proposal of Nekrasov, Rosly and Shatashvili that describes the effective twisted superpotential obtained from a class S theory geometrically as a generating function in terms of certain complexified length-twist coordinates, and extend it to higher rank. First, we introduce a higher rank analogue of Fenchel-Nielsen type spectral networks in terms of a generalized Strebel condition. We find new systems of spectral coordinates through the abelianization method and argue that they are higher rank analogues of the Nekrasov-Rosly-Shatashvili Darboux coordinates. Second, we give an explicit parametrization of the locus of opers and determine the generating functions of this Lagrangian subvariety in terms of the higher rank Darboux coordinates in some specific examples. We find that the generating functions indeed agree with the known effective twisted superpotentials. Last, we relate the approach of Nekrasov, Rosly and Shatashvili to the approach using quantum periods via the exact WKB method.  
\end{abstract}

\setcounter{page}{1}

 \tableofcontents

\section{Introduction and summary}

Suppose $T$ is a four-dimensional $\mathcal{N}=2$ quantum field theory. In this paper we restrict ourselves to quantum field theories obtained from compactification of the six-dimensional $(2,0)$ theory of type $\mathfrak{g}$ on a possibly punctured Riemann surface $C$. Then $T$ is called a theory of class $\cS$. For simplicity, we assume $\mathfrak{g}= A_{K-1}$.

The low energy dynamics of $T$ is described in terms of the prepotential $\mathcal{F}_0(a;m,q)$, a holomorphic function of the Coulomb moduli $a = (a_1,\ldots, a_{K-1})$, the mass parameters $m=(m_1,\ldots,m_n)$ and the UV gauge couplings $q=(q_1,\ldots,q_{3g-3+n})$. 

The prepotential $\mathcal{F}_0(a;m,q)$ is related to a classical algebraically integrable system \cite{donagi1996supersymmetric}. It may be interpreted as a generating function of a Lagrangian submanifold $\mathbf{L}_0$ relating the Coulomb parameters $a$ to the dual Coulomb parameters $a^D=\partial_{a} \mathcal{F}_0$. For theories of class $\cS$ this integrable system is a Hitchin system associated to $C$ \cite{GAIOTTO2013239,Gaiotto:2009we, Neitzke:2014cja}.

Consider $T$ in the background 
\begin{align}
\mathbb{R}^4 = \mathbb{R}^2_{1,2} \oplus \mathbb{R}^2_{3,4}
\end{align} 
where we turn on the so-called $\Omega$-deformation with complex parameters $\eps_1$ and $\eps_2$, both with the dimension of mass, corresponding to the two isometries rotating the planes $\mathbb{R}^2_{1,2}$ and $\mathbb{R}^2_{3,4}$, respectively. The low energy dynamics of the resulting theory $T_{\eps_1,\eps_2}$ is described in terms of the $\eps_1,\eps_2$-deformed prepotential $\mathcal{F}(a;m,q,\eps_1,\eps_2)$. Then $\epsilon_{1}\epsilon_{2}\mathcal{F}(a;m,q,\eps_1,\eps_2)$ is analytic in $\eps_1$, $\eps_2$ near zero and becomes the prepotential $\mathcal{F}_0(a;m,q)$ in the limit $\eps_1,\eps_2 \to 0$ \cite{Nekrasov:2002qd,Nekrasov:2003rj,Nakajima:2003uh}. That is,  
\begin{align}
\mathcal{F}(a;m,q,\eps_1,\eps_2) = \frac{1}{\eps_1 \eps_2} \mathcal{F}_0(a;m,q) + \ldots, 
\end{align}
with $\ldots$ denoting terms regular in $\eps_1$ and $\eps_2$. 

\subsection{Effective twisted superpotential}

Instead, consider $T$ in the background $\mathbb{R}^4$ where we only turn on the $\Omega$-deformation with parameter $\eps_1=\eps$. This is sometimes called the Nekrasov-Shatashvili limit. The resulting theory $T_\eps$ preserves a two-dimensional $\mathcal{N}=(2,2)$ super-Poincare invariance. 

In \cite{Nekrasov:2009rc} it is proposed that, in the infrared limit, $T_\eps$ has an effective two-dimensional description in terms of $K-1$ abelian gauge multiplets coupled to an effective twisted superpotential $\widetilde{W}^{\mathrm{eff}}(a;m,q,\eps)$ for the twisted chiral fields in these gauge multiplets. This effective twisted superpotential should be obtained from the four-dimensional prepotential as
\begin{align}\label{nekshat}
\widetilde{W}^{\mathrm{eff}}(a;m,q,\eps) =  \lim_{\eps_2 \to 0} \eps_2 \, \mathcal{F}(a;m,q,\eps_1=\eps,\eps_2),
\end{align} 
where we denote the complex vevs of the twisted chiral fields by $a=(a_1, \ldots a_{K-1})$. In particular, this implies that
\begin{align}
\widetilde{W}^{\mathrm{eff}}(a;m,q,\eps)  = \frac{1}{\eps} \mathcal{F}_0(a;m,q) + \ldots,
\end{align}
where $\ldots$ are terms regular in $\eps$. 

Since the gauge theory partition function 
\begin{align}
Z(a;m,q,\eps_1,\eps_2) = \exp \mathcal{F}(a;m,q,\eps_1,\eps_2),
\end{align} 
also known as the Nekrasov partition function, is the product of a one-loop perturbative contribution and a series of exact instanton corrections, we find that the effective twisted superpotential $\widetilde{W}^{\mathrm{eff}}(a;m,q,\eps)$ can similarly be written in the form
\begin{align}
\widetilde{W}^{\mathrm{eff}}(a;m,q,\eps)  = \widetilde{W}^\mathrm{eff}_{\mathrm{clas}}(a;m,\eps) \log q +  \widetilde{W}^\mathrm{eff}_{\mathrm{1-loop}}(a;m,\eps) + \widetilde{W}^\mathrm{eff}_{\mathrm{inst}}(a;m,\eps).
\end{align}
The perturbative part has a classical contribution, proportional to $\log q$, and a 1-loop-term, which is independent of $q$. The instanton part has an expansion in powers of $q$ of the form  
\begin{align}
\widetilde{W}^\mathrm{eff}_{\mathrm{inst}}(a;m,q,\eps) = \sum_{k=1}^\infty \widetilde{W}^\mathrm{eff}_{k}(a;m,\eps) \, q^k.
\end{align}
For theories $T$ with a known Lagrangian description, these terms have been computed explicitly, and the effective twisted superpotential $\widetilde{W}^{\mathrm{eff}}(a;m,q,\eps)$ has a known expression.

Note that for a superconformal theory $T$ the function $\widetilde{W}^{\mathrm{eff}}(a;m,q,\eps)$ can simply be recovered from~$\widetilde{W}^{\mathrm{eff}}(a;m,q,1)$ by scaling the Coulomb and mass parameters with $\eps^{-\#}$, where $\#$ is their mass dimension. In the following we often leave out $\eps$ from the notation, knowing that we can simply reintroduce it by scaling the Coulomb and mass parameters.

\begin{example}

Let $T_\eps$ be the four-dimensional $\mathcal{N}=2$ superconformal $SU(2)$ theory coupled to four hypermultiplets in the partial $\Omega$-background with parameter $\epsilon$.

The classical contribution to its effective twisted superpotential is simply
\begin{align}\label{eq:Wtree}
\widetilde{W}^{\mathrm{eff}}_{\mathrm{clas}}(a;m,\eps) = \frac{a^2}{4 \eps}.
\end{align}

The 1-loop contribution $\exp \widetilde{W}^{\mathrm{eff}}_{\mathrm{1-loop}}$ may be computed as a product of determinants of differential operators. There is a certain freedom in its definition due to the regularization of divergencies, which implies that it is only determined up to a phase \cite{Pestun:2007rz,Vartanov:2013ima}. For a distinguished choice of phase $\exp \widetilde{W}^{\mathrm{eff}}_{\mathrm{1-loop}}$ may be identified with the square-root of the product of two Liouville three-point functions in the Nekrasov-Shatashvili (or $c \to \infty$) limit. In this ``Liouville scheme'' the 1-loop contribution is of the form
\begin{align}\label{eq:W1-loop}
\widetilde{W}^{\mathrm{eff}}_{\mathrm{1-loop}}(a,m;\eps) &= \widetilde{W}^{\mathrm{eff}}_\mathrm{vector}(a;\eps) + \widetilde{W}^{\mathrm{eff}}_\mathrm{hyper}(a;m,\eps) 
\end{align}
with
\begin{align}
\widetilde{W}^{\mathrm{eff}}_\mathrm{vector}(a;\eps) &= - \frac{1}{2} \, \Upsilon (-a) - \frac{1}{2} \, \Upsilon (a)   \\
\widetilde{W}^{\mathrm{eff}}_\mathrm{hyper}(a;m,\eps) &= \frac{1}{2} \sum_{l=1}^4 \Upsilon \left(\frac{\eps+a+m_l}{2} \right) +  \frac{1}{2} \sum_{l=1}^4  \Upsilon \left(\frac{\eps-a+m_l}{2}\right),  
\end{align}
where 
\begin{align}
\Upsilon\left( x \right) = \int_\frac{1}{2}^{ x } du \log \frac{\Gamma(u)}{\Gamma(1-u)}.
\end{align}

The instanton contributions may be written as a sum over Young tableaux \cite{Nekrasov:2002qd,Nekrasov:2003rj}. In particular, the 1-instanton contribution is given by
\begin{align}\label{eq:1-inst}
\widetilde{W}^{\mathrm{eff}}_1(a;m,\eps) &= \prod_{l=1}^4 \frac{(a + m_l +\eps)}{16a(a+\eps)}  + \prod_{l=1}^4  \frac{(a - m_l -\eps)}{16 a(a-\eps)}.
\end{align}

\end{example}

The effective twisted superpotential $\widetilde{W}^\mathrm{eff}(a;m,q,\eps)$ not only characterizes the low energy physics of the theory $T_\eps$. According to the philosophy of \cite{Nekrasov:2009rc}, it may also be identified with the Yang-Yang function governing the spectrum of a quantum integrable system. This quantum integrable system is the quantization of the classical algebraic integrable system describing the low energy effective theory of the four-dimensional $\mathcal{N}=2$ theory $T$. The deformation parameter $\eps$ plays the role of the complexified Planck constant. For theories of class $\cS$ it is thus a quantization of a Hitchin system associated to $C$. 

 Nekrasov, Rosly and Shatashvili \cite{Nekrasov:2011bc} found yet a different, geometric, interpretation of $\widetilde{W}^{\mathrm{eff}}$ as a generating function ${W}^\mathrm{oper}$ of the space of so-called opers on the surface $C$.\footnote{This geometric description of the Yang-Yang function was found independently by Teschner as the $c \to \infty$ limit of Virasoro conformal blocks \cite{teschner2011}. The two are related by the AGT correspondence \cite{Alday:2009aq}. } Let us try to motivate this next.

\subsection{Nekrasov-Rosly-Shatashvili correspondence}

The Nekrasov-Rosly-Shatashvili correspondence has its roots in the close relation between theories of class $\cS$ and Hitchin systems.  

Suppose $T$ is a theory of class $\cS$.  If we compactify $T$ further down to three dimensions on a circle of radius $R$, the resulting theory has an effective description at low energies as a three-dimensional $\N=4$ sigma model, whose target space is the moduli space $\cM$ of solutions $(A,\varphi)$ to the Hitchin equations (with suitable boundary conditions at the punctures)
\begin{align}
F + R^2 [ \varphi, \overline{\varphi} ] &=0 \notag\\
\overline{\partial}_A \varphi &= 0 \\
\partial_A \overline{\varphi} &= 0.  \notag
\end{align}
on $C$. Here, $A$ is a $G$-connection in a topologically trivial $G$-bundle $E$ (endowed with a holomorphic structure via $\overline{\partial}_{A}$) on $\overline{C}$ and $\varphi \in \mathrm{End}(E) \otimes K_C$ the Higgs field  with appropriate boundary conditions at the punctures. Furthermore, $\overline{\varphi}$ denotes the hermitian conjugate of $\varphi$ with respect to a metric for which $A$ is the Chern connection. 

Hitchin's moduli space $\cM$ is equipped with a natural hyperkahler structure. In particular, this means that it has a $\PP^1$ worth of complex structures $J^{(\zeta)}$, parametrized by $\zeta \in \PP^1$, and a holomorphic symplectic form $\Omega_\zeta$ that is holomorphic at each fixed $\zeta$. 

The moduli space $\cM^{\zeta=0}$ can be identified with the moduli space of Higgs bundles (or its complex conjugate for $\zeta=\infty$). Higgs bundles are tuples $(E,\overline{\partial},\varphi)$ where $(E,\overline{\partial})$ is a holomorphic vector bundle and $\varphi$ is the Higgs field as above.

The Hitchin fibration $\cM^{\zeta=0} \to \mathbf{B}$ is a proper holomorphic map obtained by mapping the Higgs bundle
\begin{align}
(E,\overline{\partial},\varphi) \mapsto \det(x-\varphi)
\end{align} 
to the characteristic polynomial of $\varphi$. This gives the moduli space of Higgs bundles the structure of an integrable system. For $G=SU(K)$, the characteristic polynomial $\det(x-\varphi)$ determines a $K$-fold ramified covering 
\begin{align}
\Sigma &=  \left\{ x \in T^*C \,\, \vline \, \, \det(x-\varphi)=0\right\}\subset T^{*}C
\end{align}
over $C$. 

The spectral curve $\Sigma$ is also known as the Seiberg-Witten curve, and the tautological 1-form $\lambda$ pulled back to $\Sigma \subset T^*C$ is called the Seiberg-Witten differential. The base $\mathbf{B}$ of the Hitchin fibration parametrizes the Coulomb vacua of the four-dimensional $\N=2$ theory $T$.

The nonabelian Hodge correspondence identifies the Hitchin moduli space $\cM^\zeta$, for $\zeta \in \mathbb{C}^*$, with the moduli space of flat $SL(K,\C)$-connections on $C$. Indeed, given a solution of the Hitchin equations $(A,\varphi)$ and $\zeta \in \mathbb{C}^*$, we can form a flat $SL(K,\C)$-connection 
\begin{align}\label{eqn:Hitchinconn}
\mathcal{A} = \frac{R}{\zeta} \varphi + A + R \zeta \overline{\varphi}.
\end{align}
We denote $\cM^{\zeta=1}=\cM_\mathrm{flat}(C,SL(K))$. 

The corresponding moduli space of flat connections $\cM_\mathrm{flat}(C,SL(K))$ is holomorphic symplectic, and furthermore supports a distinguished complex Lagrangian subspace, the space $\mathbf{L}$ of $SL(K)$ \emph{opers} on $C$ \cite{beilinson1991quantization}.

An $SL(K)$ oper $(E, \{ E_i\}, \nabla)$ is defined as a rank $K$ holomorphic vector bundle $E$ over $\overline{C}$, equipped with a flat meromorphic connection $\nabla$ and a filtration $0=E_0 \subset \ldots \subset E_{K-1} \subset E_K = E$ satisfying \cite{2005math......1398B,BenZvi:1999kx}
\begin{enumerate}[(i)]
\item $\nabla E_i \subset E_{i+1} \otimes \Omega^1_C(D)$, where $D$ is the divisor of poles;
\item the induced maps $\overline{\nabla}: E_i/E_{i+1} \to (E_{i+1}/E_i) \otimes \Omega^1_C(D)$ are isomorphisms;
\item $E$ has trivial determinant (with fixed trivialization) and $\nabla$ induces the trivial connection on it.
\end{enumerate}
It can be shown that any $(E,\nabla)$ admits at most one oper structure, and thus we can indeed identify opers with a subspace of  $\cM_\mathrm{flat}(C,SL(K))$.

More concretely, any $SL(K)$ oper can locally be written as a $K$th order linear differential operator
\begin{align}\label{eqn:Doper}
\mathbf{D} = \partial_z^K + t_2(z) \, \partial_z^{K-2} + \ldots + t_K(z),
\end{align}
whose $(K-1)$th derivative vanishes.
More precisely, we consider families of $SL(K)$-valued $\eps$-opers whose coefficients are dependent on the complex parameter $\eps$. These may be obtained in the conformal limit $R,\zeta \to 0$, while $R/\zeta = \eps$ is kept fixed, of the family of flat connections~(\ref{eqn:Hitchinconn}) \cite{Gaiotto:2014bza,2016arXiv160702172D}.

Let us choose a Darboux coordinate system on $\cM_\mathrm{flat}(C,SL(K))$, say  $x_i, y_i$ with 
\begin{align}
\{ x_i, y_j \} = \delta_{ij}. 
\end{align}
Since the opers on $C$ define a complex Lagrangian submanifold of $\cM_\mathrm{flat}(C,SL(K))$, we can guarantee they posses a \emph{generating function} in this coordinate chart. This function $W^\mathrm{oper}$ is defined through the equation 
\begin{align}
y_i = \frac{\partial W^\mathrm{oper} (x,\eps)}{\partial x_i}.
\end{align}
and determined uniquely up to a constant in $x$.

Let us now go back to the beginning of this introduction, and consider $T$ in the four-dimensional background
\begin{align}\label{eqn:partOmega}
M=D^2 \times \R^2,
\end{align}
where $D^2$ is topologically a disk with the cigar metric $ds^2 = dr^2 + f(r) d \phi^2$. Here $f(r) \sim r^2$ for $r \to 0$ and $f(r) \sim R^2$ for $r \to \infty$, for some constant $R>0$. We should think of $D^2$ as a ``cigar'', a degenerate $S^1$ fibration over the positive axis $\mathbb{R}_+$, parametrized by $r>0$. Suppose we furthermore turn on a $\Omega$-deformation with complex parameter $\eps$, with the dimension of mass, corresponding to the isometry generated by $\partial/\partial \phi$.  

The resulting theory $T_\eps$ similarly preserves a $\mathcal{N}=(2,2)$ super-Poincare algebra, and is described by the same effective twisted superpotential $\widetilde{W}^{\mathrm{eff}}(x;m,q,\eps)$. Furthermore, the Omega-deformation can be undone away from the tip of the cigar, at $r=0$, in exchange for a field redefinition \cite{Nekrasov:2010ka}. 

If compactified to three dimensions along the $S^1$-fiber of the cigar $D^2$, the resulting theory may be studied in the infrared limit as a three-dimensional $\N=4$ sigma model with worldsheet $\mathbb{R}_+ \times \mathbb{R}^2$ into the Hitchin moduli space $\cM$. The boundary condition at $r=0$ is known to be specified by the space of opers $\mathbf{L}$ \cite{Nekrasov:2010ka}.

Nekrasov, Rosly and Shatashvili proposed that, as a consequence of this picture, 
\begin{align}\label{eqn:NRScorr}
\boxed{
\widetilde{W}^{\mathrm{eff}}(a;m,q,\eps) = W^\mathrm{oper}(a;m,q,\eps)
}
\end{align}
when we identified the Darboux coordinates $x_i$ with the two-dimensional scalars $a_i$ \cite{Nekrasov:2011bc}. More precisely, they studied this conjecture for $K=2$, where they introduced a particular Darboux coordinate system on $\cM_\mathrm{flat}(C,SL(2))$, which we will refer to as the NRS Darboux coordinates. They found that the correspondence~(\ref{eqn:NRScorr}) holds provided the generating function of the space of $SL(2)$ opers is expressed in the NRS Darboux coordinates.

\subsection{Summary of results}

In this paper we refine the methods to verify the NRS correspondence for $SU(2)$ gauge theories and find the ingredients to extend the NRS correspondence to any superconformal theory of class $\mathcal{S}$. That is, we find a way to construct the NRS Darboux coordinates, to describe the relevant spaces of opers, and to compute the generating functions of these spaces of opers. Our main two examples are the superconformal $SU(2)$ theory with four hypermultiplets and the superconformal $SU(3)$ theory with six hypermultiplets. In these examples we calculate the generating function $W^\mathrm{oper}(a;m,q)$ analytically in a perturbation in $q$ and compare to the known effective twisted superpotential $\widetilde{W}^{\mathrm{eff}}(a;m,q)$. \\

The first part of this paper describes the realization of the NRS Darboux coordinates as spectral coordinates through the abelianization method \cite{Gaiotto:2012rg,Hollands:2013qza}, and the construction of the desired generalization of the NRS Darboux coordinates in higher rank. 

Given a spectral network $\cW$ on $C$ and a generic $SL(K)$ flat connection $\nabla$ on $C$, together with some ``framing" data, abelianization is a way of bringing the flat connection $\nabla$ in an almost-diagonal form, such that it may be lifted to a $GL(1)$ connection on the spectral cover $\Sigma$. The spectral coordinates attached to $\nabla$ can then be read off as the abelian holonomies along the 1-cycles on $\Sigma$. 

The spectral networks that feature in this paper are a higher rank generalization of the Fenchel-Nielsen networks introduced in \cite{Hollands:2013qza}. They are dual to a pants decompositions of $C$ and may be generated by a generalized Strebel condition, which we formulate around equation~(\ref{eqn:strebelconstraint}). If the Riemann surface $C$ is built out of three-punctured spheres with one minimal and two maximal punctures, by gluing the maximal punctures, there is an essentially unique generalized Fenchel-Nielsen network. We call this a generalized Fenchel-Nielsen network of length-twist type.

The relevant moduli space $\cM^{\mathcal{C}}_\mathrm{flat}(C,SL(K))$ is the moduli space of flat connections on $C$ with fixed conjugacy classes at each puncture. We require that each conjugacy class is semisimple, with $K$ distinct eigenvalues for a maximal puncture and $K-1$ equal eigenvalues for a minimal puncture (and more generally, a partition of $K$ eigenvalues corresponding to a puncture labeled by any Young diagram). 

Abelianization for generalized Fenchel-Nielsen networks is a generalization of abelianization for Fenchel-Nielsen networks as described in \cite{Hollands:2013qza}. The framing data for generalized Fenchel-Nielsen network of length-twist type consists of an ordered choice of eigenlines at each puncture and pants curve. We spell out the resulting abelianization map in our two main examples and argue that it is 1-1. 

Any Fenchel-Nielsen network comes with two ``resolutions", representing how we think of the walls as infinitesimally displaced. Let us denote the spectral coordinates for the British resolution as $X^{+}_i, Y^+_i$ and the spectral coordinates for the Americal resolution as $X^{-}_i, Y^-_i$. In \cite{Hollands:2013qza} it was already established that both sets of spectral coordinates are examples of exponentiated complexified Fenchel-Nielsen length-twist coordinates (for a general choice of twist). In this paper we additionally find that the exponentiated NRS Darboux coordinates can be realized as ``averaged'' spectral coordinates
\begin{align}
X_i = X^{+}_i = X^{-}_i \quad \mathrm{and}  \quad Y_i = \sqrt{ Y_i^+  Y_i^- }.
\end{align}
which in particular fixes the twist ambiguity present in their definition.

Similarly, abelianization for generalized Fenchel-Nielsen networks leads to the desired higher rank versions of the NRS Darboux coordinates. We compute the associated trace functions in our $SU(3)$ example in \S\ref{section:spectral-monodromy}. \\

In the second part of the paper we describe the relevant spaces of opers  
\begin{align}
\mathbf{L} \subset \cM_\mathrm{flat}^\mathcal{C}(C,SL(K))
\end{align}
and obtain an explicit description of its generating function in the (generalized) NRS Darboux coordinates. 

We find that the opers associated to a theory of class $\cS$ with regular defects are locally described as Fuchsian differential operators with fixed semi-simple conjugacy classes at the punctures. Whereas for a surface $C$ with only maximal punctures there are no further constraints, the space of opers on a surface $C$ with other types of regular punctures is obtained by restricting the local exponents at the punctures, while keeping the conjugacy matrices semi-simple. This is analogous to the way that the space of differentials for a surface $C$ with arbitrary regular punctures may be obtained from the space of differentials for the surface $C$ with only maximal punctures, although the condition is different.

In particular, this implies that the locus of opers for the superconformal $SU(2)$ theory coupled to four hypers is described by the family of Heun's opers, characterized by the differential equation~(\ref{eqn:oper4-punctured}), whereas the locus of opers for the superconformal $SU(3)$ theory coupled to six hypers is described by the family of ``generalized Heun's opers'', characterized by the differential equation~(\ref{eqn:genHeun}). These families reduce in the limit $q \to 0$ to the hypergeometric and generalized hypergeometric oper, respectively.

We describe how to calculate the monodromy representation explicitly for the family of (generalized) Heun's opers as a perturbation in the parameter $q$, and compute the result up to first order corrections in $q$. This is a generalization of the leading order computations of \cite{Lay4347,Kazakov2001}, and a non-trivial extension of the work of \cite{Menotti:2014kra} which computes the monodromy matrix around the punctures at $z=0$ and $z=q$ in a perturbation series in $q$. The computation may be generalized to any family of opers that depends on a small parameter.

We then calculate the generating function $W^\mathrm{oper}(a;m,q)$ in the (generalized) NRS Darboux coordinates by comparing the monodromy representation for the opers to the monodromy representation in terms of the spectral coordinates. For the superconformal $SU(2)$ theory with $N_f=4$ we find 
\begin{align}
W^\mathrm{oper}(a;m,q) &= W^\mathrm{oper}_{\mathrm{clas}}(a;m) \log q  + W^\mathrm{oper}_{\mathrm{1-loop}}(a;m)  +  W^{\mathrm{oper}}_{\mathrm{1}}(a;m)\, q + \mathcal{O}(q^2),\label{eq:opergeneratingfunction}
\end{align}
where the classical and the 1-loop contribution are computed in equation~(\ref{eqn:leadingW}), and the 1-instanton contribution in equation~(\ref{eqn:1instanton}). For the superconformal $SU(3)$ theory with $N_f=6$ we find a similar expansion, where the classical and the 1-loop contribution are computed in equation~(\ref{eqn:leadingW3}).

We find that $ W^{\mathrm{oper}}_{\mathrm{1-loop}}(a;m)$ in the $SU(2)$ example equals the field theory expression~(\ref{eq:W1-loop}). This computation is similar to and in agreement with the computation in \cite{Vartanov:2013ima}. Furthermore, we find that the 1-instanton correction $W^{\mathrm{oper}}_{\mathrm{1}}(a;m)$ is equal to (\ref{eq:W1-loop}), the four-dimensional 1-instanton correction in the Nekrasov-Shatashvili limit $\eps_2 \to 0$.

The interpretation of generating function $W^\mathrm{oper}(a,m,q)$ in the $SU(3)$ example is similar. In particular, $\exp W^{\mathrm{oper}}_{\mathrm{1-loop}}(a;m)$ computes the square-root of the product of two Toda three-point functions with one semi-degenerate primary field in the Nekrasov-Shatashvili limit.

We conclude that our computation of the generating function of opers $W^\mathrm{oper}(a;m,q)$, expressed in the generalized Nekrasov-Rosly-Shatashvili Darboux coordinates, indeed agrees with the known effective twisted superpotential $\widetilde{W}^{\mathrm{eff}}(a;m,q)$. Particularly interesting is that, while the computation of the generating function $W^\mathrm{oper}(a;m,q)$ is a perturbation series in $q$, it is exact in $\eps$.\\

Given an $SL(2)$ $\eps$-oper $\nabla^\mathrm{oper}(\eps)$ there is yet another method to compute its monodromy representation. This is sometimes referred to as the exact WKB method (see \cite{exactWKB} for a good introduction). In the last part of this paper we compare abelianization to the exact WKB method. 

We argue that the monodromy representation for the oper $\nabla^\mathrm{oper}(\eps)$ computed using the abelianization method is equal to its monodromy representation computed using the exact WKB method, when the spectral network is chosen to coincide with the Stokes graph, and with an appropriate choice of framing data. In this correspondence the so-called Voros symbols are identified with the spectral coordinates.

As a consequence it follows that the spectral coordinates $X_\gamma$, when evaluated on the $\eps$-oper $\nabla^{\mathrm{oper}}(\eps)$, have good WKB asymptotics in the limit $\eps \to 0$. In this limit $X_\gamma$ is computed by what is sometimes called the quantum period $\Pi_\gamma(\eps)$ associated to $\nabla^{\mathrm{oper}}(\eps)$. The asymptotic expansion in $\eps$ of the generating function $W^\mathrm{oper}(\eps)$ may thus be simply found from the equation
\begin{align}
\log \Pi_B= \frac{\partial W^\mathrm{oper}(\Pi_A, \eps)}{\partial \log \Pi_A }.
\end{align}
This relates the Nekrasov-Rosly-Shatashvili correspondence to other approaches for computing the effective twisted superpotential \cite{Mironov:2009uv}.

We emphasize though that while the quantum periods are not particularly sensitive to the choice of Stokes graph, the exact resummed expressions are. In particular, the exact expression for the twisted effective superpotential $\widetilde{W}^{\mathrm{eff}}(a,\eps)$ can only be found by applying the exact WKB method to the oper $\nabla^\mathrm{oper}(\eps)$ where the phase of $\eps$ (and of other parameters) is chosen such that the corresponding Stokes graph is of Fenchel-Nielsen type. The results~(\ref{eq:opergeneratingfunction}) then show that there are no non-perturbative corrections to $\widetilde{W}^{\mathrm{eff}}(\eps)$, in agreement with \cite{Nekrasov:2003rj}.\\

Part of the importance of our approach lies in the fact that the geometric problem makes perfect sense regardless of the types or number of defects present, circumventing the need for a Lagrangian description of the theory to determine the superpotential. Thus, while the superpotentials in examples we study are well-known, our perspective suggests the possibility of going further and analyzing non-Lagrangian theories.

From a mathematical perspective, we have given a description of the monodromy representation of opers on a (punctured) surface $C$ in a series expansion in its complex structure parameters and verified a prediction for the generating function of a particular interesting Lagrangian subspace inside the moduli space of flat connections in a few important examples. It is reasonable to expect that known gauge theory results can predict their description in more general cases.

\subsection{Outline of the paper}

This paper is organized as follows. 

We start in \S\ref{sec:SWgeometry} with a brief review of Seiberg-Witten geometry to set notations and to introduce our two main examples, the superconformal $SU(2)$ theory with $N_f=4$ and the superconformal $SU(3)$ theory with $N_f=6$. 

In \S\ref{sec:K=3FNnetwork} we recall the definition of a spectral network and of a Fenchel-Nielsen type spectral network. We find a higher rank generalization of  Fenchel-Nielsen networks and relate this to a generalized Strebel condition on the differentials. We use this to generate examples of generalized Fenchel-Nielsen networks on the four-punctured sphere. 

In \S\ref{sec:length-twist} we define the moduli space of flat connections $\cM^{\mathcal{C}}_\mathrm{flat}(C,SL(K))$ with fixed conjugacy classes $\mathcal{C}$ at the punctures, and specify these conjugacy classes for the different kinds of punctures relevant to this paper. We then review the definition of the Fenchel-Nielsen length-twist coordinates, and generalize these length-twist coordinates to higher rank. 

In \S\ref{sec:nonab} and \S\ref{sec:ab-l-t} we show how to realize the higher rank length-twist coordinates as spectral coordinates through the abelianization method. Section~\ref{sec:nonab} contains some general background on abelianization, whereas \S\ref{sec:ab-l-t} focuses on our two main examples. In particular, we show in the latter section that the abelianization and non-abelianization mappings are 1-1. We then collect the resulting monodromy representations in terms of higher rank length-twist coordinates in \S\ref{section:spectral-monodromy}. 

Section~\ref{sec:opers} starts off with a gentle
introduction to opers, after which we introduce the relevant families of opers to our main examples. This is the family of Heun's opers for the superconformal $SU(2)$ theory and the family which we term generalized Heun's opers for the superconformal $SU(3)$ theory. In \S\ref{sec:monopers} we compute the monodromies of these opers in a perturbation series in the complex structure parameter $q$. 

The final computations of the generating function of opers are contained in \S\ref{sec:genfunopers}. Indeed, we find that in our two example the generating function agrees with the effective twisted superpotential in an expansion in the parameter $q$, up to a spurious factor that does not depend on the Coulomb parameter $a$. 

In \S\ref{sec:WKB} we comment on the relation of the abelianization method with the exact WKB method and relate the NRS conjecture to other proposals for computing the effective twisted superpotential.

\subsection*{Notational conventions}

Throughout, by punctured surface we will mean either the compact surface $\overline{C}$ equipped with the corresponding divisor of poles, or the noncompact $C$ (i.e. $\overline{C}$ with points removed) -- it should be clear from the context which is meant. $D$ will always denote the reduced divisor of poles. The number $n$ always refers to the number of punctures.

Mass parameters will usually be omitted from notation, but are present everywhere and assumed fixed from the outset, and satisfying necessary genericity assumptions.

We will sometimes shorten the surface equipped with defects $(C,\mathcal{D})$ to $C_{z_1, \ldots, z_n}$, leaving masses implicit, and underlining so-called ``minimal" punctures.

$K_C$ will always denote the canonical bundle of the compactified curve, and we always assume from the outset that we have fixed a choice of $K_C^{1/2}$.

\subsection*{Acknowledgements}

We thank Greg Moore, Philipp R\"uter, Joerg Teschner, and in particular Andrew Neitzke for very helpful discussions. LH's work is supported by a Royal Society Dorothy Hodgkin fellowship. OK's work is supported by a Royal Society Research Grant and an NSERC PGS-D award.

\section{Class S geometry}\label{sec:SWgeometry}

Fix a positive integer $K$, sometimes called the ``rank'', and a possibly punctured Riemann surface $C$. We equip the Riemann surface with a collection 
\begin{align}
\mathcal{D} = \{ \mathcal{D}_l \}
\end{align}
of ``regular defects" at each puncture $z_l$. Each such defect $\mathcal{D}_l$ is labeled by a Young diagram $Y_l$ with $K$ boxes and a collection of compatible ``mass'' parameters 
$(m_{l,i})_{i=1,\ldots,K}$ with $\sum_{i=1}^K m_{l,i}=0$. The height of each column in the Young diagram encodes the multiplicities of coincident mass parameters.

To each choice $(K,C,\mathcal{D})$ corresponds a four-dimensional $\mathcal{N}=2$ superconformal field theory $S[A_{K-1},C,\mathcal{D}]$ of type $A_{K-1}$ with defects $\mathcal{D}$ of ``regular" type, which will henceforth denote more compactly as 
\begin{align}
T = T_K[C,\mathcal{D}].
\end{align}
This is a so-called ``theory of class $\mathcal{S}$'' \cite{Gaiotto:2009we}. 

The surface $C$ is known as the UV curve of the theory~$T$. It encodes the microscopic definition of the theory. 
Complex structure parameters correspond to gauge couplings of the theory, whereas the data at the punctures encodes flavour symmetries. The flavour symmetry associated to a puncture labeled by the Young diagram $Y$ is
\begin{align}
S[U(n_1) \times \cdots \times U(n_k)],
\end{align}
where $n_1, \ldots, n_k$ count columns of $Y$ with the same height.

The Coulomb branch $\mathbf{B}=\mathbf{B}(T)$ of the theory $T$ is equal to the corresponding Hitchin base, parametrized by tuples 
\begin{align}
(\varphi_2, \ldots, \varphi_K) \in \mathbf{B} = \bigoplus_{i=2}^K H^0(C,K_C(D)^{\otimes i})
\end{align}
of $k$-differentials $\varphi_k$ on $C$, with regular singularities of the appropriate pole structure at the punctures. Here 
$D=\sum_{i=1}^{n}{1\cdot z_i}$
denotes the divisor of punctures, and the residues are taken to be fixed at each puncture.

Thus, $\mathbf{B}$ is an affine space for the space of differentials with strictly lower order poles (possibly with restrictions as described below). Concretely, $\varphi_k$ is given locally by 
\begin{align}
\varphi_k(z) = u_k(z) dz^{\otimes k}
\end{align}
where the function $u_k(z)$ has at most a pole of order $k$ at each puncture.

To each tuple $(\varphi_2, \ldots, \varphi_K)$ we can associate a spectral curve $\Sigma \subset T^*C$, which is defined by the equation
\begin{align}\label{eqn:swcurveSU}
\lambda^{K} + \lambda^{K-2}\varphi_{2} + \ldots +\varphi_{K} = 0,
\end{align} 
where $\lambda$ is the tautological 1-form on $T^*C$, locally given by $\lambda = w dz$. 

The spectral curve~$\Sigma$ is known as the Seiberg-Witten curve, and the restriction of $\lambda$ to $\Sigma$ is called the Seiberg-Witten differential. The $K$ residues of $\lambda$ at each puncture $z_i$ are fixed to be the mass parameters $m_{i,j}$. The Seiberg-Witten curve is a possibly ramified $K$-fold branched covering 
\begin{align}
\pi: \Sigma \rightarrow C.
\end{align}
Together with the Seiberg-Witten differential this curve captures the low-energy data of the theory $T$.
	
Just as any Riemann surface $C$ can be glued out of three-punctured spheres, the basic building blocks of theories of class $\mathcal{S}$ are those corresponding to three-punctured spheres. The possible building blocks are specified by the integer $K$ and the choice of defects. 

Some building blocks have an elementary field theory description in terms of well-known matter multiplets of the $\mathcal{N}=2$ algebra, others are described as intrinsically strongly coupled (non-Lagrangian) SCFTs. 

In particular, none of the building blocks involve any gauge multiplets. These are only introduced when gluing the three-punctured spheres. On the level of the $\mathcal{N}=2$ theory this corresponds to gauging the corresponding flavour symmetry groups. 

Our main examples in this paper are the theories $T_K[C,\mathcal{D}]$ where $C$ is the four-punctured sphere $\mathbb{P}^1_{0,q,1,\infty}$, with $q \in \mathbb{C}\setminus\{0,1\}$, and the rank of the bundle is either $K=2$ or $K=3$. In the following we briefly review their geometry.

\subsection{$K=2$}

For $K=2$ there is only one possible regular defect, labeled by the Young diagram 
\begin{align}
\young(\hfil\hfil)
\end{align}
consisting of one row with two boxes. The mass parameters corresponding to this defect are generic, with $m_{l,2} = -m_{l,1}$. In the corresponding four-dimensional quantum field theory this decoration corresponds to an $SU(2)$ flavour symmetry group.  In particular, this implies that there is a single building block $T_2[\PP^1_{0,1,\infty}]$.  

\begin{example} 
The theory $T_2[\PP^1_{0,1,\infty}]$ describes a half-hypermultiplet in the trifundamental representation of $SU(2)_0 \times SU(2)_1 \times SU(2)_\infty$. Its Coulomb branch $\mathbf{B}$ is a single point corresponding to the quadratic differential 
\begin{equation}\label{eqn:K=2phi2}
\varphi_2(z) =  - \frac{m_\infty^2 z^2 - (m_\infty^2 + m_0^2 - m_1^2)z + m_0^2}{4 z^2 (z-1)^2 } (dz)^2, 
\end{equation}
for fixed values of the parameters $m_0$, $m_1$ and $m_\infty$. The combinations $\frac{\pm m_0 \pm m_1 \pm m_{\infty}}{2}$ correspond to the (bare) masses.
\end{example}

\insfigscaled{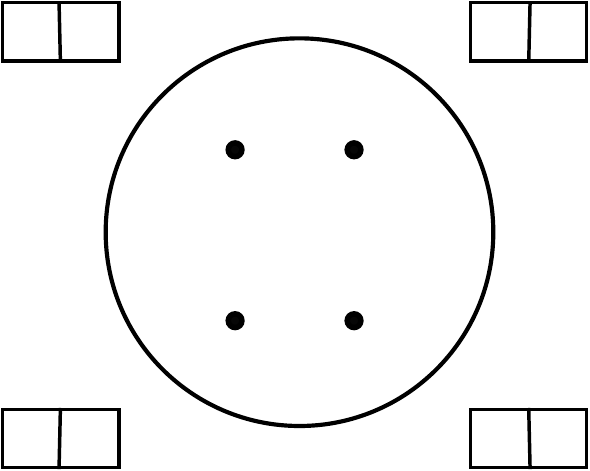}{0.8}{The UV curve for the theory $T_2[\mathbb{P}^1_{0,q,1,\infty}]$.}

Gauge fields are introduced by gluing three-punctured spheres. The corresponding complex structure parameters $q$ are identified with the gauge couplings $e^{2\pi i \tau }$. The limit $q \to 0$ corresponds to the weakly coupled description of the gauge theory at a cusp of the moduli space. For every pants cycle $\alpha$ there is a Coulomb parameter $a_0$, which is defined as the period integral $a_0=\oint_{A} \lambda$ along a lift $A$ of the pants cycle.

\begin{example}
The theory $T_2[\PP^1_{0,q,1,\infty}]$ corresponds to the superconformal $SU(2)$ gauge theory coupled to four hypermultiplets, see Figure~\ref{fig:4psphere}. Its Coulomb branch $\mathbf{B}$ is 1-dimensional and parametrized by the family of quadratic differentials 
\begin{align}\label{eqn:K=2phi24-punctured}
\varphi_2(z) &= - \Bigg( \frac{m_0^2}{4z^2} + \frac{m^2}{4(z-q)^2} + \frac{m_1^2}{4(z-1)^2} \\
& \quad + \frac{m_0^2+ m^2 + m_1^2-m_\infty^2}{4z(z-1)} - \frac{u}{z(z-q)(z-1)}  \Bigg)(dz)^2,  \notag  
\end{align}
where the parameter $u$ is free and the parameters $m_0$, $m$, $m_1$ and $m_\infty$ are fixed. The combinations $\frac{m_0 \pm m}{2}$ and $\frac{m_1 \pm m_\infty}{2}$ correspond to the (bare) masses of the four hypermultiplets. 

The corresponding Seiberg-Witten curve $\Sigma$ is (after compactifying) a genus one covering of $\mathbb{P}^1_{0,q,1,\infty}$ with four simple branch points. Let $A$ be the lift of the 1-cycle $\alpha$ going counterclockwise around the punctures at $z=0$ and $z=q$. The Coulomb parameter $a_0=a_0(u)$ is defined as the period integral $a_0=\oint_A \lambda$.

\end{example}

\subsection{$K=3$}\label{sec:SU3bifund}

For $K=3$ there are two types of punctures, which we will refer to as ``maximal'' and ``minimal'' punctures. For a maximal puncture $z_l$ the mass parameters $m_{l,i}$ are generic with $m_{l,1} \neq m_{l,2}$, whereas for a minimal puncture $m_{l,1}=m_{l,2}$. (Sometimes the maximal puncture is called a ``full'' puncture, and the minimal puncture a ``partial'' puncture.)

A \emph{maximal puncture} is labeled by the Young diagram 
\begin{align}
\young(\hfil\hfil\hfil)
\end{align}
consisting of one row with three boxes. In the corresponding quantum field theory this defect corresponds to an $SU(3)$ flavour symmetry group. 

A \emph{minimal puncture} is labeled by the Young diagram 
\begin{align}
\young(\hfil\hfil,\hfil)
\end{align}
consisting of one row with two boxes and one row with a single box. In the corresponding quantum field theory this decoration corresponds to a $U(1)$ flavour symmetry group.

In terms of the Seiberg-Witten differential, a maximal puncture at $z=z_l$ turns into a minimal puncture if it satisfies two requirements:
\begin{enumerate}[(i)]
\item Two of the masses at the puncture should coincide:
\begin{align}\label{eqn:bifmassconstr}
m_{l,1}=m_{l,2}=m_l.
\end{align}
\item The discriminant of 
\begin{align}\label{eqn:bifdiscrcontr}
\lambda^3 + (z-z_l)^2 \varphi_2 \, \lambda + (z-z_l)^3 \varphi_3
\end{align} 
should vanish up to order $(z-z_l)^2$. This enforces two simple branch points of type $(ij)$ of the covering to collide with the puncture at $z=z_l$. 
\end{enumerate}

\begin{figure} \centering \includegraphics[scale=0.8]{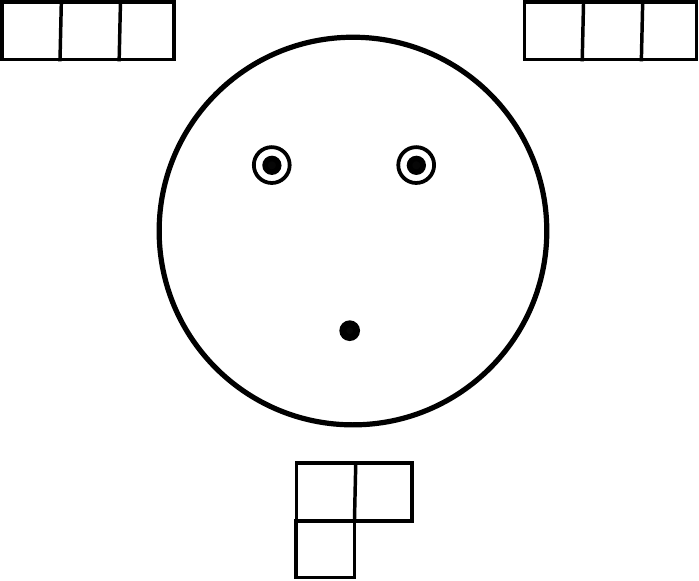} \hspace{20mm} \includegraphics[scale=0.8]{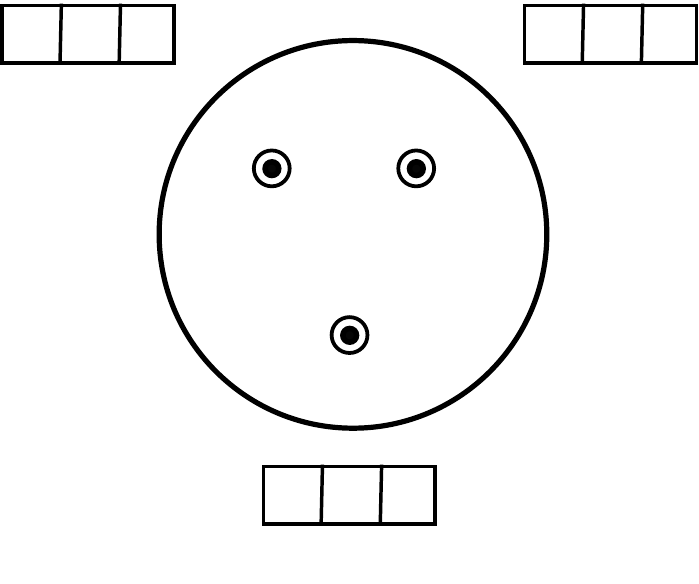} \caption{Left: the UV curve for the free bifundamental hypermultiplet $T_3[\mathbb{P}^1_{0,\underline{1},\infty}]$. Right: the UV curve for the non-Lagrangian $E_6$ theory $T_3[\mathbb{P}^1_{0,1,\infty}]$. } \label{fig:SU3theories}\end{figure}

\begin{example}
The theory $T_3[\PP^1_{0,1,\infty}]$ with three maximal punctures, see on the right of Figure~\ref{fig:SU3theories}, is the so-called $E_6$ Minahan-Nemeschansky theory. Microscopically its flavour symmetry group is $SU(3)_1 \times SU(3)_2 \times SU(3)_3$. In the low energy limit this group is enhanced to $E_6$. 

The Coulomb branch $\mathbf{B}$ of the theory $T_3[\PP^1_{0,1,\infty}]$ is described by the 1-dimensional family of differentials
\begin{align}
\varphi_2 &= \frac{c_\infty z^2 - (c_0-c_1+c_\infty) z + c_0}{z^2(z-1)^2} (dz)^2 \label{eqn:K=3phi2}\\
\varphi_3 &= \frac{d_\infty z^3 + u z^2 + (d_0+d_1-d_\infty-u) z - d_0}{z^3(z-1)^3} (dz)^3,\label{eqn:K=3phi3}
\end{align}
where $u$ is a free parameter, whereas the parameters $c_l$ and $d_l$ are fixed and can be written as combinations of $SU(3)_1 \times SU(3)_2 \times SU(3)_3$ mass parameters. If we choose 
\begin{align}
c_l &= \frac{1}{4}({-m_{l,1}^2-m_{l,1}m_{l,2}-m_{l,2}^2})\\
d_l &= \frac{1}{8}\left(m_{l,1} m_{l,2} (m_{l,1}+m_{l,2})\right)
\end{align}
then the residues at the punctures $z=l$ are $\{\frac{m_{l,1}}{2},\frac{m_{l,2}}{2},\frac{-m_{l,1}-m_{l,2}}{2}\}$, respectively. 

The Seiberg-Witten curve $\Sigma$ defines a 3-fold ramified covering over the UV curve $\PP^1_{0,1,\infty}$, with generically six simple branch points. This implies that $\Sigma$ is a punctured genus one Riemann surface. In contrast to weakly coupled gauge theories, the Seiberg-Witten curve has no distinguished A-cycle. 
\end{example}

As a matter of notation, let us henceforth write $(C,\mathcal{D})$ as $C_{z_1,\ldots z_n}$, and denote a minimal puncture by underlining the position of the puncture. Mass parameters are left implicit.

\begin{example} The theory $T_3[\PP^1_{0,\underline{1},\infty}]$ with two maximal and one minimal puncture, see on the left of Figure~\ref{fig:SU3theories}, corresponds to a free hypermultiplet in the bifundamental representation of $SU(3)_0 \times SU(3)_\infty$. 
We find its Coulomb vacua by applying the constraints~(\ref{eqn:bifmassconstr}) and~(\ref{eqn:bifdiscrcontr}) to the family of $T_3[\PP^1_{0,1,\infty}]-$differentials described in equation~(\ref{eqn:K=3phi2}) and~(\ref{eqn:K=3phi3}) at $z=1$. 
Concretely, the latter constraint implies that
\begin{align}\label{eqn:uconstraint}
u =  \left(\frac{m_1}{2}\right)^3 - d_0 -2 d_{\infty}  + \frac{m_1}{2} (c_0 - c_\infty).
\end{align} 
The Coulomb branch is thus reduced to a single point $\varphi^{\mathrm{bif}} = (\varphi_2^{\mathrm{bif}},\varphi_3^{\mathrm{bif}})$. 

The resulting Seiberg-Witten curve $\Sigma$ determines a 3-fold ramified covering of the UV curve $\PP^1_{0,\underline{1},\infty}$ with four simple branch-points. It is therefore a punctured genus zero surface.
\end{example}

These two examples provide the possible building blocks for $K=3$ theories \cite{Gaiotto:2009we,Chacaltana:2010ks}. Gauge fields can be introduced by gluing three-punctured spheres at maximal punctures. The gauge coupling corresponds to the complex structure parameter $q$, where the gluing is performed in a standard way according to the transition $z_1 z_2 = q$.

\begin{example} The theory $T_3[\PP^1_{0,\underline{q},\underline{1},\infty}]$ is the superconformal $SU(3)$ gauge theory coupled to $N_f=6$ hypermultiplets. It may be obtained by gluing two three-punctured spheres with two maximal and one minimal puncture. Its Coulomb branch $\mathbf{B}$ is parametrized by two parameters $u_{1}$ and $u_{2}$.

The explicit form of the differentials $\varphi_2$ and $\varphi_3$ can be obtained as before. First we write down the most generic quadratic and cubic differential with regular poles at the punctures. Eight of the twelve parameters are fixed by writing the residues at each punctures in terms of the mass parameters. Two more parameters are fixed by additional requirements at both minimal punctures, analogous to equation~(\ref{eqn:bifmassconstr}) and (\ref{eqn:uconstraint}). The resulting differentials can be written in the form  
\begin{align}
\varphi_2 &= \frac{c_0}{z^2} + \frac{c}{(z-q)^2} + \frac{c_1}{(z-1)^2} + \frac{c_\infty-c_0-c-c_1}{z(z-1)} + \frac{u_1}{z(z-q)(z-1)}\label{eqn:K=3phi24punctures2}\\
\varphi_3 &= \frac{d_0}{z^3} + \frac{d}{(z-q)^3}  + \frac{d_1}{(z-1)^3} + \frac{d_\infty- d_0 - d-d_1}{z(z-q)(z-1)} +\label{eqn:K=3phi24punctures3}
\\ & \quad + \frac{(1-q) (4 c_0 - 3 m^2 - 3 m_1^2 - 4 c_\infty)m_1}{8 z (z-1)^2 (z-q)}  + \frac{u_2}{z^2(z-q)(z-1)} \\
& \quad - \frac{u_1 }{z(z-1)^{2}(z-q)^{2}} \left(\frac{m_1}{2}(z-q)+\frac{m}{2}(z-1) \right) 
\end{align}
where $c, d$ are as above.

The resulting Seiberg-Witten curve $\Sigma$ is a genus two covering of $\mathbb{P}^1_{0,\underline{q},\underline{1},\infty}$ with eight simple branch points. The two Coulomb parameters $a_0^1$ and $a_0^2$ are defined as the period integrals $a_0^{1}=\oint_{A^{(1)}} \lambda$ and $a_0^{2}=\oint_{A^{(2)}} \lambda$ along two independent lifts of the pants cycle $\alpha$ to the Seiberg-Witten curve.

 \end{example}

\section{Generalized Fenchel-Nielsen networks}\label{sec:K=3FNnetwork}

Fix a pants decomposition of the punctured curve $C$. In this section we define a type of spectral network on $C$ that respects this pants decomposition, called a Fenchel-Nielsen network when $K=2$ and a generalized Fenchel-Nielsen network when $K>2$. We begin with a brief introduction to spectral networks. More precisely, we describe a subclass of networks known as WKB spectral networks. A more general definition can be found in \cite{Gaiotto:2012rg,Williams2016}.

Fix some phase $\vartheta \in \mathbb{R}/2 \pi \mathbb{Z}$ and a tuple $\varphi =(\varphi_2, \ldots, \varphi_K) \in \mathbf{B}$. Write the corresponding spectral curve $\Sigma$ in terms of a tuple $(\lambda_1, \ldots, \lambda_K)$ of meromorphic $1$-differentials on $C$ as
\begin{align}
\Sigma: \quad \lambda^K + \varphi_2 \, \lambda^{K-2} + \ldots + \varphi_K = \prod_{j=1}^K \,(\lambda - \lambda_j) =0.
\end{align}

We define an \emph{$ij$ trajectory (of phase $\vartheta$)} , for $i \neq j \in \{ 1, \ldots, K\}$, as an open path on $C$ such that 
\begin{align}\label{eq:ijtrajectory}
(\lambda_i - \lambda_j) (v) \in e^{i \vartheta} \mathbb{R},
\end{align}
for every nonzero tangent vector $v$ to the path. The \emph{WKB spectral network} $\cW_\vartheta(\varphi)$ consists of a certain collection of such $ij$-trajectories on $C$, as follows. 

Call any $ij$-trajectory that has an endpoint on a branch-point of the covering $\Sigma \to C$ a wall. We orient the wall such that it starts at the branch-point. Any other $ij$-trajectory that has its endpoint at the intersection of previously defined walls is another wall. We orient this wall such that it starts at the intersection. The spectral network $\cW_\vartheta (\varphi)$ is the collection of all walls. 

We label the walls as follows. The two sheets $i$ and $j$ of $\Sigma$ over a wall correspond to the two differentials $\lambda_i$ and $\lambda_j$. Given a positively oriented tangent vector $v$ to the wall, the quantity $e^{-i \vartheta} (\lambda_i - \lambda_j)(v)$ is real. If it is positive we label the wall by the ordered pair $ji$, and if negative we label the wall by $ij$.\footnote{As we will see in \S\ref{sec:WKBab}, this choice of labeling is motivated by WKB properties of the S-matrices.}

Generically, the spectral network in the neighbourhood of a simple branch-point of the covering $\Sigma \to C$ is depicted in Figure~\ref{fig:localbranchpoint}. In a neighbourhood of a simple intersection of walls the spectral network is illustrated in Figure~\ref{fig:intersection}. Generically, each wall ends at a puncture of $C$.

\insfigscaled{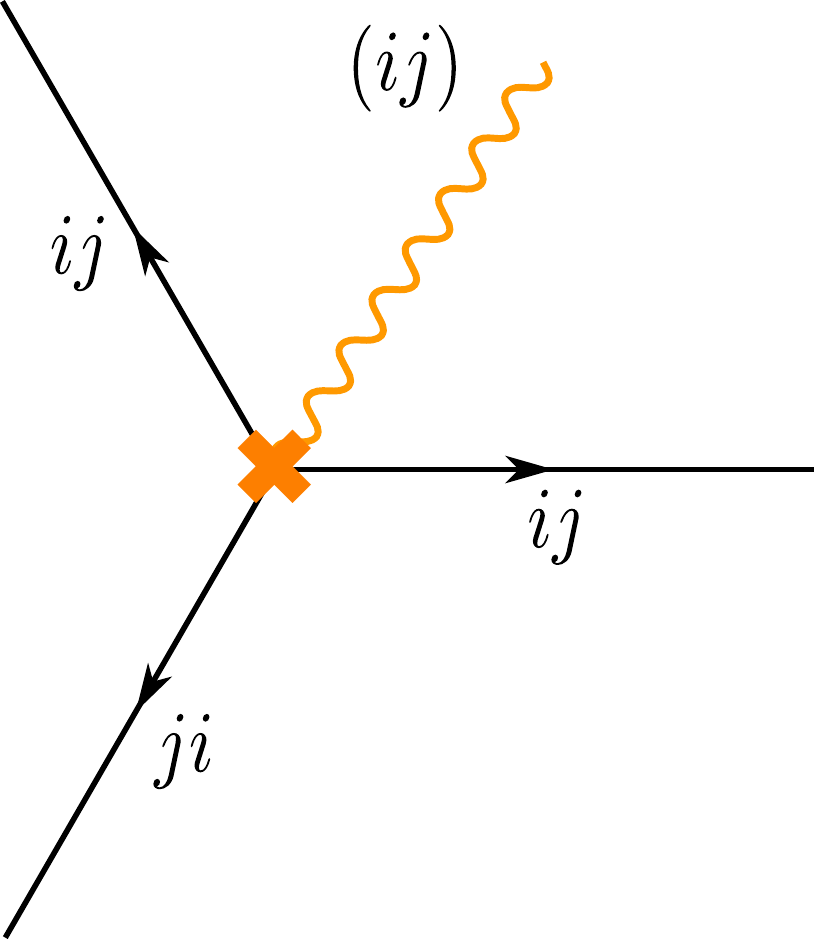}{0.45}{Configuration of walls around a simple branch-point.}

\insfigscaled{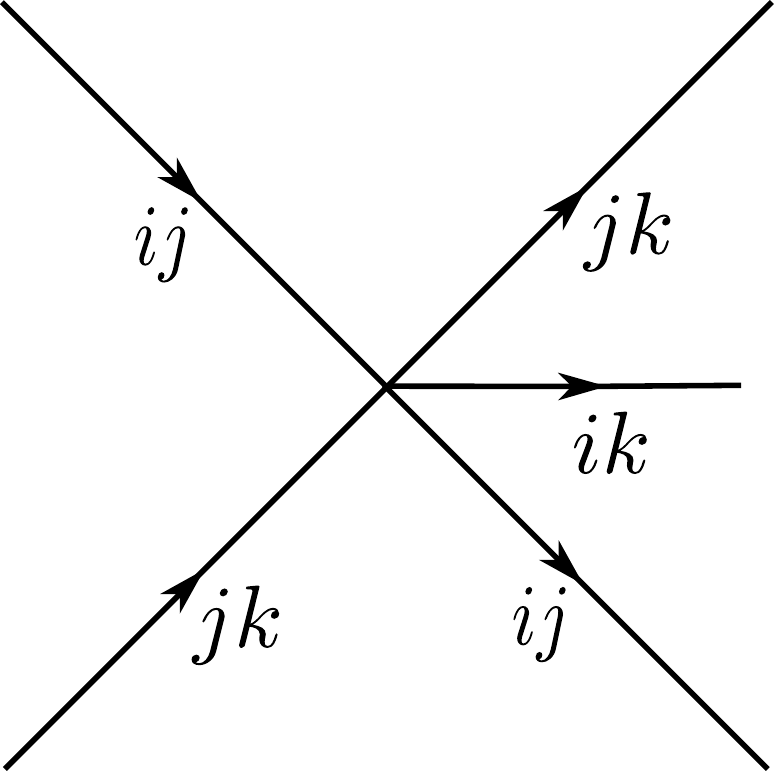}{0.45}{A wall with label $ik$ is born at the intersection of two walls with label $ij$ and $jk$.}

We decorate a puncture with incoming walls as follows. Each root $\lambda_i$ has a simple pole at the puncture with residue $m_i$. We decorate the puncture with an ordered tuple $i_1 \ldots i_K$ such that $\mathrm{Re}(e^{-i \vartheta} m_{i_j})> \mathrm{Re}(e^{-i \vartheta} m_{i_k})$ for each $j < k$. One then checks that the only walls which fall into the puncture are the ones whose labeling matches the decoration. 

At special values for the differentials $\varphi$ and the phase $\vartheta$ it might happen that two walls $ij$ and $ji$, with opposite orientations, overlap. This is illustrated in Figure~\ref{fig:doublewall}. We say that the locus where the two walls overlap is a \emph{double wall}. If there is at least one double wall, the spectral network must be decorated with a choice of a \emph{resolution}, which is either ``British'' or ``American''. We think of the resolution as telling us how the two constituents of a double wall are infinitesimally displaced from one another, and draw the walls as such.

\insfigscaled{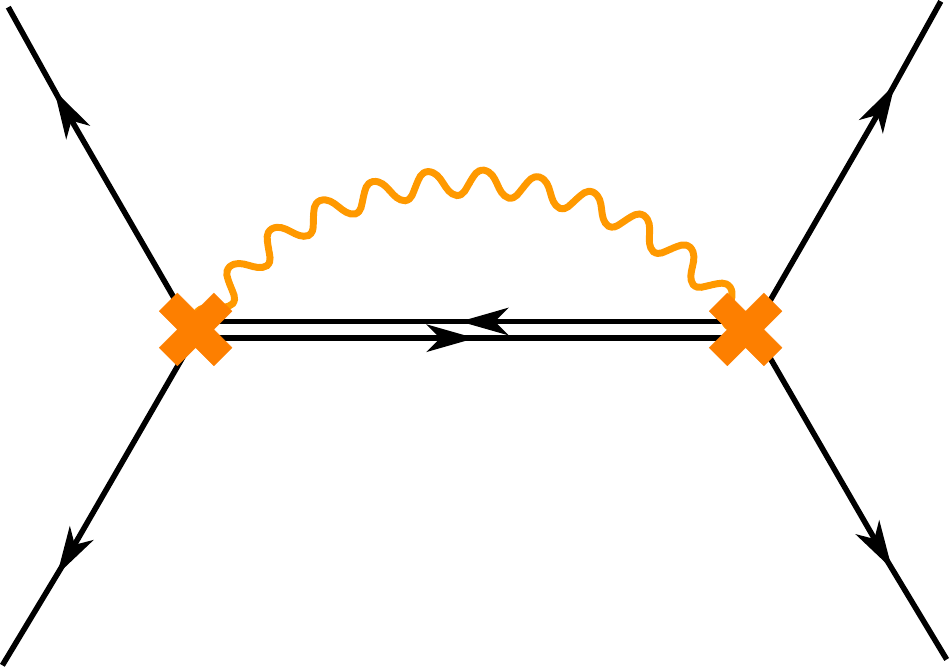}{0.55}{Local configuration with a double wall in the American resolution.}

We say that a spectral network $\cW_\vartheta (\varphi)$ is a \emph{Fenchel-Nielsen} network (for $K=2$) or \emph{generalized Fenchel-Nielsen} network (for $K>2$) if it consists of only double walls and respects some pants decomposition of $C$, i.e. that the restriction to every three-punctured sphere in the decomposition is itself a network of only double walls. In particular, for such networks each wall both begins and ends on a branch-point of the covering $\Sigma \to C$, and there are no incoming walls at any puncture. We will discuss the decoration at such punctures, as well as along the pants curves, later in this section.   

In \cite{Hollands:2013qza} it was found that when $K=2$ the corresponding differential $\varphi_2$ satisfies the Strebel condition. In the following we will argue that for $K>2$ there is a natural generalization of the Strebel condition.

Since by definition a Fenchel-Nielsen network respects a pants decomposition, we can glue it from Fenchel-Nielsen networks on the individual pairs of pants. In this section we analyze the possible Fenchel-Nielsen networks on the three-punctured sphere for $K=2$ and $K=3$ and detail the gluing procedure.  

Even though in the above we have fixed a complex structure on $C$ and described a spectral network in terms of the tuple $\varphi$ of differentials, we will later only be interested in the isotopy class of the spectral network on the topological surface $C$. We thus define two spectral networks $\cW$ and $\cW'$ to be equivalent if one can be isotoped into the other. 

\subsection{$K=2$}

Let $\varphi_2$ be a meromorphic quadratic differential on $\overline{C}$, holomorphic away from the punctures $z_l$. Locally such a differential is of the form 
\begin{align}
\varphi_2 = u(z) (dz)^2
\end{align}
It is well-known that given a phase $\vartheta$, the differential $\varphi_2$ canonically determines a singular foliation $\mathcal{F}_{\vartheta}(\varphi_2)$ on $C$. Its leaves are real curves on $C$ such that, if $v$ denotes a nonzero tangent vector to the curve,
\begin{align} 
e^{-2 i \vartheta} \varphi_2 (v^2) \in \mathbb{R}_+. 
\end{align}
The differential $e^{-2 i \vartheta} \varphi_2$ is called \emph{Strebel} if all leaves of the foliation $\mathcal{F}_\vartheta(\varphi_2)$ are either closed trajectories or saddle connections (i.e. trajectories that begin and end at a simple zero of $\varphi_2$). 

Suppose that the singular foliation $\mathcal{F}_\vartheta(\varphi_2)$ respects a given pants decomposition of the surface $C$. That is, suppose that each pants curve $\alpha_k$ is homotopic to a closed trajectory of $\mathcal{F}_\vartheta(\varphi_2)$. Then the Strebel condition implies that the period of $\sqrt{-\varphi_2}$ around each pants curve $\alpha_k$ as well as around a small loop $\gamma_l$ around each puncture $z_l$ has phase $\vartheta$, that is 
\begin{align}\label{eqn:strebel}
e^{-i \vartheta}  \oint_{\alpha_k} \sqrt{-\varphi_2} \in \mathbb{R} \quad \mathrm{and} \quad 
e^{-i \vartheta} \oint_{\gamma_l} \sqrt{-\varphi_2} \in \mathbb{R}.
\end{align}

Conversely, given any pants decomposition consisting of simple closed curves $\{ \alpha_k\}$ of a punctured Riemann surface $C$ and arbitrary $h_k>0$ and $m_j>0$, there is a unique Strebel differential $\varphi_2$ whose foliation consists of punctured discs centered at the punctures and characteristic annuli homotopic to $\alpha_k$, such that 
\begin{align}
 \oint_{\alpha_k} \sqrt{-\varphi_2} = h_k \quad \mathrm{and} \quad 
\oint_{\gamma_l} \sqrt{-\varphi_2}= m_j,
\end{align}
for a suitable choice of branch of the root $\sqrt{-\varphi_2}$ \cite{liu2008jenkins}.

As explained in \cite{Hollands:2013qza} a rank $K=2$ spectral network $\cW_\vartheta(\varphi_2)$ can be obtained from the critical locus of the singular foliation $\mathcal{F}_\vartheta(\varphi_2)$. The resulting network $\cW_\vartheta(\varphi_2)$ is Fenchel-Nielsen if and only if the foliation respects a given pants decomposition of $C$, has no leaves ending on punctures and only compact leaves. This is equivalent to saying that $e^{-2 i \vartheta} \varphi_2$ is a Strebel differential. 

\begin{example}
Recall from equation~(\ref{eqn:K=2phi2}) that any meromorphic quadratic differential $\varphi_2$ on the three-punctured sphere $\mathbb{P}_{0,1,\infty}$, with regular singularities and prescribed residues $-m_i^2$ can be written as  
\begin{equation}
\varphi_2 =  - \frac{m_\infty^2 z^2 - (m_\infty^2 + m_0^2 - m_1^2)z + m_0^2}{z^2 (z-1)^2 } \,(dz)^2.
\end{equation}
The above differential is a Strebel differential if and only if all parameters $m_l$ have the same phase $\vartheta - \frac{\pi}{2}$. Without loss of generality we can assume all $m_l$ are real and $\vartheta = \frac{\pi}{2}$. 

The isotopy class of the corresponding spectral network $\cW_\frac{\pi}{2}(\varphi_2)$ depends on the precise values of the parameters $m_0$, $m_1$ and $m_\infty$. The spectral network changes its isotopy class when one of the four hyperplanes defined by the equations
\begin{align} \label{eqn:hyperplane}
m_\infty &= \pm m_0 \pm m_1 
\end{align}
in parameter space is crossed, which is when two branch-points of the covering $\Sigma \to C$ collide. The spectral networks on either side of such a hyperplane are related by a ``flip move'' (in the terminology of \cite{Gabella:2017hpz}), where two branch points approach each other, collide and then move away in perpendicular directions, as is illustrated in Figure~\ref{fig:branchpointmove}.

\insfigscaled{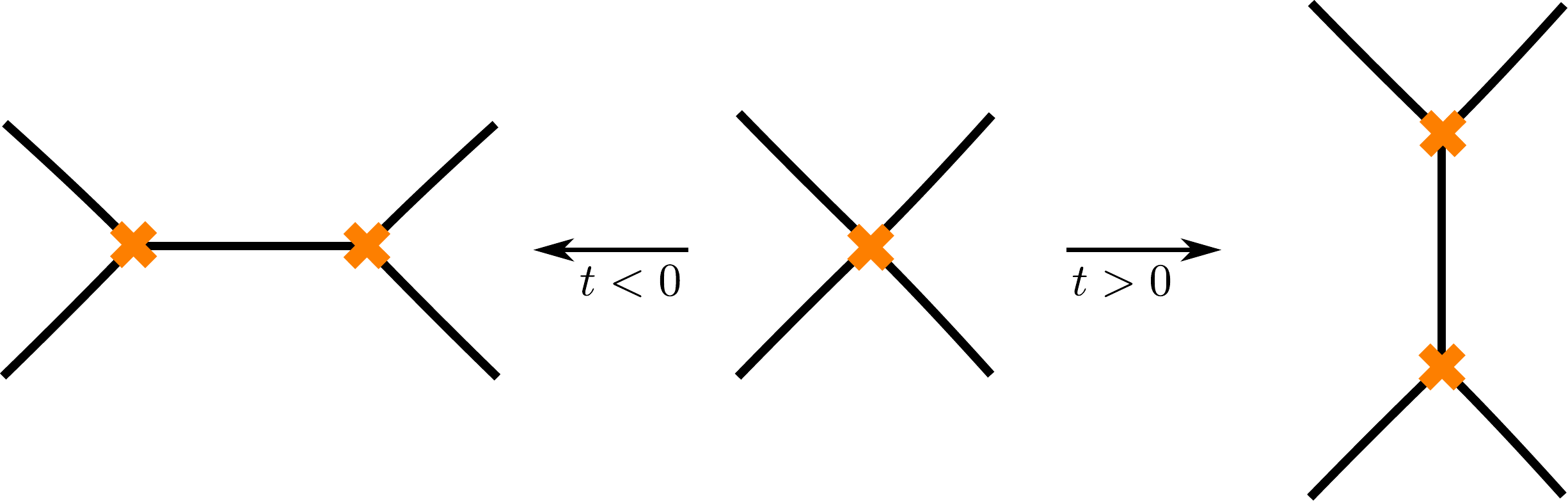}{0.45}{Flip move: when varying a real parameter $t$ in $\varphi$ from a small negative value to a small positive value two branch points come closer until they collide and then move away from each other in a perpendicular direction. All walls in this figure are double walls.}

If we do not distinguish the three punctures on $\mathbb{P}_{0,1,\infty}$ there are only two inequivalent spectral networks, named ``molecule I'' and ``molecule II'', which are plotted in Figure~\ref{fig:SU2-nw-mols-new} for $m_{\infty} =1$ and $m_0=m_1=0.45$ and $m_{\infty} = m_0 = m_1 =1$, respectively. The illustrated molecules are related by varying the parameter $t = m_0 + m_1-m_{\infty}$ from $t=-0.1$ to $t=1$ (while keeping $m_\infty > -m_0 + m_1$). 

\insfigscaled{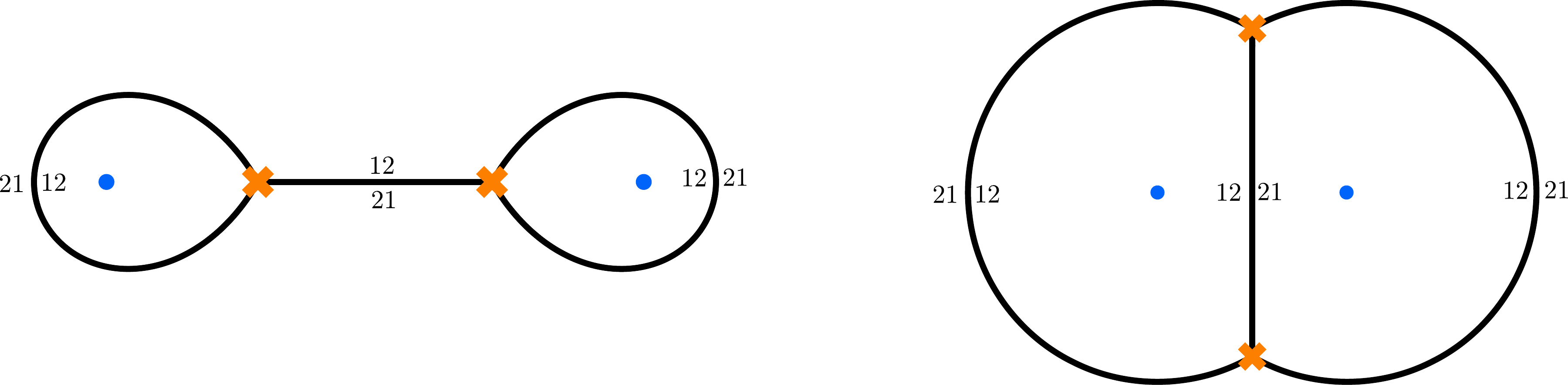}{0.45}{Fenchel-Nielsen networks on the three-punctured sphere. On the left is molecule I with $m_{\infty} =1$ and $m_0=m_1=0.45$. On the right is molecule II with $m_{\infty} = m_0 = m_1 =1$. The blue dots are the punctures, and the orange crosses are branch points of the covering $\Sigma \to C$. All walls are double walls.}

In applications we often need to study the two limits $\vartheta \to 0^{\pm}$. These correspond to the two ``resolutions'' of the network. In each of the two resolutions each double wall is split into two infinitesimally separated walls. The two resolutions of molecule I are shown in Figure~\ref{fig:SU2-nw-molI-res}. By drawing the branch cuts in this figure we have moreover fixed a local trivialization of the covering $\Sigma \to C$.

\insfigscaled{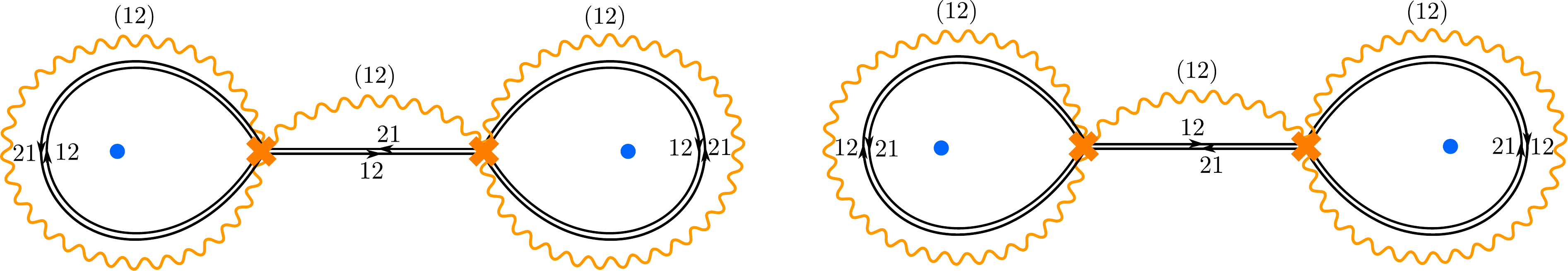}{0.43}{The two resolutions of molecule I: on the left the American resolution and on the right the British one. The wavy orange lines illustrate a choice of branch cuts of the covering $\Sigma \to C$.}

\end{example}

A Fenchel-Nielsen network on a general Riemann surface $C$ is defined with respect to a pants decomposition of $C$ and can be found by gluing together molecules (in the same resolution) on the individual pairs of pants. The molecules are glued together along the boundaries of the pairs of pants while inserting a circular branch cut around each pants curve.  

Any puncture in a molecule is surrounded by a polygon of double walls. The decoration at a puncture is an assignment of an ordering of the sheets of the spectral curve $\Sigma$ over the puncture to each direction around the puncture, compatible with the labelings of the double walls surrounding it, in such a way that reversing the direction reverses the ordering. In Figure~\ref{fig:SU2-nw-molI-res} we have chosen the branch cuts such that the 12-walls run in the clockwise direction around each puncture. The decoration thus assigns the sheet ordering 12 to the clockwise orientation. 

Similarly, any pants curve in a Fenchel-Nielsen network is surrounded on either side by a polygon of walls. We thus also associate a decoration to each pants curve. This is an assignment of an ordering of the sheets to each direction around the pants curve, compatible with the labelings of the double walls surrounding it, in such a way that reversing the direction reverses the ordering. 

\subsection{$K=3$}\label{sec:FNnetwork-K=3}

We generalize the Strebel condition to $K>2$ by saying that a tuple of differentials $\varphi=(\varphi_2, \ldots, \varphi_K)$ is \emph{generalized Strebel} if there exists a canonical homology basis for $\overline{\Sigma}$, i.e.~a choice of A and B-cycles on the compactified spectral cover $\overline{\Sigma}$, such that 
\begin{align}\label{eqn:strebelconstraint}
e^{-i \vartheta} \oint_{A_k} \lambda \in \R \quad \mathrm{and} \quad 
e^{-i \vartheta} \oint_{\widetilde{\gamma}_l} \lambda \in \R,
\end{align}
for each A-cycle $A_k$ and each lift $\widetilde{\gamma}_l$ of a small loop around each puncture $z_l$ to $\Sigma$, where $\lambda$ is the tautological 1-form on $\Sigma$. We say that the generalized Strebel tuple $\varphi$ \emph{respects a pants decomposition} of $C$ if the generalized Strebel condition~(\ref{eqn:strebelconstraint}) holds for the lifts of each pants curve in the decomposition to $\Sigma$. 

Recall that a spectral network $\mathcal{W}_\vartheta(\varphi)$ is called a generalized Fenchel-Nielsen network if it respects some pants decomposition and consists of only double walls. In the examples below we find that generalized Fenchel-Nielsen networks $\mathcal{W}_\vartheta(\varphi)$ correspond to generalized Strebel tuples $\varphi$ that respects the pants decomposition. 

(In \cite{Longhi:2016wtv,Gabella:2017hpz} a related class of networks, called BPS graphs, were given an interpretation in terms of BPS quivers. In the terminology of \cite{Hollands:2013qza} we would call them generalized fully contracted Fenchel-Nielsen networks. In particular, they do not respect any pants decomposition. )

\begin{example}

The differentials $\varphi_2, \varphi_3$ on the three-punctured sphere $\mathbb{P}^1_{0,\underline{1},\infty}$ with two maximal and one minimal puncture were discussed in \S\ref{sec:SU3bifund}.
After we apply a Mobius transform to move the punctures to $z_a=1$, $z_b=\omega$ and $z_c=\omega^2$, where $\omega$ is the third root of unity, these differentials can be written in the form
\begin{align}
\varphi_2^\mathrm{bif}(z) & = \frac{ -9 m_a^2}{(1-z)(1 - z^3)} +  \frac{3(1-z)^2}{(1-z^3)^2} (m_{b,1}^2 + m_{b,1} m_{b,2} + m_{b,2}^2) \\
\varphi_3^\mathrm{bif}(z) & = \frac{ 9 (1+z)m_a^3}{(1-z)(1 - z^3)^2} - \frac{9(1-z)^2(1+z)}{(1-z^3)^3} m_{b,1} m_{b,2} (m_{b,1} + m_{b,2}), 
\end{align}
where $m_a$ is the single mass parameter at the minimal puncture at $z=1$, and where we have set the mass parameters $m_{b,1}$ and $m_{b,2}$ at the maximal puncture at $z_b=\omega$ to be minus the ones at $z_c=\omega^2$.

The spectral network $\c(\varphi^\mathrm{bif},\vartheta)$ is a generalized Fenchel-Nielsen network if and only if all mass parameters $m_a$, $m_{b,1}$ and $m_{b,2}$ have the same phase $\vartheta - \frac{\pi}{2}$. This is precisely when the corresponding tuple $\varphi^\mathrm{bif}$ is generalized Strebel. Without loss of generality we can assume that the mass parameters are real and $\vartheta = \frac{\pi}{2}$. 

Just like in the previous example, it is possible to classify the different isotopy classes by writing down the equations for the hyperplanes corresponding to the collision of two or more branch-points of the covering $\Sigma \to C$. We refer to any of these isotopy classes as a $K=3$ generalized Fenchel-Nielsen molecule with two maximal and one minimal puncture. Any two such molecules are related by a sequence of elementary local transformations, such as the flip move. Some molecules are shown in Figure~\ref{fig:ltnets}.

\insfigscaled{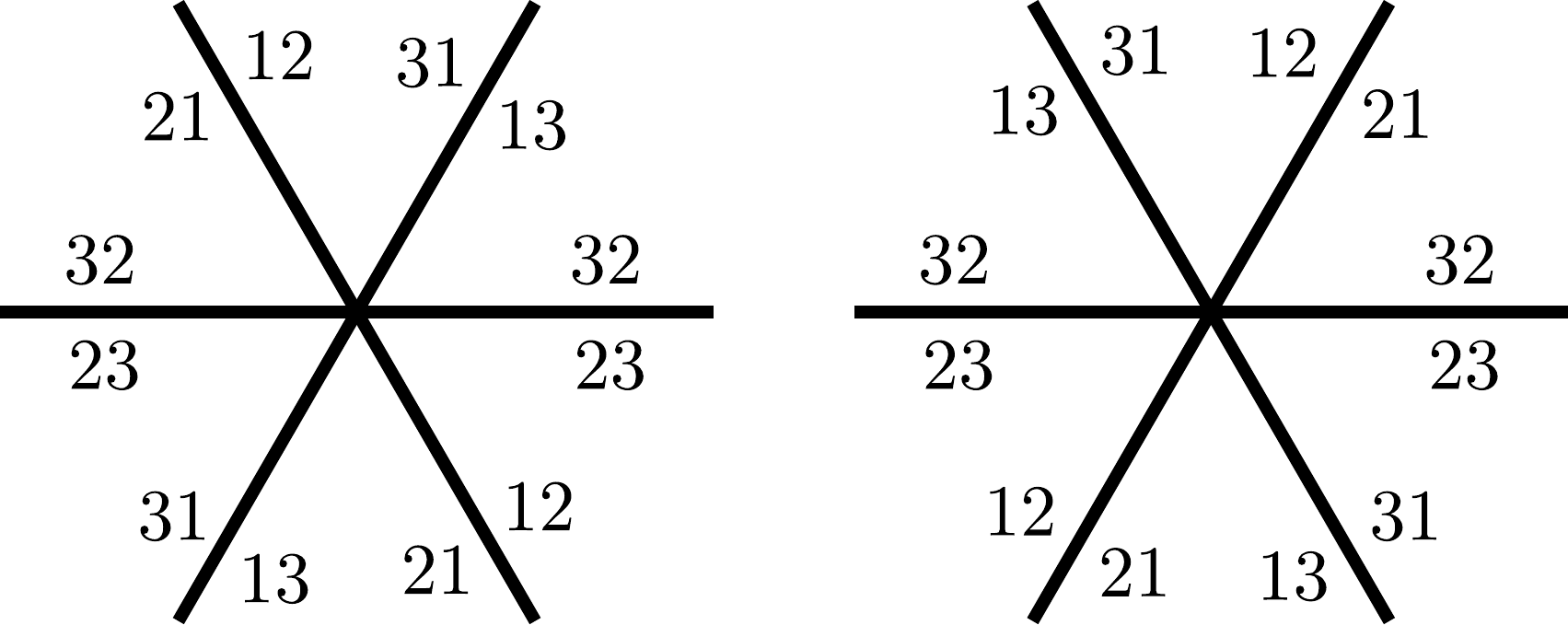}{0.4}{The two possible joints in which six double walls can intersect.}

The generalized Fenchel-Nielsen molecules with two maximal and one minimal puncture share a number of features. They are built out of two (rank 2) Fenchel-Nielsen molecules, intersecting each other in (both of) the 6-joints illustrated in Figure~\ref{fig:junctions}. Maximal punctures are surrounded by a polygon of double walls, whereas minimal punctures lie on top of a double wall.

\insfigscaled{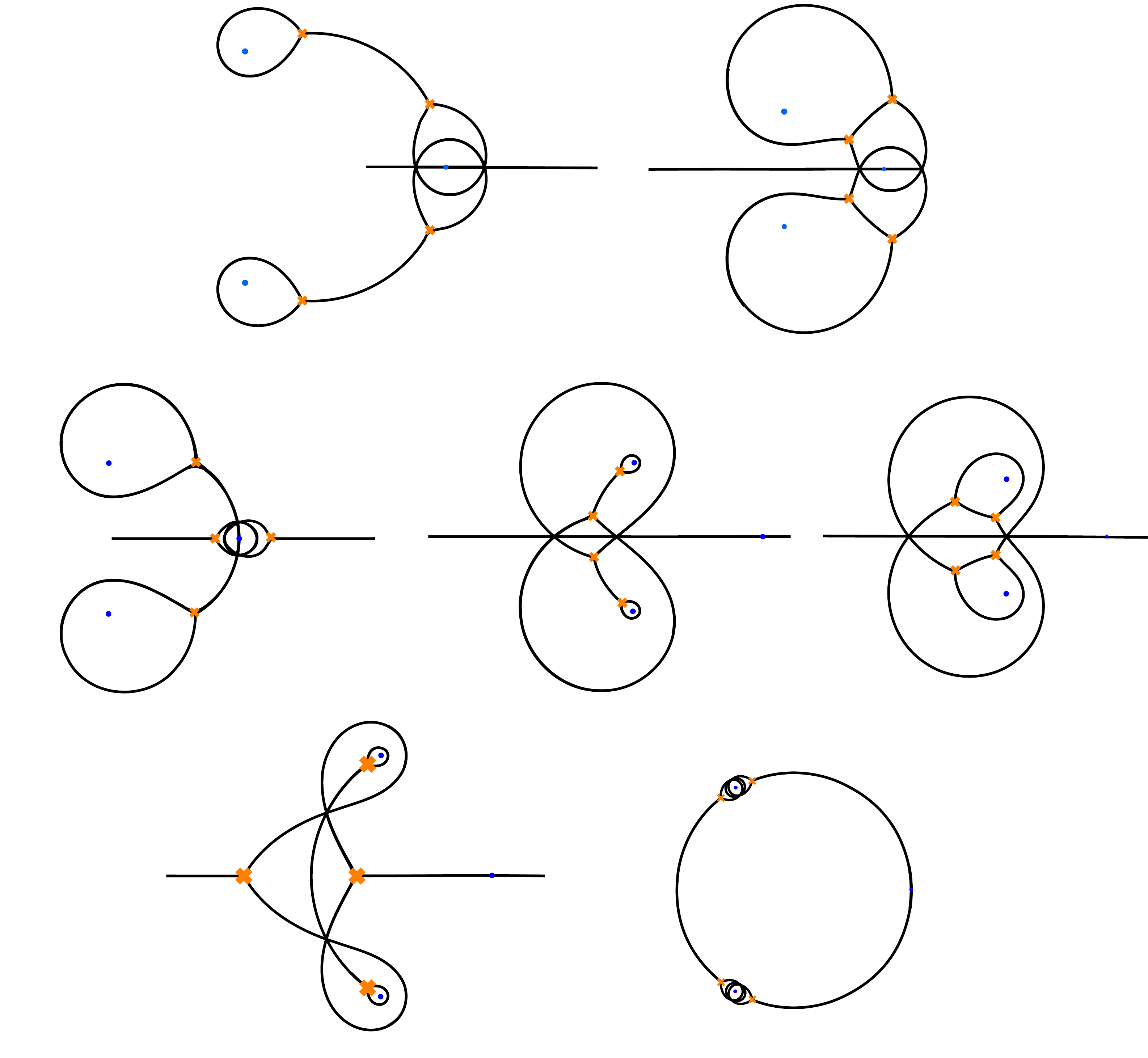}{0.53}{Examples of non-isotopic generalized Fenchel-Nielsen molecules with two maximal and one minimal puncture, symmetric about the horizontal. All walls are double walls. We obtained these pictures using~\cite{neitzkeswn}.}

Each $K=3$ molecule comes with two resolutions, in which each double wall is split into two infinitesimally separated walls. For instance, the two resolutions of the molecule at the top-left in Figure~\ref{fig:ltnets} are illustrated in Figure~\ref{fig:SU3bifund-nw-res}. Note that a minimal puncture is in between two single opposite walls. In Figure~\ref{fig:SU3bifund-nw-res} we have also chosen a local trivialization of the spectral cover $\Sigma$.

Each $K=3$ molecule can be represented with several choices of wall labelings. For instance, for the $K=3$ molecule in Figure~\ref{fig:SU3bifund-nw-res} the wall labelings are completely determined if we fix the labels for the double wall surrounding the maximal puncture at $z=\omega$ as well as one of the two possible combinations of joints around the minimal puncture at $z=1$. All different choices can be obtained from the representation in Figure~\ref{fig:SU3bifund-nw-res} by introducing additional branch cuts around the punctures. 

Each choice of wall labelings determines a decoration at the punctures and along the pants curves. As before, the decoration assigns an ordering of the sheets of the spectral curve $\Sigma$ over the puncture or over the pants curve to each direction, in such a way that reversing the direction reverses the ordering. For instance, for the $K=3$ molecule in Figure~\ref{fig:SU3bifund-nw-res} the decoration at the maximal puncture at $z=\omega$ assigns the sheet ordering $(123)$ to the clockwise direction and $(321)$ to the anti-clockwise direction, whereas the decoration at the maximal puncture at $z=\omega^2$ assigns the sheet ordering $(321)$ to the clockwise direction and $(123)$ to the anti-clockwise direction. The decoration at the minimal puncture at $z=1$ assigns the sheet ordering $(31;2)$ to the clockwise direction and $(13;2)$ to the anti-clockwise direction, where $2$ is the distinguished sheet that does not appear in the label of the double wall intersecting the minimal puncture. 
\end{example}

\vspace*{3mm}

\insfigscaled{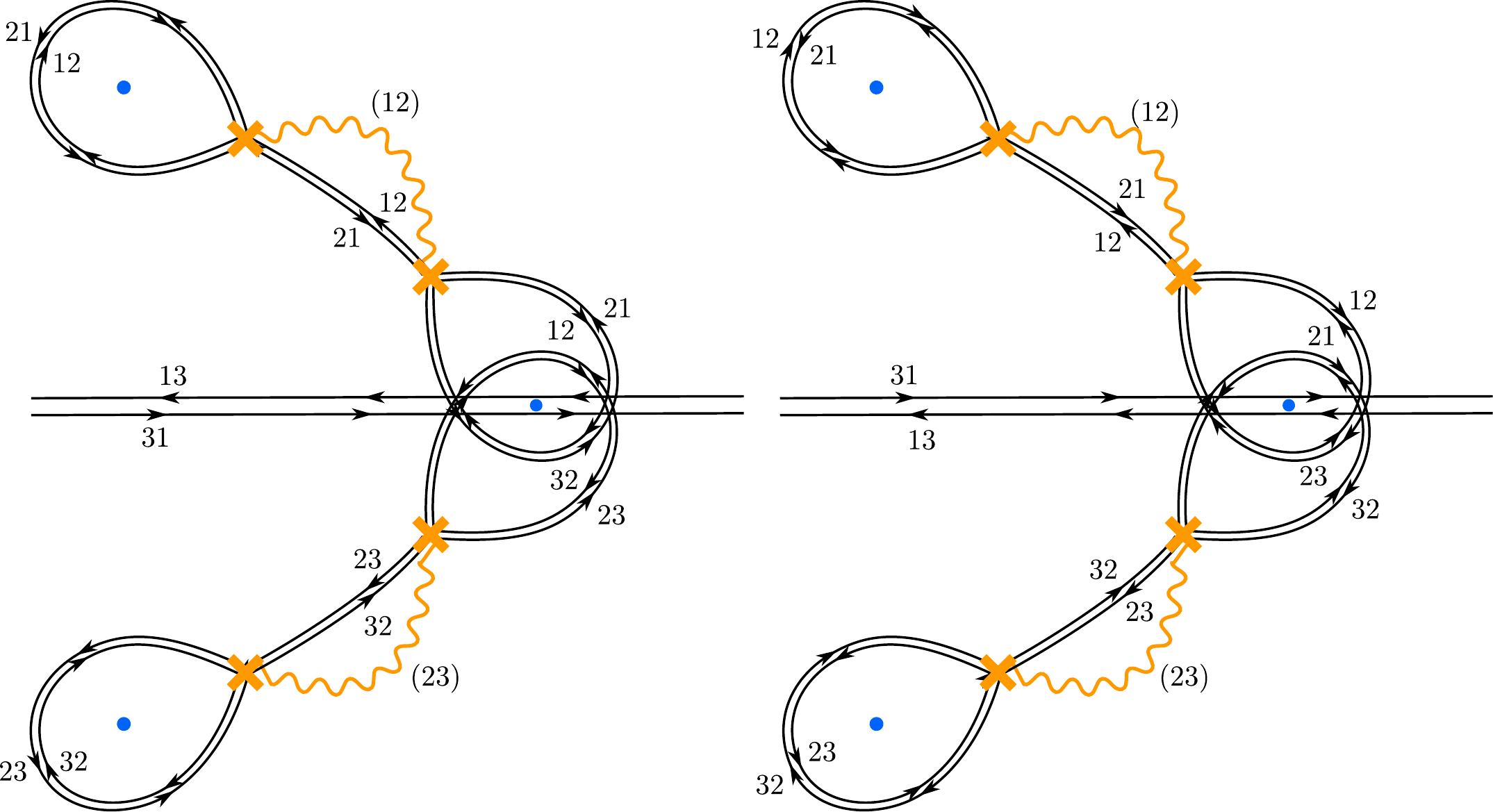}{0.64}{The two resolutions of the generalized Fenchel-Nielsen molecule at the top-left in Figure~\ref{fig:ltnets} together with a choice of local trivialization of the spectral cover $\Sigma$.}

\begin{example}
Equations~(\ref{eqn:K=3phi2}),~(\ref{eqn:K=3phi3}) characterize the 1-dimensional family of tuples $(\varphi_2,\varphi_3)$ on the three-punctured sphere $\mathbb{P}^1_{0,1,\infty}$ with three maximal punctures. Each tuple defines a spectral cover $\Sigma$ over $C$ whose compactification has genus 1. This implies that the possible generalized Strebel tuples are labeled by a choice of A-cycle on $\overline{\Sigma}$. The generalized Strebel condition~(\ref{eqn:strebelconstraint}) fixes the parameters $u$ and $m_{i,j}$ relative to the choice of the phase $\vartheta$. 

Generalized Fenchel-Nielsen networks on the three-punctured sphere $\mathbb{P}^1_{0,1,\infty}$ with three maximal punctures were classified in \cite{Hollands:2016kgm} in the limit in which all parameters $m_{i,j}$ are sent to zero. In this limit we may just as well set $u=1$. It was found that there is a single generalized Fenchel-Nielsen network at each phase $\vartheta_{[p,q]}$ with 
\begin{align}
\tan \vartheta_{[p,q]} = \frac{\sqrt{3} q}{q-2p}, 
\end{align}
for any pair of coprime integers $(p,q)$. Each generalized Fenchel-Nielsen network indeed corresponds to a generalized Strebel differential $\lambda$ with 
\begin{align}
e^{-i \vartheta_{[p,q]}} \oint_{A_{p,q}} \lambda \in \mathbb{R}. 
\end{align}
where $A_{p,q} = p \gamma_1 + q \gamma_2$, for a certain basis of 1-cycles $\gamma_1$ and $\gamma_2$ on $\Sigma$.

\end{example}

Generalized Fenchel-Nielsen networks on a (punctured) Riemann surface $C$  are defined with respect to a pants decomposition of $C$ and can be found by gluing together generalized Fenchel-Nielsen molecules on the individual pairs of pants. Not only the type of punctures should match, but also the decorations along the pants curves (possibly by inserting additional branch cuts).

In the following we restrict ourselves to Fenchel-Nielsen networks obtained from gluing Fenchel-Nielsen molecules with two maximal and one minimal puncture along maximal boundaries. We call this subset of generalized Fenchel-Nielsen networks \emph{of length-twist type}. Figure~\ref{fig:SU3Nf6} gives an example of such a length-twist type network on the four-punctured sphere (where we have replaced the two maximal punctures by boundaries).

\section{Higher length-twist coordinates}\label{sec:length-twist}

Let $\nabla$ be a flat $SL(K)$-connection on $C$ with a fixed \emph{semi-simple} conjugacy class 
\begin{align}
\mathcal{C}_l=\mathrm{diag }\{M_{l,1},\ldots M_{l,K}\}
\end{align}
at each puncture with $M_{l,i} \in \mathbb{C}^\times$. 

The partition of the $K$ eigenvalues can be read off from the Young diagram assigned to the puncture: the height of each column in the Young diagram encodes the multiplicities of coincident eigenvalues. In particular, for generic values of the eigenvalues ($M_{l,1}=\ldots =M_{l,K-1}$ not equal to a $K$-th root of unity), a conjugacy class at a minimal puncture is a scalar multiple of a so-called complex reflection matrix. The latter is defined as a matrix $A$ that satisfies $\mathrm{rk} (A-I) = 1$.  

We denote the moduli space of such flat connections by 
\begin{align}
\mathcal{M}_\mathrm{flat}^\mathcal{C}(C,SL(K)),
\end{align}
where $\mathcal{C} = \{ \mathcal{C}_l \}$ is the collection of conjugacy classes. 

We will restrict ourselves to Riemann surfaces $C$ that can be obtained by gluing spheres with two maximal and one minimal puncture along maximal boundaries. Since a generic flat $SL(K)$-connection on the sphere with two maximal and one minimal puncture is completely specified (up to equivalence) by the eigenvalues of the monodromy around the punctures, the moduli space of flat $SL(K)$-connections on any such surface $C$ is $(K-1)$~$(3g-3+n)$-dimensional, where $3g-3+n$ is the number of pants curves.

In this section we define a generalization of the standard Fenchel-Nielsen length-twist coordinates on the moduli space 
$$\mathcal{M}_\mathrm{flat}^\mathcal{C}(C,SL(K),\cW)$$ 
of so-called $\cW$-framed flat $SL(K)$ connections. In section~\ref{sec:ab-l-t} we show that these coordinates are realized as spectral coordinates through the abelianization method. The $\mathcal{W}$-framing will be crucial in proving that an abelianization of $\nabla$ exists and is unique.

\subsection{Framing}\label{sec:framing}

In this section we fix a (possibly punctured) surface $C$ together with a pants decomposition into pairs of pants with two maximal and one minimal puncture. We also fix a generalized Fenchel-Nielsen network $\cW$ of length-twist type relative to this pants decomposition. The individual molecules of the network $\cW$ are glued together along a collection of maximal boundaries.

We define a $\cW$-framed connection $\nabla$ on $C$ to be a flat $SL(K)$ connection on $C$ together with a \emph{framing} of $\nabla$ at each maximal puncture and maximal boundary.\footnote{The reason for only fixing a framing at the maximal punctures and maximal boundaries of $C$ will become clear in \S\ref{sec:ab-l-t}, where we also discuss framings at other types of punctures and boundaries.} The framing is just an ordered tuple of eigenlines $(l_{\alpha_1},\ldots,l_{\alpha_K})$ of the monodromy $M_+$ (in the $+$ direction) around the maximal boundary or maximal puncture. We require furthermore that $l_{\alpha_i} \neq l_{\alpha_j}$ for $i \neq j$ and also that each of $l_{\alpha_i}$ for any puncture or boundary is distinct from each of $l_{\alpha_j}$ for any adjacent puncture or boundary (that is, a puncture or boundary belonging to the same pair of pants). Note that a $\cW$-framing of $\nabla$ exists only if all of the $M_\pm$ are diagonalizable.

\subsection{Higher length-twist coordinates}\label{sec:higher-length-twist}

A $\cW$-framed $SL(K)$ flat connection $\nabla$ on $C$ is completely specified (up to equivalence) by $2K-2$ parameters at each pants curve (or maximal boundary) $\alpha$. 

Half of this set of parameters, say $\ell_1, \ldots, \ell_{K-1}$, are the eigenvalues of the monodromy $M_+$. The indexing of these parameters is determined by the decoration as well as the framing data. If the decoration at the boundary assigns the sheet ordening $(i_1, \ldots, i_K)$ to the $+$ direction and the framing of $\nabla$ at the boundary in the $+$ direction is given by the ordered tuple of eigenlines $(l_{\alpha_1}, \ldots, l_{\alpha_K})$, then we define
\begin{align}
L_{i_j} = - e^{\pi i \, \ell_{i_j}}
\end{align}
as the eigenvalue corresponding to the eigenline $l_{\alpha_{j}}$.\footnote{A rationale for the slightly odd conventions is given in \S\ref{sec:opers}.}

The other half of the parameters, say $\tau_1, \ldots, \tau_{K-1}$, have a more indirect definition. One approach is in terms of how they transforms under the following modification of the flat connection $\nabla$. Suppose we cut the surface $C$ into two pieces along a pants curve $\alpha$.\footnote{Here we suppose that $\alpha$ is a separating loop, a similar discussion holds if it is nonseparating.} We obtain two surfaces with boundary, say $C_1$ and $C_2$ carrying flat connections $\nabla_1$ and $\nabla_2$, as well as an isomorphism $\iota$ that relates $\nabla_1$ to $\nabla_2$. Let us now change $\nabla_1$ by a gauge transformation $\kappa$ that preserves the monodromy $M$ around $\alpha$, and then glue $C$ back along the boundary $\alpha$. 

If the monodromy $M_+$ is diagonalized by the gauge transformation $g$, then the transformation $\kappa$ can be written as 
\begin{align}
\kappa= g^{-1} \circ \mathrm{diag}\left(e^{\lambda_{1}},\ldots, e^{\lambda_{K}} \right) \circ g
\end{align}
with $\sum_{i=1}^K \lambda_i = 0$. After gluing back we thus obtain a 1-parameter family of modified connections $\nabla(\lambda)$. This operation is sometimes called the (generalized) twist flow (see for instance \cite{Goldman1986} in the real-analytic setting, which builds on \cite{10.2307/2007011,10.2307/2007075,GOLDMAN1984200}).

Any choice of parameters $\tau_1, \ldots, \tau_{K-1}$ with the property that they change under the twist flow as
\begin{align}
\tau_j & \mapsto  \tau_j + \frac{\lambda_j}{2} - \frac{\lambda_K}{2}
\end{align}
are called twist parameters. The twist parameters $\tau_i$ are thus only defined up to an additive function in the length parameters $\ell_1, \ldots, \ell_K$.

The construction of the length-twist coordinates $\ell_1, \ldots, \ell_{K-1}$ and $\tau_1, \ldots, \tau_{K-1}$ guarantees that they are Darboux coordinates on the moduli space of $SL(K)$ flat $\cW$-framed connections.
We refer to them as (complex) higher length and twist coordinates, respectively.\footnote{This is a rather straight-forward higher rank generalization of the definition of Fenchel-Nielsen length-twist coordinates in \cite{Hollands:2013qza}.}
In \S\ref{sec:ab-l-t} we will realize these coordinates explicitly as spectral coordinates associated to the generalized Fenchel-Nielsen network $\mathcal{W}$ of length-twist type. 

\subsection{Standard twist coordinate}\label{sec:distinguishedtwist}

The twist coordinate defined as above is only determined up to a canonical transformation $\tau' = \tau + f(\ell)$. A distinguished choice for the twist is given by the so-called complex Fenchel-Nielsen twist $\tau^\mathrm{FN}$ \cite{tan_1994,kourouniotis_1994,goldman2009trace,Dimofte:2011jd}. This twist parameter is identical to the NRS Darboux coordinate $\beta/2$ \cite{Nekrasov:2011bc}.

\begin{example} 

On the four-punctured sphere $\mathbb{P}^1_{0,q,1,\infty}$ fix the presentation of the fundamental group as illustrated in Figure~\ref{fig:pi1new}, generated by the paths $\delta_0$, $\delta$, $\delta_1$ and $\delta_\infty$ with the relation 
\begin{align}
\langle \delta_0, \delta, \delta_1, \delta_\infty | \, \delta_0 \, \delta \, \delta_\infty \, \delta_1 = 1 \rangle. 
\end{align}
If the conjugacy class around the path $\delta_l$ is fixed to be a diagonal matrix with eigenvalues $M_l$ and $M_l^{-1}$, we have that the traces of the monodromy matrices $\mathbf{M}_{\alpha} = M_{\delta_0} M_{\delta}$ and $\mathbf{M}_{\beta} =  M_{\delta_0} M_{\delta_\infty}$ are given by
\begin{align}
\Tr \mathbf{M}_{\alpha} &= L + L^{-1},\\
\Tr \mathbf{M}_{\beta} &= \sqrt{N(L)} \left( T + T^{-1} \right) + N_\circ(L),
\end{align}
where  
\begin{align}
 L + L^{-1} &= - 2 \cos ( \pi \ell ),\notag \\
T + T^{-1} & = - 2 \cosh ( 2 \tau^\mathrm{FN} ),\\
M_l + M_l^{-1} & = - 2 \cos ( \pi m_l ), \notag 
\end{align}
and
\begin{align}
N(L) &= \dfrac{c_{0q}(L) c_{1\infty}(L)}{\sin^4(\pi \ell)},  \\
c_{kl} &= \cos ( \pi \ell )^2 + \cos ( \pi m_k )^2 + \cos ( \pi  m_l )^2 + \cos ( \pi \ell, \pi m_k, \pi m_l ) -4, \notag \\ 
N_\circ(L) &= \frac{\cos( \pi \ell ) \left( \cos(\pi m_0, \pi m_1) + \cos ( \pi m, \pi m_\infty)\right) + \cos ( \pi m, \pi m_1) + \cos ( \pi m_0, \pi m_\infty) }{\frac{1}{2}\sin^2(\pi \ell)},  \notag
\end{align}
where we defined $\cos(x,y) = \cos(x) \cos(y)$. We realize the Fenchel-Nielsen length-twist coordinates $\ell$ and $\tau^\mathrm{FN}$ as spectral coordinates in \S\ref{sec:monFNfoursphere} by averaging over the two resolutions of a Fenchel-Nielsen network.

\end{example}

\insfigscaled{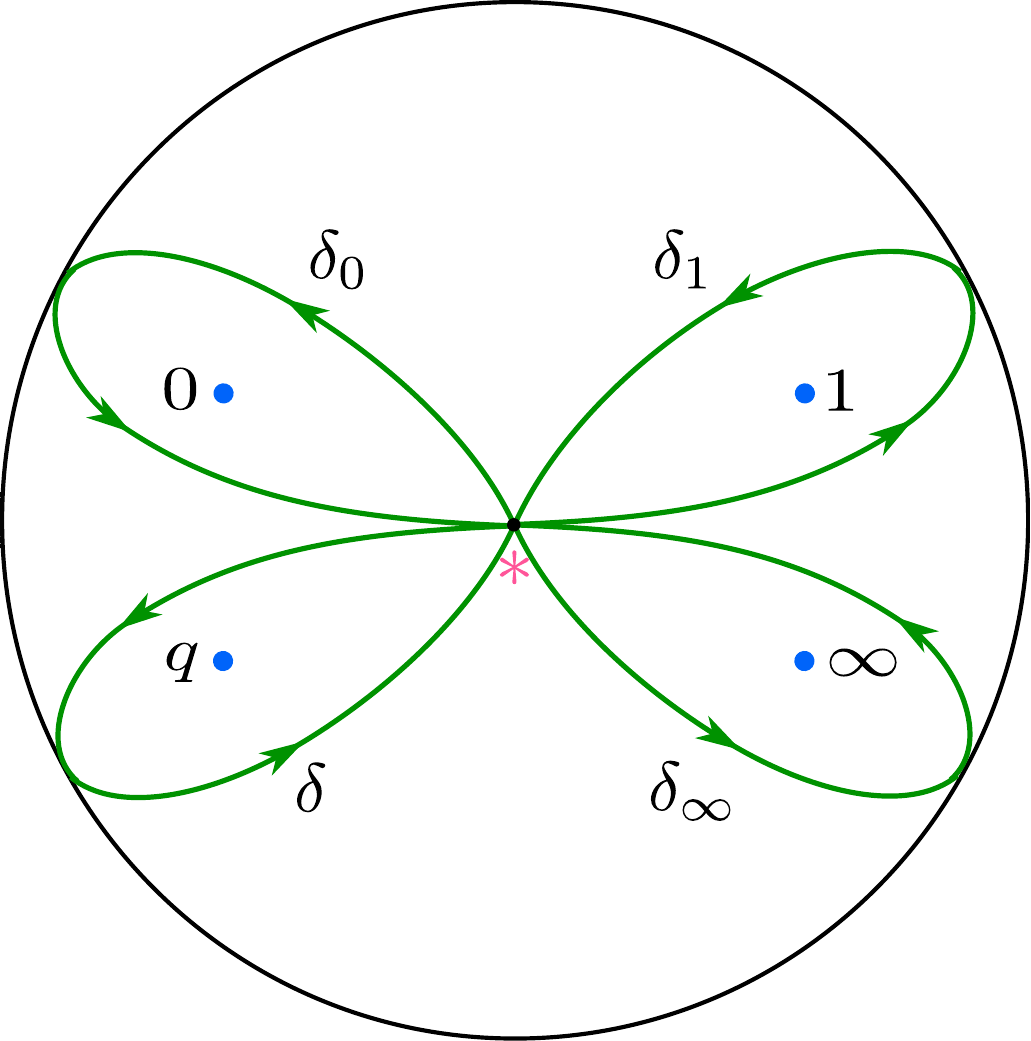}{0.5}{Generators for the fundamental group $\pi_1(C,*)$ of the four-punctured sphere $\mathbb{P}^1_{0,q,1,\infty}$.   }

\section{Abelianization and spectral coordinates}\label{sec:nonab}

One of the mathematical applications of spectral networks is that they induce holomorphic Darboux coordinate systems on moduli spaces of flat connections, called spectral coordinates \cite{Gaiotto:2012rg}. These are very special coordinate systems, subsuming a range of previously known examples. In particular, in \cite{Gaiotto:2012db} it was found that for certain types of spectral networks the resulting spectral coordinates are the same as coordinates introduced earlier by Fock and Goncharov. In \cite{Hollands:2013qza} this was detailed in the special case of rank $K=2$, and it was found that other types of spectral networks, namely the Fenchel-Nielsen networks, lead to (complexified) Fenchel-Nielsen length-twist coordinate systems. In this section we simply extend the techniques from that work to describe the higher length-twist coordinates as spectral coordinates.

In the following we replace all maximal punctures in a generalized Fenchel-Nielsen network by holes.

\subsection{Abelianization}

The key to the construction of spectral coordinate systems is the notion of ``abelianization'' \cite{Hollands:2013qza,Gaiotto:2012rg}. Let $C$ be a punctured Riemann surface. Fix a branched covering $\pi: \Sigma \to C$ and a spectral network $\cW$ subordinate to this covering. Given a generic $\cW$-framed flat $SL(K)$-connection $\nabla$ in a complex vector bundle $E$ over $C$, a $\cW$-abelianization of $\nabla$ is a way of putting $\nabla$ in almost-diagonal form, by locally decomposing $E$ as a sum of line bundles, which are preserved by $\nabla$.
We may define $\cW$-abelianization of $\nabla$ in terms of $\cW$-pairs \cite{Hollands:2013qza}. Let $C'$, $\Sigma'$ denote $C$, $\Sigma$ with the (preimages of) branch points removed.

\begin{definition}
A \emph{$\mathcal{W}$-pair} $(E,\nabla,\iota,\mathcal{L}',\nabla^{\mathrm{ab}})$ for a network $\mathcal{W}$ subordinate to the branched covering $\pi: \Sigma \rightarrow C$ is a collection of data:

\begin{enumerate}[(i)]
\item A flat rank $K$ bundle $(E,\nabla)$ over $C$
\item A flat rank $1$ bundle $(\mathcal{L}',\nabla^{\mathrm{ab}})$ over $\Sigma'$
\item An isomorphism $\iota: E\rightarrow \pi_{*}\mathcal{L}'$ defined over $C' \setminus \mathcal{W}$
\end{enumerate}
such that 
\begin{enumerate}[(a)]
 \item the isomorphism $\iota$ takes $\nabla$ to $\pi_* \nabla^\ab$,
 \item at each single wall $w \subset \cW$, $\iota$ jumps by a map $\cS_w = 1 + e_w \in \End(\pi_* \cL')$
where $e_w: \cL_i' \to \cL_j'$ if $w$ carries the label $ij$.  (Here by $\cL_i$ we mean the summand of $\pi_* \cL'$ associated to sheet $i$.
Relative to diagonal local trivializations of $\pi_* \cL$, this condition says
$\cS_w$ is upper or lower triangular). At each double wall $w' w$ $\iota$ jumps by a map $\cS_{w'} \cS_{w}$, with the ordering determined by the resolution as in \cite{Hollands:2013qza}. 
\end{enumerate}

We call two $\cW$-pairs \ti{equivalent} and write $(E_{1}, \nabla_{1},\iota_{1}, \cL_{1}'\nabla_{1}^\ab) \sim (E_{2}, \nabla_{2},\iota_{2}, \cL_{2}',\nabla_{2}^\ab)$,
if there exist maps $\varphi: \cL_{1}' \to \cL_{2}'$ and $\psi: E_{1} \to E_{2}$ which take $\iota_{1}$ into $\iota_{2}$,
$\nabla_{1}^\ab$ into $\nabla_{2}^\ab$, and $\nabla_{1}$ into $\nabla_{2}$.
In particular, in this case we have equivalences $(\cL_{1}
, \nabla_{1}^\ab) \sim (\cL_{2}', \nabla_{2}^\ab)$
and $(E_{1}, \nabla_{2}) \sim (E_{2}, \nabla_{2})$.

\end{definition}

\begin{definition}Given a flat $SL(K)$-connection $\nabla$ on a complex rank $K$ bundle $E$ over $C$,
a \emph{$\cW$-abelianization} of $\nabla$ is any extension of 
$(E, \nabla)$ to a $\cW$-pair $(E, \nabla,\iota, \cL', \nabla^\ab)$.

In the other direction, given an equivariant flat $GL(1)$-connection $\nabla^\ab$ on a complex line bundle $\cL'$ over $\Sigma'$,
a \emph{$\cW$-nonabelianization} of $\nabla^\ab$ is any extension of 
$(\cL', \nabla^\ab)$ to a $\cW$-pair $(E, \nabla,\iota, \cL', \nabla^\ab)$.
\end{definition}

In fact, to abelianize a $\cW$-framed flat $SL(K)$-connection $\nabla$, it is sufficient to define the flat $GL(1)$-connection $\nabla^\mathrm{ab}$ on $\cL'$  restricted to $\Sigma' \backslash \pi^{-1}(\cW)$. Then $\nabla^\mathrm{ab}$ automatically extends from $\Sigma' \backslash \pi^{-1}(\cW)$ to $\Sigma'$. The argument for this is a straightforward generalization of the argument in \S5.1 of \cite{Hollands:2013qza}. 

\subsection{Boundary}

If $C$ has boundary, it is useful to consider connections and $\cW$-pairs with extra structure. We fix a marked point on each boundary component of $C$. Then, a $\cW$-pair with boundary \cite{Hollands:2013qza} consists of
\begin{itemize}
\item A $\cW$-pair $(E, \nabla, \cL, \nabla^\ab, \iota)$,
\item a basis of $E_z$ for each marked point $z$,
\item a basis of $\cL_{z_i}$ for each preimage $z_i \in \pi^{-1}(z)$ for a marked point $z$,
\item a trivialization of the covering $\Sigma$ over in a neighbourhood of each marked
point $z$,
\end{itemize}
such that $\iota$ maps the basis of $E_z$ to
the basis of $\pi_* \cL_{z}'$ induced from those 
of $\cL_{z_i}'$ and $\Sigma$.

Given two surfaces $C$, $C'$ with boundary we can glue along a boundary 
component, in such a way that the marked points are identified. Suppose that we have a $\cW$-pair with boundary on each of $C$ and $C'$,
and that the monodromies around the glued component are the same (when
written relative to the given trivializations at the marked points).
Then using the trivializations we can glue the two $\cW$-pairs to 
obtain a $\cW$-pair over the glued surface.

\subsection{Equivariant $GL(1)$ connections}

The abelianization of a flat $SL(K)$-connection $\nabla$ amounts to choosing a basis $(s_1, \ldots, s_K)$ of $E$ at any point in $C \backslash \cW$, satisfying certain constraints ensuring the correct form of transition across walls. Any $GL(1)$ connection $\nabla^{\mathrm{ab}}$ that is obtained by $\cW$-abelianizing a flat $SL(K)$-connection $\nabla$ automatically carries some additional structure. We will capture this by saying that the $GL(1)$ connection $\nabla^{\mathrm{ab}}$ is equivariant on $\Sigma$ \cite{Hollands:2013qza}.

Suppose we are given a $\cW$-pair $(E,\nabla,\iota,\cL,\nabla^{\mathrm{ab}})$ with $\nabla$ a flat $SL(K)$-connection. Then the underlying bundle $E$ carries a $\nabla$-invariant, nonvanishing volume form $\varepsilon_E \in \Lambda^K(E)$. If we introduce a cyclic covering permutation $\rho: \Sigma \to \Sigma$ with $\rho^K = \mathrm{id}$, we have a $\nabla^{\mathrm{ab}}$-invariant nondegenerate linear mapping 
\begin{align}
\mu: \cL \otimes \cdots \otimes (\rho^{K-1})^* \cL \to \C
\end{align} 
given by 
\begin{align}
\mu(s_1,\ldots,s_K) = \iota^{-1}(s_1) \wedge \ldots \wedge \iota^{-1}(s_K) /\varepsilon_E,
\end{align}
which clearly extends over $\pi^{-1}(\cW)$. If we let 
\begin{align}
\tau: \cL \otimes \cdots \otimes (\rho^{K-1})^* \cL \to \rho^* \left(\cL \otimes \cdots \otimes (\rho^{K-1})^* \cL \right),
\end{align} 
then we find that 
\begin{align}
\rho^* \mu = (-1)^{K-1} \, \mu \circ \tau
\end{align}
We call any connection $\nabla^\mathrm{ab}$ in a line bundle $\cL$ equipped with such a pairing \emph{equivariant}.

Equivariance just says that the parallel transport of the vector $(s_1, \ldots, s_K)$ over a path in $C \backslash \cW$ (not crossing a branch cut) is given by a diagonal matrix with determinant 1. It also implies that the holonomy of $\pi_* \nabla^\mathrm{ab}$ around a simple branch point of type $(ij)$ can be represented by the matrix whose only vanishing diagonal, and whose only non-vanishing off-diagonal, entries are
\begin{align}
\begin{pmatrix} d_{ii} & d_{ij} \\ d_{ji} & d_{jj} \end{pmatrix} = \begin{pmatrix} 0 & d \\ - d^{-1} & 0 \end{pmatrix}. 
\end{align}
This says that the holonomy of $\nabla^\mathrm{ab}$ around a simple branch point of type $(ij)$ is $-1$. A connection $\nabla^\mathrm{ab}$ with this property is called an almost-flat connection over $\Sigma$ in \cite{Hollands:2013qza}. 

Last, equivariance implies that the $GL(1)$ connection $\nabla^\mathrm{ab}$ carries additional structure at the punctures, characterized by the type of the puncture. In particular, since the monodromy of $\nabla$ around minimal puncture is a multiple of a reflection matrix, this implies that the monodromy of $\pi_* \nabla^\mathrm{ab}$ around a minimal puncture is given by a diagonal matrix with $K-1$ equal eigenvalues.

\subsection{Moduli spaces}\label{sec:modulispaces}

Consider the following moduli spaces:
\begin{itemize}
\item let $\cM_\mathrm{flat}(C, SL(K), \cW)$ be the moduli space parametrizing flat
$\cW$-framed $SL(K)$-connections over $C$, up to equivalence,
\item let $\cM_\mathrm{eq}(\Sigma, GL(1))$ be the moduli space parametrizing
equivariant $GL(1)$-connections over $\Sigma$, up to equivalence,
\item let $\cM_\mathrm{pair}(\cW)$ be the moduli space parameterizing $\cW$-pairs,
up to equivalence.
\end{itemize}

The abelianization and nonabelianization constructions lead to the following diagram relating these spaces:
\begin{center}
\begin{tikzcd}
& \cM_\mathrm{pair}(\cW) \arrow[ld, "\pi_1"] \arrow[rd, "\pi_2", swap] & \\
\cM_{\mathrm{eq}}(\Sigma, GL(1)) \arrow[ru, start anchor={[xshift=-1.9ex]}, end anchor={[yshift=1ex]}, "\psi_1"] & & \cM_\mathrm{flat}(C, SL(K), \cW) \arrow[lu, start anchor={[xshift=1.9ex]}, end anchor={[yshift=1ex]}, "\psi_2", swap]
\end{tikzcd}
\end{center}
where $\pi_1$ and $\pi_2$ are the forgetful maps which map a $\cW$-pair to the underlying equivariant $GL(1)$-connection or $\cW$-framed flat $SL(K)$-connection respectively, whereas $\psi_1$ is the $\cW$-nonabelianization map and $\psi_2$ the $\cW$-abelianization map. From this description it is evident that $\pi_1 \circ \psi_1$
and $\pi_2 \circ \psi_2$ are the identity maps. 

To avoid notational clutter we have not explicitly mentioned the restricted boundary monodromies in the above. Yet, all remains true if we consider flat $\cW$-framed $SL(K)$-connections with fixed conjugacy classes at the boundaries and punctures, and interpret their eigenvalues as the boundary monodromies for the equivariant $GL(1)$ connections. 

In \cite{Hollands:2013qza} it was established that all of these mappings are bijections for $K=2$ Fenchel-Nielsen networks $\cW$. In particular, it was established that $\cW$-abelianizations are in one-to-one correspondence with $\cW$-framings for Fenchel-Nielsen networks $\cW$, and that there is there is a unique nonabelianization for any equivariant $GL(1)$-connection. In particular, this shows that the mapping $\Psi = \pi_1 \circ \psi_2$ is a bijection (in fact a diffeomorphism).

In the next section we will show that this result extends to $K=3$ Fenchel-Nielsen networks $\cW$ of length-twist type. We expect it to even hold for any generalized Fenchel-Nielsen networks of length-twist type. $\cW$-abelianizations for generic generalized Fenchel-Nielsen networks (not of length-twist type) are more subtle, however, and will be discussed in \cite{Hollands:toappear}.

\subsection{Spectral coordinates}\label{sec:spectral-coord}

Let $\Sigma'$ denote $\Sigma$ with the preimages of branch points removed. Given an equivariant $GL(1)$ connection $\nabla^\mathrm{ab}$ we can construct the holonomies
\begin{align}
\cX_{\gamma} = \mathrm{Hol}_\gamma(\nabla^\mathrm{ab}) \in \mathbb{C}^\times
\end{align}
where $\gamma \in H^1(\Sigma',\mathbb{Z})$. Together these form a coordinate system on the moduli space of equivariant $GL(1)$ connections. (Because of the equivariance, we will really only need a subset of $\gamma$'s.) Through the abelianization map, these complex numbers also determine a coordinate system on the moduli space of $\cW$-framed flat connections\footnote{We discuss framings at general regular punctures in \S\ref{sec:ab-l-t}}. The resulting coordinates are called spectral coordinates.

Spectral coordinates have a number of good properties. For instance, they are ``Darboux" coordinates with respect the holomorphic Poisson structure on the moduli space of flat  connections
\begin{align}
\{ \cX_\gamma, \cX_{\gamma'} \} = \langle \gamma, \gamma' \rangle \cX_{\gamma + \gamma'},
\end{align}
where $\langle . , . \rangle$ denotes the intersection pairing on $H_1(\Sigma',\mathbb{Z})$. 

\subsection{Higher length-twist coordinates as spectral coordinates}

The coordinate system one obtains from abelianization depends in general on the isotopy class of  the spectral network $\cW$. The spectral coordinates for generalized Fenchel-Nielsen networks of length-twist type are higher length-twist coordinates, in the sense that they satisfy the twist flow property described in \S\ref{sec:higher-length-twist}. The proof of this is a straight-forward generalization of the argument in \cite{Hollands:2013qza}, where it was shown that the spectral coordinates corresponding to Fenchel-Nielsen networks are Fenchel-Nielsen length-twist coordinates.

Indeed, let us fix an annulus $A$, corresponding to a maximal boundary, and construct the corresponding basis of equivariant 1-cycles $A_j$ and $B_j$, corresponding to a choice of $A$ and $B$-cycles on the cover $\Sigma$. 

Suppose that the decoration at the annulus $A$ in the $+$ direction is $(i_1, \ldots i_K)$ and that the framing of $\nabla$ in the $+$ direction is $(l_{\alpha_1},\ldots, l_{\alpha_K}).$ Fix a path $\wp$ going around the annulus $A$ in the $+$ direction. Consider the lift $A_j \in H_1(\Sigma,\mathbb{Z})$ of $\wp$ onto sheet $j$. The spectral coordinate $\cX_{A_{j}}$ is equal to the eigenvalue corresponding to the eigenline $l_{\alpha_{j'}}$ with $i_{j'}=j$, which according to \S\ref{sec:higher-length-twist} equals the higher length coordinate $L_j$. 

Fix a 1-cycle $B_j$ that crosses the $j$-th and $K$-th lift of the annulus $A$. Under the twist flow parametrized by $(\Lambda_1,\ldots, \Lambda_{K})$ the section $(s_1, \ldots, s_K) \mapsto (\Lambda_{1}  s_1, \ldots,  \Lambda_{K} s_K).$ This shows that the twist flow acts on $\cX_{B_j}$ as $$\cX_{B_j} \mapsto  (\Lambda_{K })^{-1} \, \Lambda_{j}  \, \cX_{B_j},$$ which according to \S\ref{sec:higher-length-twist} implies that $\cX_{B_j} $ is a higher twist coordinate. 

The 1-cycles $A_j$ and $B_j$ satisfy $$\langle A_j, B_k \rangle =\delta_{jk}$$ and thus indeed correspond to a choice of $A$ and $B$-cycles on the cover $\Sigma$.

\subsection{Representations}

Instead of working directly with flat connections, it will be convenient to work with the corresponding integrated objects, namely the parallel transport maps. As a straight-forward generalization of \S6 of\cite{Hollands:2013qza} we replace flat $SL(K)$-connections of $C$ by $SL(K)$-representations of a groupoid $\cG_C$ of paths on $C$ and equivariant $GL(1)$-connections by equivariant $GL(1)$-representations of a groupoid $\cG_{\Sigma'}$ of paths on $\Sigma'$. The objects of the paths groupoid $\cG_C$ are basepoints on either side of a single wall in the spectral network $\cW$, whereas the objects of the paths groupoid $\cG_{\Sigma'}$ are lifts of these basepoints to the cover $\Sigma'$. Morphisms of the groupoid $\cG_C$ (and $\cG_{\Sigma'}$) are homotopy classes of oriented paths $\wp$ which begin and end at basepoints on $C$ (and their lifts to $\Sigma'$ respectively). Examples of such path groupoids are illustrated in Figure~\ref{fig:SU2-nw-molI-res-paths} and \ref{fig:SU3bifund-nw-res-paths}. Paths that do not cross any walls are coloured light-blue, whereas paths that connect the two basepoints attached to a single wall are coloured red.

\section{Abelianization for higher length-twist networks}\label{sec:ab-l-t}

Let $C$ be a (possibly punctured) surface together with a pants decomposition into pairs of pants with two maximal and one minimal puncture. Also, choose a generalized Fenchel-Nielsen network $\mathcal{W}$ of length-twist type on $C$ relative to the pants decomposition. Our aim in this section is to show that the $\cW$-nonabelianization mapping $\psi_1$ as well the $\cW$-abelianization mapping $\psi_2$ both are bijections. We use the following strategy.    
 
Fix a length-twist type network $\cW$ and a $\cW$-framed flat $SL(K)$ connection $\nabla$. Suppose that we are given a $\cW$-abelianization of $\nabla$. That is, suppose that we are given local bases $$(s^\mathcal{R}_1,\ldots,s^\mathcal{R}_K)$$ on all domains $\mathcal{R}$ of $C \backslash \cW$, and that the transformation $\cS_w$ that relates the bases in adjacent domains, divided by a wall of type $ij$ is of the form
\begin{align}\label{eqn:Swmatrix}
\cS_w = 1 + e_w,
\end{align}
where $e_w$ lies in the 1-dimensional vector space Hom$(\cL_i,\cL_j)$. 

The transformations $e_w$ cannot be completely arbitrary. Going around each branch-point of the covering $\Sigma \to C$ gives a constraint on these coefficients, just like going around each joint of the spectral network $\cW$. We show that these constraints admit a unique solution for the transformations $\cS_w$, up to abelian gauge equivalence, depending on the choice of $\cW$-framing of $\nabla$. 

Remember that the $\cW$-framing of $\nabla$ is specified by a choice of framing at the maximal punctures and boundaries (see \S\ref{sec:framing}). Instead, we find that the local basis at each minimal puncture may be expressed uniquely in terms of the transformation $\cS_w$ and the framing data at any one of the maximal punctures or boundaries. 

We conclude that there is a unique $\cW$-abelianization of $\nabla$ for every choice of $\cW$-framing of $\nabla$. But at the same time we deduce that there is a unique $\cW$-nonabelianization of the corresponding equivariant $GL(1)$ connection $\nabla^\mathrm{ab}$. Furthermore, we find that all $\cW$-abelianizations of $\nabla$ are obtained in this way: if we have an abelianization of $\nabla$, then the $K$ lines $\iota^{-1}(\mathcal{L}_i)$ for $1 \le i \le K$ at each maximal puncture or maximal boundary must all be eigenlines of the monodromy of $\nabla$ around that maximal puncture or maximal boundary. The only freedom is the choice which of these lines is which eigenline, i.e.~the choice of framing. Hence we find that both mappings $\psi_1$ and $\psi_2$ (from \S\ref{sec:modulispaces}) are bijections.

In the following we first spell out the details for a Fenchel-Nielsen and a generalized Fenchel-Nielsen molecule. We then use the gluing formalism to complete the argument for Fenchel-Nielsen and generalized Fenchel-Nielsen networks of length-twist type on any surface $C$.\footnote{To be precise, we show uniqueness only for networks glued from the molecule illustrated in Figure~\ref{fig:SU3bifund-nw-res-dec}, but we expect it to hold for any network built from the $K=3$ Fenchel-Nielsen molecules of length-twist type.}

\subsection{$K=2$ molecule}\label{sec:molecules}

Fix the Fenchel-Nielsen molecule $\cW$ from Figure~\ref{fig:SU2-nw-molI-res-dec} on the three-holed sphere $C$. This was one of the examples from \cite{Hollands:2013qza} (although we discuss the $\cW$-abelianization in a slightly different way). Say that $\nabla$ is a $\cW$-framed flat $SL(2)$ connection on $C$. The framing corresponds to a choice of eigenlines $l_+$ and $l_-$ at each annulus $A$. Say that $M$ is the eigenvalue corresponding to the eigenline $l_+$ and $M^{-1}$ the eigenvalue corresponding to the eigenline $l_-$.

\insfigscaled{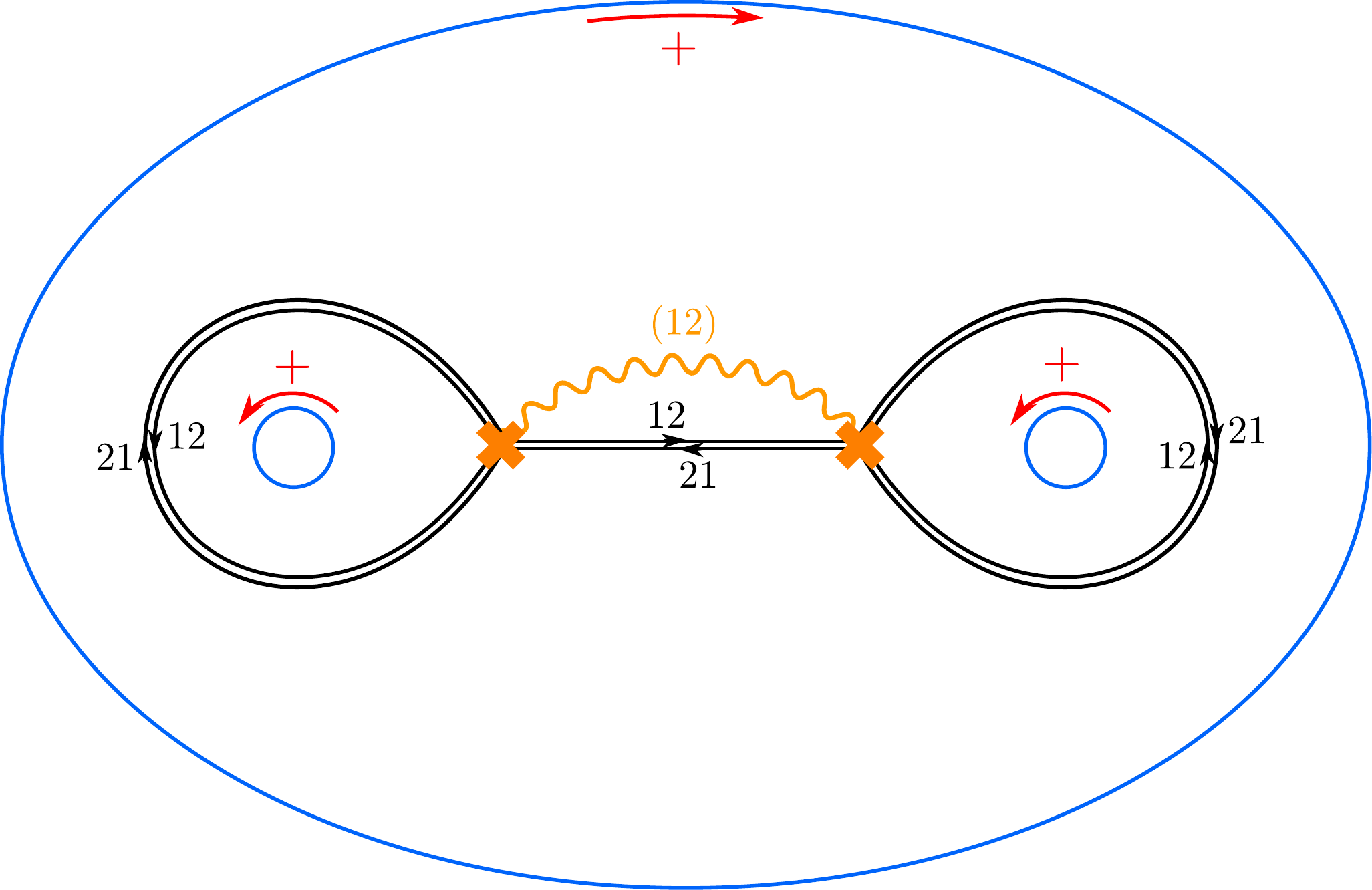}{0.45}{Length-twist network on the three-holed sphere together with a choice of $+$ direction around each hole.}

Choose a trivialization of the covering $\pi: \Sigma \to C$, and suppose that $\nabla$ admits a $\cW$-abelianization. We require that the corresponding $\cW$-abelianization singles out the basis of eigenlines $l_i = l_+$ and $l_j = l_-$ if the decoration in the $+$ direction is $ij$. We will now show that this uniquely determines the $\cW$-abelianization.

The $\cW$-abelianization corresponds to choosing a basis $(s_1,s_2) \in l_1 \oplus l_2$ in each annulus, such that the bases in adjacent domains are related by a transformation $\cS_w$.
Choose base-points and generators of the path groupoids $\cG_C$ and $\cG_{\Sigma'}$ as in Figure~\ref{fig:SU2-nw-molI-res-paths}. The section $s_i$ changes by a constant when parallel transporting it along a light-blue path that does not cross a branch-cut. If the path does cross branch-cuts it furthermore changes sheet accordingly. We encode the parallel transport of the basis $(s_1,s_2)$ along light-blue paths $\wp$ in matrices $D_\wp$ and along red paths $w$, connecting the red dots across walls, in matrices $\cS_w$. 

\insfigscaled{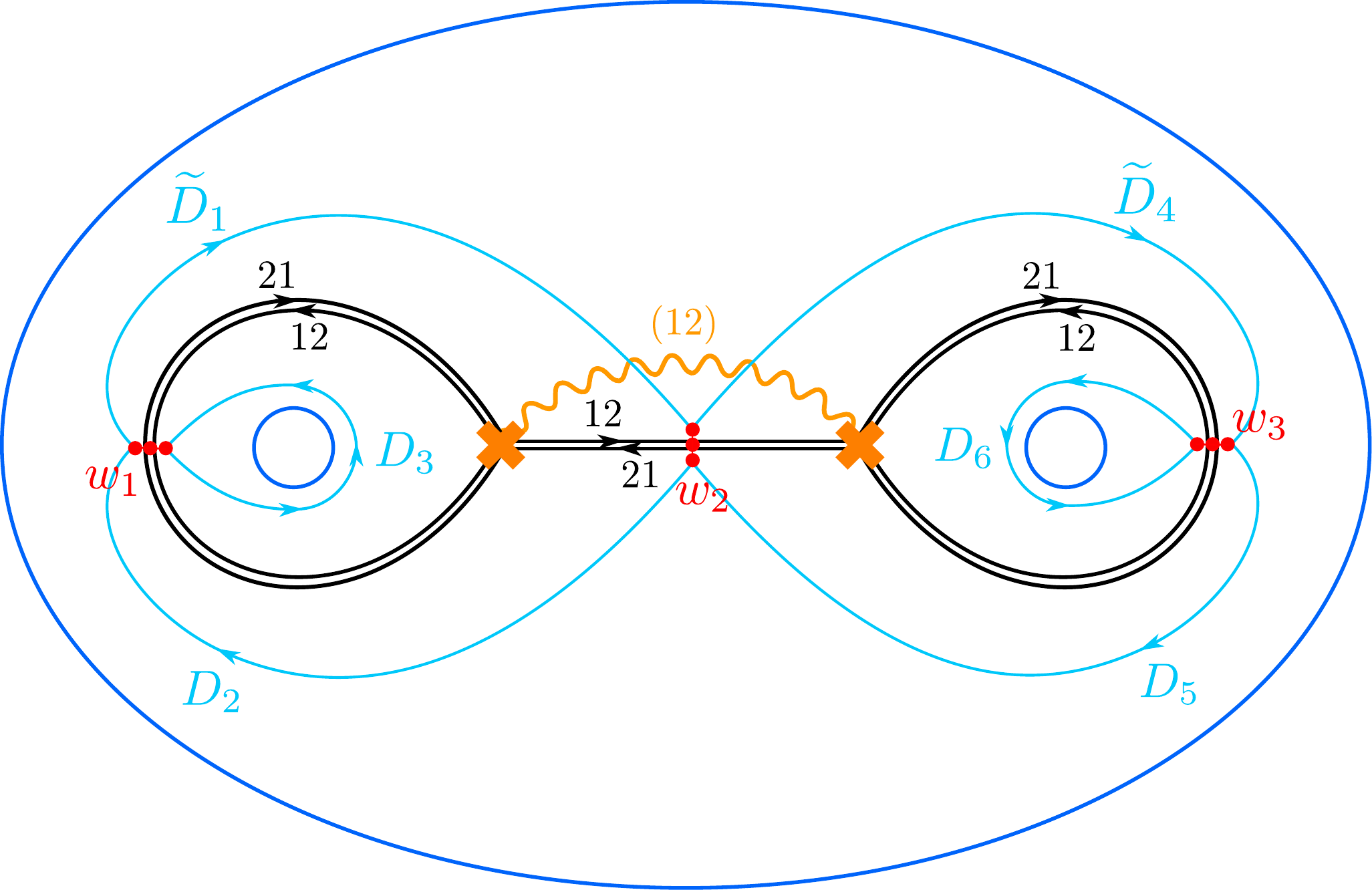}{0.45}{Length-twist network on the three-punctured sphere together with a choice of basepoints (the red dots) and a choice of paths (in light-blue and red). The red paths are admittedly almost invisible in this figure, but connect the red dots across walls. }

The matrices $D_\wp$ are not quite arbitrary, as they encode the parallel transport coefficients of the equivariant flat connection $\nabla^{\mathrm{ab}}$. When going around the boundaries we find the constraints
\begin{align}
D_3 &= \diag(M_0,M_0^{-1})\\
D_6 &= \diag(M_1,M_1^{-1})\\
D_2 D_5 \widetilde{D}_4 \widetilde{D}_1 &= \diag(M_\infty^{-1},M_\infty).
\end{align} 
When going around the branch-points we find the almost-flatness constraints
\begin{align}
(D_3 D_2 \widetilde{D}_1)^2 &= -1\\
(D_6 D_5 \widetilde{D}_4)^2 &= -1.
\end{align}

Together, these abelian flatness constraints determine the matrices $D_\wp$ up to transformations $G_z = \diag(g_z,g_z^{-1})$ at the basepoints $z$ that act on the matrices $D_\wp$ by
\begin{align}
D_\wp \mapsto G_{f(\wp)} D_\wp \, G^{-1}_{i(\wp)},
\end{align}
where $i(\wp)$ is the initial point of the path $\wp$ and $f(\wp)$ its end point. That is, up to abelian gauge transformations, $\cW$-abelization determines a unique equivariant $GL(1)$ connection $\nabla^\mathrm{ab}$ on the cover~$\Sigma$.

It remains to check that there is a unique solution to the transformations $\cS_w$ (up to an abelian gauge transformation).
The transformations $\cS_w$ are constrained by the requirement that following any contractible loop on the base $C$ should result in the identity. Going around either branch-point gives the constraints 
\begin{align}
D_2 \, \cS_{w_2} \, \widetilde{D}_1 \, \cS_{w_1}^{-1} \, D_3 \, \cS_{w_1} &= 1 \\
\widetilde{D}_4 \, \cS_{w_2}^{-1} \,  D_5 \, \cS_{w_3}^{-1} \, D_6  \, \cS_{w_3} &= 1,
\end{align}
where 
\begin{align}
\cS_{w_1} &= \left( \begin{array}{cc} 1 & 0 \\ \widetilde{c}_1  & 1\end{array} \right) \left( \begin{array}{cc} 1 & c_1 \\ 0 & 1\end{array} \right) \\
\cS_{w_2} &= \left( \begin{array}{cc} 1 & c_2 \\ 0 & 1\end{array} \right) \left( \begin{array}{cc} 1 & 0 \\ \widetilde{c}_2 & 1\end{array} \right)  \\
\cS_{w_3} &= \left( \begin{array}{cc} 1 & 0 \\ \widetilde{c}_3 & 1\end{array} \right) \left( \begin{array}{cc} 1 & c_3 \\ 0 & 1\end{array} \right) 
\end{align}
Indeed, this has a unique solution with, for instance,
\begin{align}
c_1 =\frac{ g_1^2  \left(1-M_0 M_1 M_\infty \right)\left(1-\frac{M_0 M_\infty}{M_1}\right)}{ \left(1-M_\infty^2 \right)}& \qquad \widetilde{c}_1 = -\frac{1}{g_1^2 (1-M_0^2)} , 
\end{align}
where $g_1$ is a coefficient of the abelian gauge transformation at basepoint $1$. 
As explained in \cite{Hollands:2013qza}, the S-wall coefficients $c_z$ and $\widetilde{c}_z$ have an interpretation as abelian parallel transport along auxiliary paths (also sometimes called detour paths). 

Note that that the unique solution to the branch-point constraints crucially depends on the chosen framing at each of the annuli, but in a simple way: changing the framing at any one of the annuli $A_l$ corresponds to replacing $M_l \mapsto M_l^{-1}$ in the expressions for the transformations $\cS_w$.  

\subsection{$K=3$ molecule}

Fix the length-twist type network $\cW$ from Figure~\ref{fig:SU3bifund-nw-res-dec} on the sphere $C$ with two (maximal) holes and one minimal puncture. Say that $\nabla$ is a $\cW$-framed flat $SL(3)$ connection on $C$. As before, the framing corresponds to an ordered tuple of three eigenlines around each boundary component. That is, an ordered tuple $(l_{0,\alpha},l_{0,\beta},l_{0,\gamma})$ at the top annulus and an ordered tuple $(l_{\infty,\alpha},l_{\infty,\beta},l_{\infty,\gamma})$ at the bottom annulus. We will show that there is a canonical $\cW$-abelianization of this framed $\nabla$.

One might ask why we did not introduce framings for minimal punctures. Suppose for the moment that there is a "framing" at a minimal puncture, given by any ordered tuple $(l_{1,\alpha},l_{1,\beta};l_{1,\gamma})$ of eigenlines, where the eigenline $l_{1,\gamma}$ corresponds to the distinguished eigenvalue of the monodromy. It seems like this would introduce a continuous family of abelianizations, but in fact we find in the following that the abelianization constraints determine the tuple $(l_{1,\alpha},l_{1,\beta};l_{1,\gamma})$ uniquely.

\insfigscaled{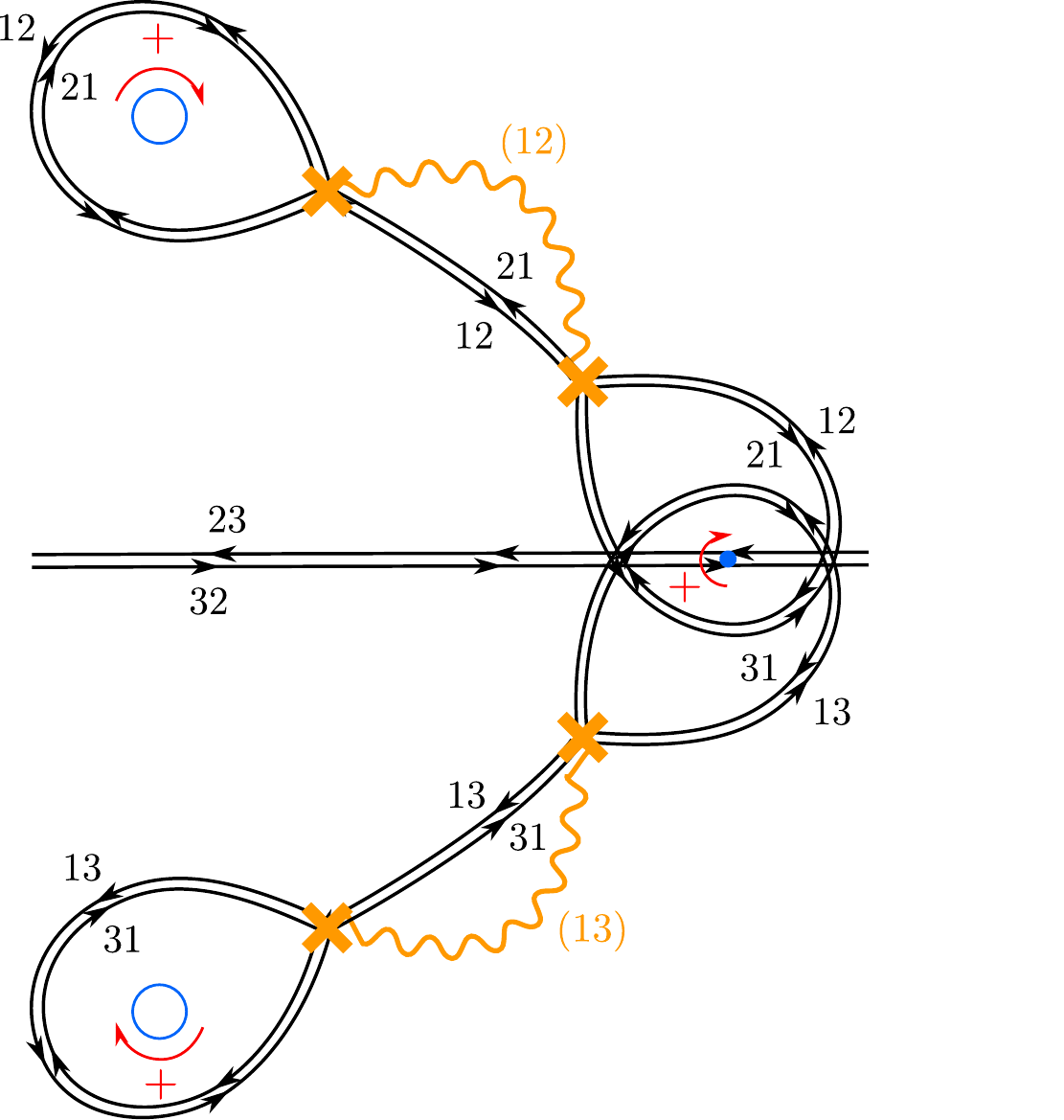}{0.75}{Rank 3 Fenchel-Nielsen molecule together with a $+$ direction at each puncture and hole. }

Choose a trivialization of the covering $\pi: \Sigma \to C$, and suppose that $\nabla$ admits a $\cW$-abelianization. Around each hole $A_l$ we require that the corresponding $\cW$-abelianization has eigenlines $l_{l,i} = l_{l,\alpha}$ and $l_{l,j} = l_{l,\beta}$ and $l_{l,k}=l_{l,\gamma}$ if the decoration in the $+$ direction is $(ijk)$.  Around the minimal puncture we require that the corresponding $\cW$-abelianization has eigenlines $l_{1,i} = l_{1,\alpha}$ and $l_{1,j} = l_{1,\beta}$ and $l_{1,k}=l_{1,\gamma}$ if the decoration in the $+$ direction is $(ij;k)$.

The $\cW$-abelianization corresponds to a basis $(s^\mathcal{R}_1,s^\mathcal{R}_2,s^\mathcal{R}_3)$ on all domains $\mathcal{R}$ of $C \backslash \cW$, such that the bases in adjacent domains are related by a transformation $\cS_w$. Choose base-points and generators of the path groupoids $\cG_C$ and $\cG_{\Sigma'}$ as in Figure~\ref{fig:SU3bifund-nw-res-paths}. As before, we encode the parallel transport of $\nabla$ in ``abelian gauge'' along light-blue paths $\wp$ in matrices $D_\wp$ and along red paths $w$ in matrices $\cS_w$. 

The matrices $D_\wp$ are not arbitrary, as they encode the parallel transport coefficients of the equivariant connection $\nabla^\mathrm{ab}$ on the cover $\Sigma'$. For instance, when going around a loop encircling the top branch-point twice we find the constraint
\begin{align}
(\widetilde{D}_3  D_1 D_2)^2 = \diag(-1,-1,1).
\end{align} 
The abelian holonomies in the $+$ direction around the holes, labeled by $0$ and $\infty$, and around the puncture labeled by $1$ are given in terms of the monodromy eigenvalues as 
\begin{align}
\mathrm{Hol}_0 \nabla^\mathrm{ab} &= \diag(M_{0,1},M_{0,2},(M_{0,1} M_{0,2})^{-1}), \\
\mathrm{Hol}_\infty \nabla^\mathrm{ab} &= \diag(M_{\infty,1},M_{\infty,2},(M_{\infty,1} M_{\infty,2})^{-1}), \\
\mathrm{Hol}_1 \nabla^\mathrm{ab} &= \diag((M_{1})^{-2},M_{1},M_{1}).
\end{align}

Note that, whereas the framing at each annulus fixes the ambiguity of which eigenvalue corresponds to which sheet, for the puncture there is no such ambiguity. The abelian holonomy around a puncture must have coefficient $M_1^{-2}$ for the distinguished sheet, and the coefficient $M_1$ for the two other sheets is the same. 

\insfigscaled{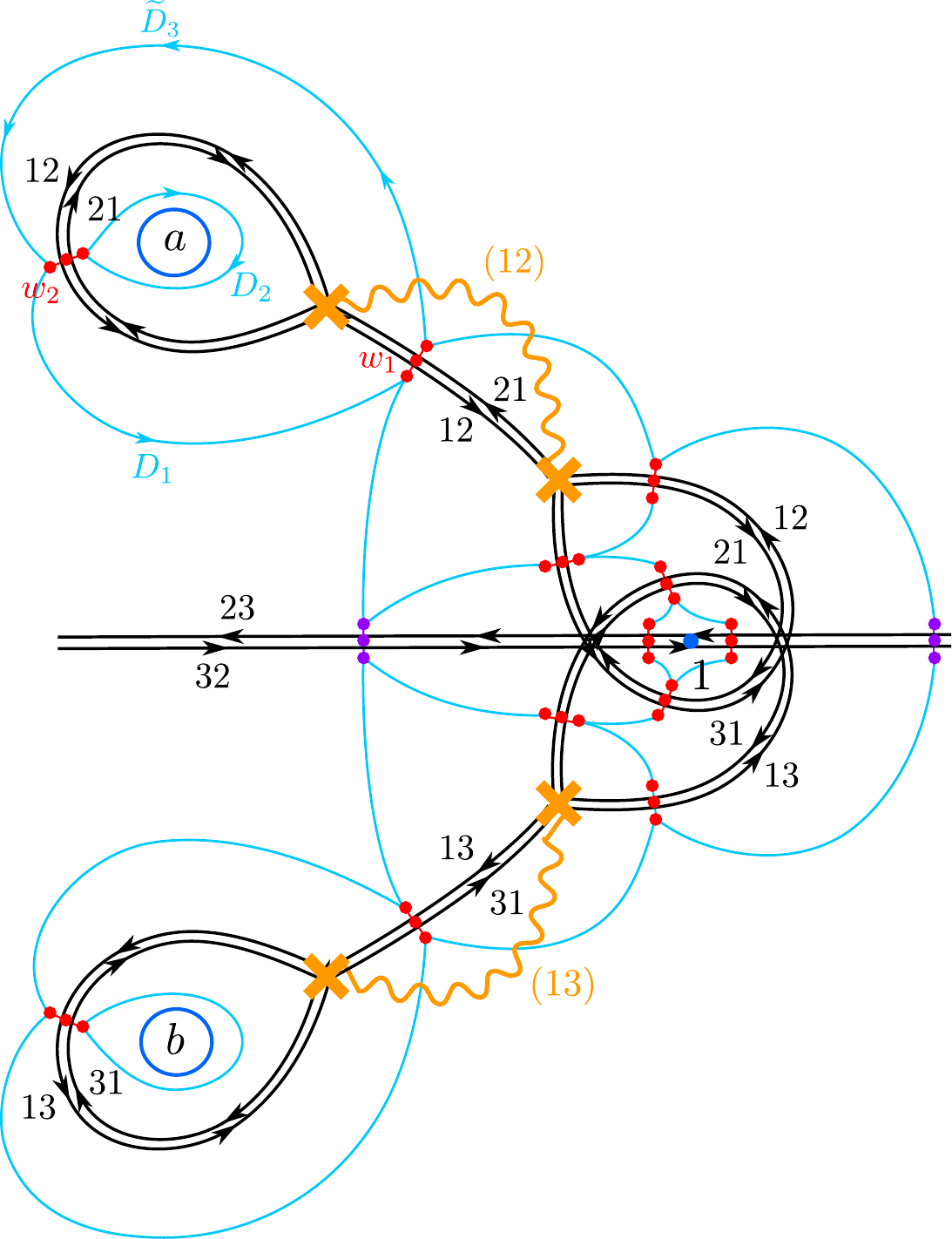}{0.75}{Rank 3 Fenchel-Nielsen molecule together with a choice of basepoints (the red and purple dots) and a choice of paths (in light-blue and red). The purple basepoints should be identified. Even though we have only oriented and labeled a few paths to avoid cluttering of the picture, all paths are oriented and labeled.}

Solving for all abelian flatness constraints shows that the matrices $D_\wp$ are uniquely determined up to abelian gauge transformations. In other words, there is a unique equivariant $GL(1)$-connection $\nabla^{\mathrm{ab}}$ on the cover $\Sigma$.

It remains to check that there is a unique solution to the transformations $\cS_w$, which are constrained by nonabelian "branch-point constraints" and "joint constraints". The former impose that $\nabla$ has trivial monodromy around the branch-points, and the latter that $\nabla$ has trivial monodromy around the joints. For instance, going around the top branch-point gives the constraint  (see Figure~\ref{fig:SU3bifund-nw-res-paths})
\begin{align}
\widetilde{D}_3 \, \cS_{w_1} D_1 \, \cS_{w_2}  D_2  \, \cS_{w_2}^{-1}  = \mathbf{1}.
\end{align} 
Furthermore, we need to enforce the boundary conditions at the punctures. For instance, going around the minimal puncture gives the constraint (see Figure~\ref{fig:SU3bifund-zoom-in})
\begin{align}
\cS_{w_{3,a}} D_4 \, D_5 \, \cS_{w_{4}}^{-1} \, D_6 \, D_7 \, \cS_{w_{3,b}} = \mathrm{Hol}_1 \nabla^\mathrm{ab}, 
\end{align}
where $\cS_{w_{4}} =\cS_{w_{4,b}}  \cS_{w_{4,a}}$.

\insfigscaled{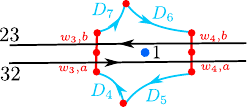}{2.1}{A close-up of Figure~\ref{fig:SU3bifund-nw-res-paths} near the minimal puncture labeled by $1$. The red short paths $w_3$ and $w_4$ are split in half, labeled by the letters $a$ and $b$.}

Solving all these constraints shows that the matrices $\cS_w$ have a canonical solution, which (just like for $K=2$ abelianizations) have an interpretation as parallel transport along auxiliary paths. 

The resulting expressions for the matrices $\cS_w$ depend on the choice of framing at the (maximal) holes through the choice of ordening the eigenvalues in the abelian holonomy matrices $\mathrm{Hol}_{0}{\nabla^\mathrm{ab}}$ and $\mathrm{Hol}_{\infty}{\nabla^\mathrm{ab}}$. In contrast, the particular choice of eigenlines $(l_{1,\alpha},l_{1,\beta};l_{1,\gamma})$ at the minimal puncture doesn't play any role in computing the $\cS_w$. 

Yet, the canonical solution for the transformations $\cS_w$ implies that the basis in any region $C\backslash \cW$, and in particular near the minimal puncture, is uniquely determined in terms of the choice of eigenlines at the boundary components. That is, the abelianization of $\nabla$ canonically determines the choice of eigenlines $(l_{1,\alpha},l_{1,\beta};l_{1,\gamma})$ at the minimal puncture. In particular, there is no framing ambiguity at the minimal puncture after all.

Note that this is consistent with the interpretation of the framing data in terms of the S-wall matrices. Indeed, whereas a change in framing at an annulus $A_l$ corresponds to a permutation of the mass parameters $M_{l,i}$, the permutation group of mass parameters at the minimal puncture is trivial. Generalizing this argument to any regular puncture, we expect that the framing ambiguity at a regular puncture with Young diagram $Y$ is given by the group $S_{n_1} \times \ldots \times S_{n_k}$, where $n_1, \ldots, n_k$ counts columns of $Y$ with the same height and $S_n$ is the permutation group with $n$ elements.

\subsection{Gluing}\label{sec:gluing-ab}

Fix a length-twist type network $\cW$ built out of two molecules. (The same argument can be extended to more molecules.) Say that $\nabla$ is a $\cW$-framed flat connection on $C$ and suppose that $\nabla$ admits a $\cW$-abelianization. 

Choose a pants cycle $\alpha$ relative to $\cW$ and fix a marked point $z_\alpha$ on $\alpha$. The monodromy $\nabla$ along $\alpha$ is diagonal in the ``abelian'' gauge. Cut the surface $C$ along $\alpha$ into two pair of pants $C_1$ and $C_2$. Say $\nabla_1$ is the restriction of $\nabla$ to $C_1$, and $\nabla_2$ the restriction to $C_2$. $\nabla_1$ and $\nabla_2$ are both flat $SL(K)$-connections with trivialization at the marked point $z_\alpha$. 

The $\cW$-abelianization of $\nabla_1$ (as well as $\nabla_2$) is almost the same as described in the previous subsection. In particular, we still find the same unique solution to the S-wall matrices $\cS_w$. The only difference that we have to introduce an additional path $p_{\alpha,1}$ connecting the base-point $z_2$ with $z_\alpha$. The parallel transport matrix $D_{\alpha_1}$ along this path is diagonal and determined by $\nabla_1$. This uniquely fixes the $\cW$-abelianization on $C_1$ (and similarly on $C_2$). 

If we glue back together the three-holed spheres $C_1$ and $C_2$, we can glue the two $\cW$-abelianizations on $C_1$ and $C_2$ to obtain a unique $\cW$-abelianization of $\nabla$. Since we need to divide out by (diagonal) gauge transformations at the marked point $z_\alpha$, the resulting equivariant $GL(1)$ connection $\nabla^\mathrm{ab}$ on $C$ is characterized by its parallel transport along the lifts of the path $p_{\alpha,1} \circ p_{\alpha,2}^{-1}$ to $\Sigma$.  

We conclude that the $\cW$-framed connection $\nabla$ admits a unique $\cW$-abelianization and that the corresponding $\nabla^\mathrm{ab}$ admits a unique $\cW$-nonabelianization. Moreover, as before, different $\cW$-abelianizations of $\nabla$ (without the $\cW$-framing) correspond to different $\cW$-framings.

\section{Monodromy representations in higher length-twist coordinates}\label{section:spectral-monodromy}

In the previous section we have explicitly constructed $\cW$-abelianizations as well as $\cW$-nonabelianizations. With the resulting description of $\nabla$ in terms of the parallel transport matrices $D$ and transformations $\cS_w$ it is a straightforward matter to write down the monodromy representation for $\nabla$ in terms of the spectral coordinates $\cX_\gamma$.  In this section we summarize these monodromy representations in a few examples. 

Recall that any length-twist type network carries a resolution, which can either be American or British. The spectral coordinates $\cX_\gamma$ corresponding to either resolution are generalized Fenchel-Nielsen length-twist coordinates. The spectral length coordinates are the same in either resolution, while the spectral twist coordinates differ (corresponding to the ambiguity in the Fenchel-Nielsen twist coordinates). 

In this section we will see that the NRS Darboux coordinates (i.e. the standard complex Fenchel-Nielsen length-twist coordinates) are only obtained by \ti{averaging} over the two resolutions. More precisely, we define the average higher length-twist coordinates as\footnote{We thank Andrew Neitzke for this suggestion.}  
\begin{align}
L_i &= \cX_{A_i}^+ = \cX_{A_i}^-\\
T_i & = \sqrt{\cX_{B_i}^+ \, \cX^-_{B_i}},
\end{align} 
where $A_i$ and $B_i$ constitute a choice of $A$ and $B$-cycles on the cover $\Sigma$, as defined in \S\ref{sec:spectral-coord}, and $+$ and $-$ refer to the American and the British resolution, respectively. Indeed, we find that the average length and twist agree with the standard length and twist of \S\ref{sec:distinguishedtwist}.

The only left-over ambiguity in the spectral coordinates is an ambiguity in defining the $B$-cycles on the cover $\Sigma$ and a choice of (generalized) Fenchel-Nielsen length-twist network. Resultingly, we find that the higher length-twist coordinates are determined up to a multiplication by a simple monomial in the (exponentiated) mass parameters. 

\subsection{Strategy}

Let us spell out our strategy for the length-twist network $\cW$ on the four-holed sphere $C$ illustrated on the left in Figure~\ref{fig:SU2Nf4}. 

First we cut the four-holed sphere $C$ into two three-holed spheres $C_1$ and $C_2$ along the pants cycle $\alpha$. Say that $C_1$ is the upper and $C_2$ the lower three-holed sphere. Any flat $SL(2)$-connection $\nabla$ restricts to flat connections $\nabla_1$ on $C_1$ and $\nabla_2$ on $C_2$ with fixed trivialization at a marked point $z_\alpha$ at the boundary. The $\cW$-abelianization of $\nabla_1$ is outlined in \S\ref{sec:molecules}. The $\cW$-abelianization of $\nabla_2$ is similar, but with opposite wall labels.  
 
Then we can construct a monodromy representation for $\nabla_1$ with base point $z_{w_2}$ from the matrices
\begin{align}
M_0 &= \cS_{w_2} \widetilde{D}_1 D_2, \\
M_1 &= D_5 \widetilde{D}_4 \cS_{w_2}^{-1}  \\
M_\alpha &= \mathrm{diag}(M_\infty^{-1},M_\infty) 
\end{align}
with
\begin{align}
M_\alpha \, M_1 \, M_0= 1.
\end{align}
Recall that the matrices $D_\wp$ encode the parallel transport coefficients along the paths $\wp$ illustrated in Figure~\ref{fig:SU2-nw-molI-res-paths}.

Applying the same recipe to $\cW_2$ yields a monodromy representation of $\nabla_2$ on $C_2$, generated by the three matrices $M_{0'}$, $M_{1'}$ and $M_{\alpha'}$ with the constraint $M_{\alpha'} \cdot M_{1'} \cdot M_{0'}= \mathbf{1}$. We have that  
\begin{align}
M_{\alpha'} &= \mathrm{diag}(M_\infty,M_\infty^{-1}).
\end{align}

Now glue the three-holed spheres $C_1$ and $C_2$ along $\alpha$ together again, while introducing the matrix 
\begin{align}
P = \mathrm{diag}(p,p^{-1}),
\end{align}
describing the parallel transport of $\nabla$ along the annulus $A$ (from basepoint $w_2'$ to basepoint $w_2$). Then we can construct a monodromy representation for $\nabla$ in terms of the matrices $M_0$, $M_1$, $M_\alpha$, $M_{0'}$, $M_{1'}$, $M_{\alpha'}$ and $\mathbf{P}$. For instance, 
\begin{align}
\mathbf{M}_\beta = M_1 \, P \, M_{1'} \, P^{-1}.
\end{align}

Since $\overline{\Sigma}$ is a torus, the monodromy representation depends on two spectral variables: the abelian monodromy along an A-cycle on $\overline{\Sigma}$, which can be expressed in terms of $M_\infty$, and the abelian monodromy along a B-cycle on $\overline{\Sigma}$, which can be expressed in terms of $p$. 

In the next section we give explicit expressions for invariants constructed from this monodromy representation.

\subsection{$K=2$, four-punctured sphere}\label{sec:monFNfoursphere}

Consider the $K=2$ Fenchel-Nielsen network on the sphere $\mathbb{P}_{0,q,1,\infty}$ with four (maximal) punctures that is illustrated on the left in Figure~\ref{fig:SU2Nf4} (where we have replaced all punctures by holes). We choose counter-clockwise abelian holonomies around the punctures and holes as
\begin{align}
\mathbf{M}^{\mathrm{ab}}_l &= \mathrm{diag}(M_l, \frac{1}{M_l}), 
\end{align}
and two spectral coordinates $\cX_A$ and $\cX_B$ as the abelian holonomies along the 1-cycles $A$ and $B$ that are illustrated on the right in Figure~\ref{fig:SU2Nf4}. These 1-cycles form a basis of 1-cycles on the compactified cover $\overline{\Sigma}$.

\insfigscaled{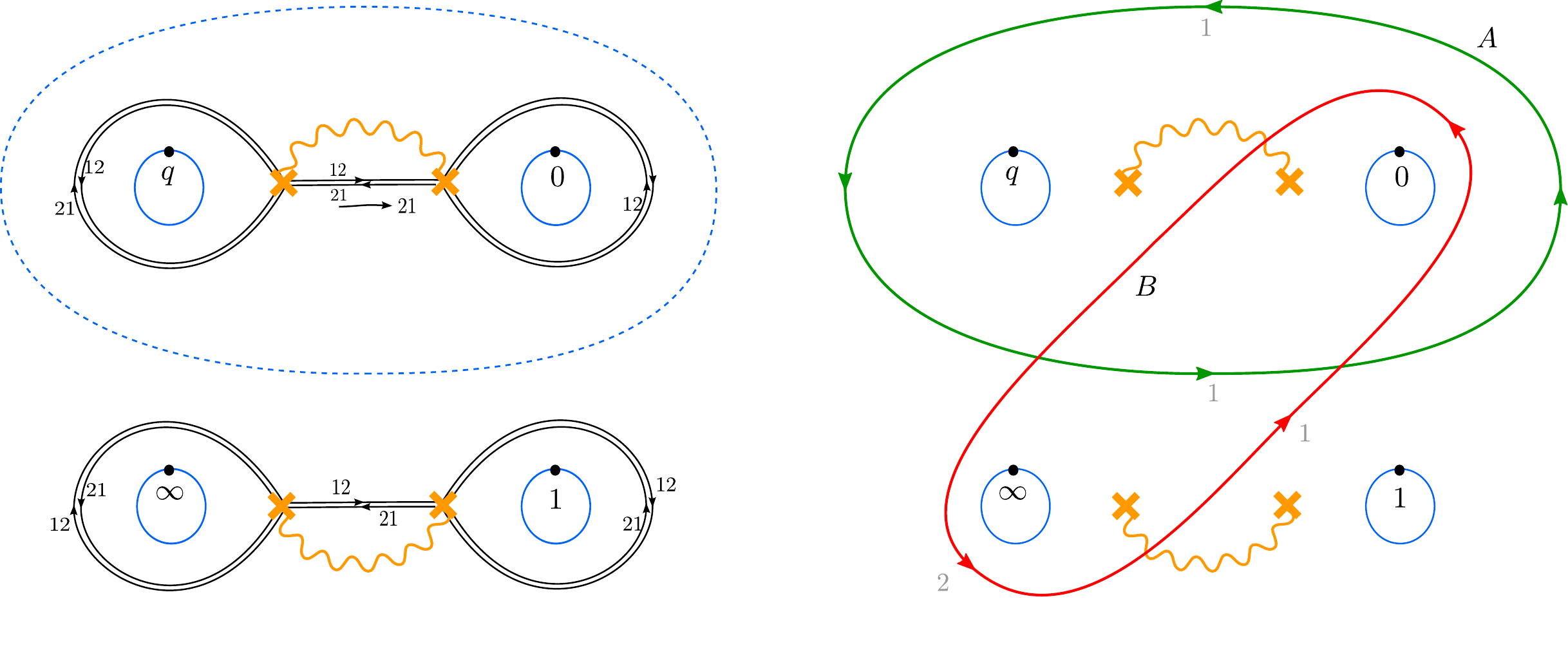}{0.65}{Left: Fenchel-Nielsen network on the four-holed sphere (in the British resolution). Right: basis of 1-cycles $A$ and $B$ on the compactified cover $\overline{\Sigma}$.}

With the $\cW$-abelianization construction the monodromy representation of a generic flat $SL(2)$ connection $\nabla$ can be expressed in terms of the spectral parameters $\cX_A$, $\cX_B$ and the mass parameters $M_0, M, M_1, M_\infty$. Choose generators $\delta_0$, $\delta$, $\delta_1$ and $\delta_\infty$ for the fundamental group of the four-punctured sphere as in Figure~\ref{fig:pi1new}. The corresponding monodromy matrices $M_{\delta_0}$, $M_{\delta}$, $M_{\delta_1}$ and $M_{\delta_\infty}$ with  
\begin{align}
M_{\delta_0} M_{\delta} M_{\delta_\infty} M_{\delta_1} = 1
\end{align}
whose conjugacy classes at the punctures are fixed such that
\begin{align}
\Tr M_{\delta_l} &= M_l + \frac{1}{M_l}. \label{eqn:Mrank2}
\end{align}

Here we focus on the monodromies $\mathbf{M}_\alpha = M_{\delta_0} M_{\delta}$ and $\mathbf{M}_\beta = M_{\delta_0} M_{\delta_\infty} $ (although other monodromies are just as easy to compute). 

In the British resolution, with spectral coordinates\footnote{Here we use the freedom in scaling the exponentiated twist coordinate to simplify the resulting monodromy traces.}
\begin{align}
L &=\cX_A^+ \\
T^+ & = - \cX_A^+ \cX^+_B, 
\end{align}
we find that   
\begin{align}
\Tr \mathbf{M}_{\alpha} &= L + \frac{1}{L} \\
\Tr \mathbf{M}_{\beta} &= N \, T^+ + N_\circ + \frac{1}{T^+}, 
\end{align}
with 
\begin{align}
N &= \frac{(f_L^2 + f_0^2 + f^2 - f_L f_0 f-4) (f_L^2 + f_1^2 + f_\infty^2 - f_L f_1 f_\infty-4)}{(L-\frac{1}{L})^4} \\ 
N_\circ &= \frac{f_L ( f_0 \, f_1 + f \, f_\infty) - 2 (f \, f_1 + f_0 \, f_\infty)}{(L-\frac{1}{L})^2}, \label{eqn:N0rank2}
\end{align}
and where  $f_L = L + \frac{1}{L} $ and $f_l = M_l + \frac{1}{M_l}$.

On the other hand, in the American resolution, with spectral coordinates
\begin{align}
L &=\cX_A^- \\
T^- & = - \cX_A^-  \cX^-_B, 
\end{align}
we find that   
\begin{align}
\Tr \mathbf{M}_{\alpha} &= L + \frac{1}{L} \\
\Tr \mathbf{M}_{\beta} &= T^- + N_\circ + \frac{N}{ T^-}.
\end{align}

Hence, in terms of the average spectral coordinates 
\begin{align}\label{eq:mon4psphereK=2avLT}
L &=\cX_A \\
T & = \cX_A \sqrt{\cX^+_B \cX_B^-}, 
\end{align}
we have that
\begin{align}
\Tr \mathbf{M}_{\alpha} &= L + \frac{1}{L},\label{eq:K=2monMAav} \\
\Tr \mathbf{M}_{\beta} &= \sqrt{N(L)} \left( T + \frac{1}{T} \right) + N_\circ(L), \label{eq:K=2monMBav}
\end{align}

Later, it will be useful that $N$ can be rewritten as 
\begin{align}
N(L) &= \frac{\prod_{i,j \in \{\pm {\frac{1}{2}}\}} \left(L^{\frac{1}{2}} M_0^{i} M^{j} - L^{-\frac{1}{2}} M_0^{-i} M^{-j} \right)\left(L^{\frac{1}{2}} M_1^{i} M_\infty^{j} - L^{-\frac{1}{2}} M_1^{-i} M_\infty^{-j} \right)}{(L-\frac{1}{L})^4}. \label{eqn:Nplus}
\end{align}

Note that the monodromy invariants expressed in terms of the average length-twist coordinates agree with those in \S\ref{sec:distinguishedtwist}. That is, the average length-twist coordinates $L$ and $T$ are the standard exponentiated complex Fenchel-Nielsen length-twist coordinates (which are equal to the NRS Darboux coordinates $\alpha$ and $\beta$).

\subsection{$K=3$, sphere with two minimal and two maximal punctures}

\insfigscaled{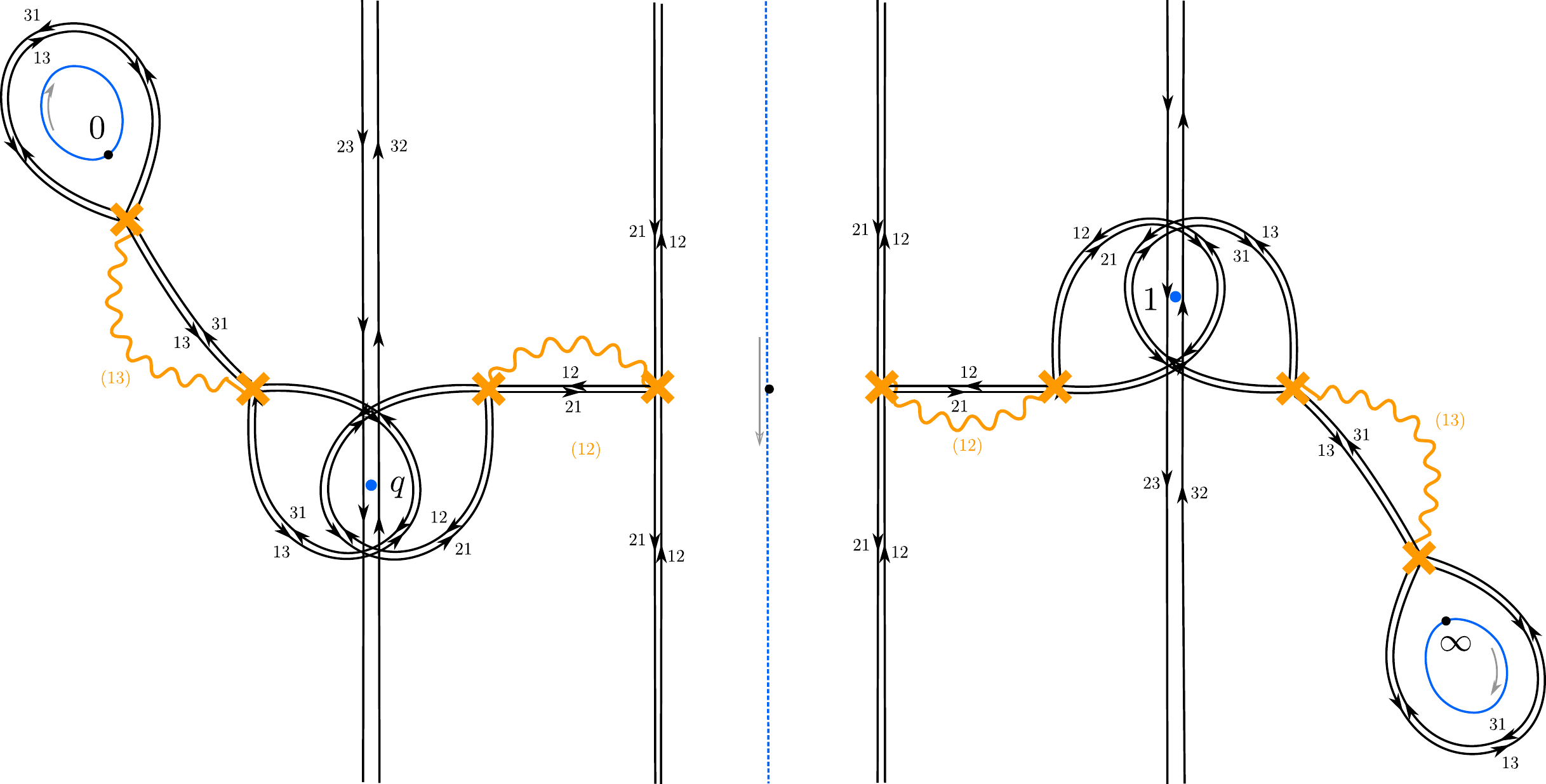}{0.55}{Fenchel-Nielsen network on an open region on the sphere $\mathbb{P}^1_{0,1,q,\infty}$ with two minimal and two maximal punctures (in the American resolution). The complete network on $C$ is found by identifying the end-points of all vertical double walls.}

Next, consider the $K=3$ Fenchel-Nielsen network on the sphere $\mathbb{P}_{0 \underline{q} \underline{1} \infty}$ with two minimal and two maximal punctures that is illustrated in Figure~\ref{fig:SU3Nf6}. We choose counter-clockwise abelian holonomies around the punctures and holes as
\begin{align}
M^{\mathrm{ab}}_0 &= \mathrm{diag}(M_{0,1},M_{0,2},\frac{1}{M_{0,1} M_{0,2}} )   \label{eqn:nonab-masses-23}  \\
M^{\mathrm{ab}} &= \mathrm{diag}(\frac{1}{M^2},M,M ) \label{eqn:nonab-masses-14} \\
M^{\mathrm{ab}}_1 &= \mathrm{diag}(\frac{1}{M_1^2},M_1,M_1 )\\
M^{\mathrm{ab}}_\infty &= \mathrm{diag}(M_{\infty,1},M_{\infty,2},\frac{1}{M_{\infty,1} M_{\infty,2}} ). 
\end{align}

We choose four spectral coordinates $\cX^{1}_A$, $\cX^{2}_A$, $\cX^{1}_B$, $\cX^{2}_B$ as the abelian holonomies along the 1-cycles $A_{1}$, $A_{2}$, $B_{1}$, $B_{2}$, respectively, that are illustrated in Figure~\ref{fig:SU3Nf6-beta-bare}. These 1-cycles form a basis of 1-cycles on the compactified cover $\overline{\Sigma}$.

\insfigscaled{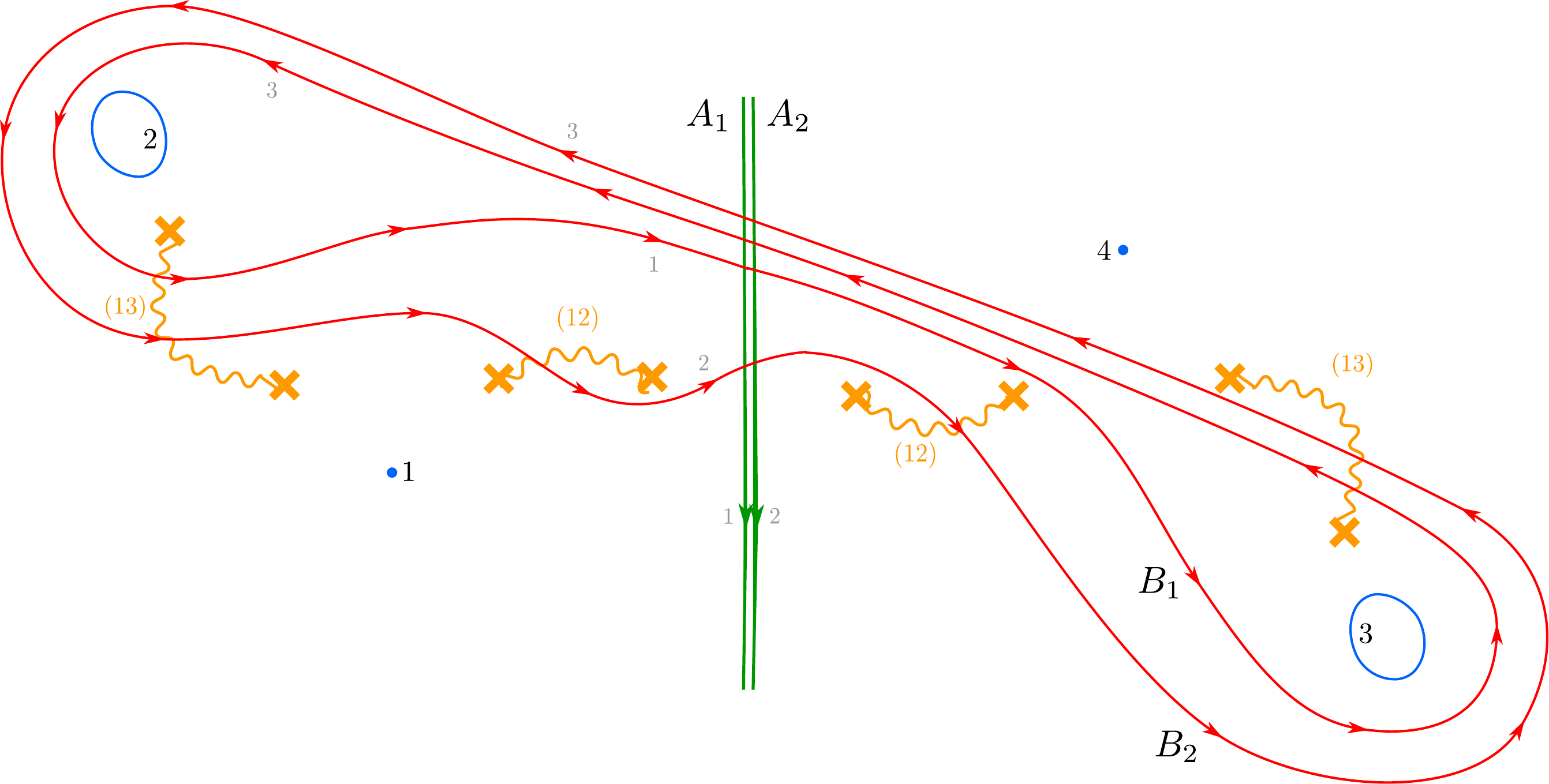}{0.45}{Illustrated in green and red are four 1-cycles $A_1, A_2, B_1,B_2$ which form a basis of 1-cycles on the compactified cover $\overline{\Sigma}$.}

Nonabelianization with respect to the spectral network in Figure~\ref{fig:SU3Nf6}, in either the American or British resolution, yields a family of $SL(3)$ flat connections, depending on the spectral parameters $\cX^{1}_A$, $\cX^{2}_A$, $\cX^{1}_B$, $\cX^{2}_B$ and the mass parameters $M_{0,1}, M_{0,2}, M, M_1, M_{\infty,1}$ and $M_{\infty,2}$. Their monodromy representations can be expressed in terms of $M_{\delta_0}$, $M_{\delta}$, $M_{\delta_1}$ and $M_{\delta_\infty}$ with  
\begin{align}
M_{\delta_0} \, M_{\delta} \, M_{\delta_\infty} \, M_{\delta_1} = 1
\end{align}
whose conjugacy classes at the punctures are fixed such that
\begin{align}
\Tr M_{\delta_0} &= M_{0,1} + M_{0,2} + \frac{1}{M_{0,1} M_{0,2}} \\
\Tr M_{\delta} &= \frac{1}{M^2}  + 2 M \\
\Tr M_{\delta_1} &= \frac{1}{M_1^2}  + 2 M_1 \\
\Tr M_{\delta_\infty} &= M_{\infty,1} + M_{\infty,2} + \frac{1}{M_{\infty,1} M_{\infty,2}}. 
\end{align}

Here we focus on the monodromies $\mathbf{M}_\alpha = M_{\delta_0} \, M_{\delta}$ and $\mathbf{M}_\beta = M_{\delta_0} \,  M_{\delta_\infty} $ (although other monodromies are just as easy to compute). In terms of the average generalized length and twist coordinates\footnote{As before, we used the freedom in scaling the exponentiated twist coordinate to simplify the resulting monodromy traces.}   
\begin{align}
L_1=\cX_{A,1} \qquad & T_1 =  \sqrt{M_{0,2} M_{\infty,2}}  \sqrt{\cX^{+}_{B_1} \cX_{B,1}^{-}}\\
L_2=\cX_{A,2} \qquad & T_2= \sqrt{M_{0,2} M_{\infty,2}}  \, \frac{\cX_{A,1} }{\cX_{A,2} } \, \sqrt{\cX^{+}_{B,2} \cX_{B,2}^{-}}
\end{align}
we find\footnote{The expressions for $\underline{N}_{\circ}$ and $\overline{N}_{\circ}$  are available upon request.} 
\begin{align}
\Tr \mathbf{M}_\alpha &= L_1 + L_2 + L_3, \label{eq:K=3monMAav}\\
\Tr \mathbf{M}_\alpha^{-1} &= \frac{1}{L_1} + \frac{1}{L_2} + \frac{1}{L_3} , \\
\Tr \mathbf{M}_\beta &= \underline{N}_{\circ} + \underline{N}(L_1,L_3) T_1 + \underline{N}(L_2,L_3) T_2 + \underline{N}(L_1,L_2) \frac{T_1}{T_2}  \label{eq:K=3monMBav} \\
& \quad + \underline{N}(L_2,L_1) \frac{T_2}{T_1}  + \frac{\underline{N}(L_3,L_2)}{T_2}+ \frac{\underline{N}(L_3,L_1)}{T_1},  \notag \\
\Tr \mathbf{M}_\beta^{-1} &= \overline{N}_{\circ} + \overline{N}(L_1,L_3) T_1 + \overline{N}(L_2,L_3) T_2 + \overline{N}(L_1,L_2) \frac{T_1}{T_2} \label{eq:K=3monMBavinv} \\
& \quad + \overline{N}(L_2,L_1) \frac{T_2}{T_1} + \frac{\overline{N}(L_3,L_2)}{T_2}+ \frac{\overline{N}(L_3,L_1)}{T_1}.  \notag
\end{align}
where we introduced $L_3 = 1/(L_1 L_2)$.
Furthermore,
\begin{align} 
\underline{N}(L_k,L_l) &= \frac{1}{\sqrt{M M_1}} \frac{N(L_k) N(L_l)}{N_{\star}(L_k, L_l)} \\
\overline{N}(L_k,L_l) &= \sqrt{M M_1} \frac{N(L_k) N(L_l)}{N_{\star}(L_k, L_l)}
\end{align}
are symmetric in $L_k$ and $L_l$, whose numerators are defined by
\begin{align}
N(L_k) &=  M^{- \frac{3}{4}} M_1^{- \frac{3}{4}} \sqrt{(L_k M M_{0,1} - 1)(L_k M M_{0,2} - 1)\left(\frac{L_k M}{M_{0,1} M_{0,2}} -1\right)}  
  \notag \\
& \quad \times \sqrt{\left(\frac{M_1 M_{\infty,1} }{L_k}-1\right)\left(\frac{M_1 M_{\infty,2} }{L_k}-1 \right)\left( \frac{M_1}{L_k M_{\infty,1} M_{\infty,2}}-1\right)}
\end{align}
and whose denominators are defined as
\begin{align}
N_{\star}(L_k,L_l) &=  L_k^{-\frac{5}{2}} L_l^{-\frac{5}{2}}
(L_k-L_l)^2 (1-L_k^2 L_l ) (1- L_k L_l^2 ).
\end{align}

\section{Opers from class $S$}\label{sec:opers}

The moduli space $\mathcal{M}^{\mathcal{C}}_{\mathrm{flat}}(C,SL(K))$ of flat $SL(K)$ connections has a distinguished complex Lagrangian submanifold
\begin{equation}
\mathbf{L} \subset \mathcal{M}_{\mathrm{flat}}^{\mathcal{C}}(C,SL(K))
\end{equation} 
of ``$SL(K)$-opers", known to physicists as the ``brane of opers''.\footnote{These objects were first formalized in \cite{2005math......1398B}, and play an important role in the geometric Langlands program \cite{beilinson1991quantization,2013arXiv1306.0876F}. They also appeared in a conjecture of Gaiotto \cite{Gaiotto:2014bza} as the ``conformal limit" of a certain canonical family of flat connections in the moduli space, which was recently proved in \cite{2016arXiv160702172D}. } 

$SL(K)$ opers are essentially a special kind of $SL(K)$ flat connections, which can locally be written in the form of a (single, scalar) differential equation
\begin{align}\label{eqn:oper}
\mathbf{D} \, y(z) = y^{(K)}(z) + \sum_{i=2}^K t_k(z) y^{(K-i)}(z) = 0,
\end{align}
where globally $y(z)$ is not just a function on $C$, but rather a $(-\frac{K-1}{2})$-differential. The latter ensures that the differential equation is globally well-defined after specifying the transformation laws for the coefficients.  So $\mathbf{D}$ is really an operator between line bundles 
\begin{align}
\mathbf{D}: K_C^{(1-K)/2} \rightarrow K_C^{(K+1)/2}\otimes \mathcal{O}(K \cdot D),
\end{align}
which in the $SL(K)$ case must have vanishing $K-1$th order term in the local form. This definition is equivalent to the one given in the introduction, and the most convenient for our purposes.

With the assumption we are acting on $K_C^{(1-K)/2}$, imposing that the differential equation~(\ref{eqn:oper}) stays invariant under holomorphic coordinate transformations yields the transformation properties of the coefficients $t_k(z)$. As we spell out in detail below, 
\begin{align}
 \frac{12 t_2}{K(K^2-1)}
\end{align}
transforms as a projective connection, whereas we can find linear combinations $w_k$ of $t_j$ ($j \le K$) and its derivatives, such that the $w_k$ transform as $k$-differentials. 
The $SL(K)$ flat connection is obtained from the oper equation~(\ref{eqn:oper}) by writing it instead as a linear rank~$K$ differential equation. 

In this section we characterize the space of opers $\mathbf{L}$ associated to any theory $T_K[C, \mathcal{D}]$ of class $\cS$ with regular defects. The fact that the defects are regular implies that the associated opers are necessarily Fuchsian (i.e. have regular singularities at the punctures of $C$). Our recipe for obtaining $\mathbf{L}$ is similar to the recipe for obtaining the space of differentials associated to $T_K[C, \mathcal{D}]$, as explained in \S\ref{sec:SWgeometry}. That is, we first describe the space of opers on $C$ with only maximal punctures, and then impose appropriate restrictions at the punctures to obtain the space of opers on $C$ with any regular punctures.

An important point is to note that the conjugacy classes are not simply the exponentiated mass parameters, but acquire a slight shift: the space of opers associated to the theory $T_K[C, \mathcal{D}]$ sits inside $\mathcal{M}_\mathrm{flat}^{\mathcal{C}}(C,SL(K))$, where the $\mathcal{C}_l$ are such that 
\begin{align}
M_{l,i}= e^{2 \pi i (m_{l,i}+\frac{K-1}{2})}.
\end{align}
This corresponds to the most symmetric choice of local exponents for an $SL(K)$ oper, and is necessary to ensure the desired equality between the generating function and the superpotential.\footnote{A related mass shift has been observed in the context of the AGT correspondence, see for instance \cite{Okuda:2010ke}.}

In particular, our characterization leads to a concrete description of the space of opers on any three-punctured sphere with regular punctures. These spaces may be seen as the building blocks for the space of opers. For instance, we find that opers on the three-punctured sphere with two maximal and one minimal puncture are characterized by the (generalized) hypergeometric equation. Furthermore, we find that the space of $SL(K)$ opers on the four-punctured surface $\mathbb{P}^1_{0,\underline{q},\underline{1},\infty}$ are characterized by the (generalized) Heun's equation. 

Although in the following we will only spell out the details for $K=2$ and $K=3$, it should be straightforward how to generalize the discussion to find the space of $SL(K)$ opers associated to any building block, and more generally any theory $T_K[C,\mathcal{D}]$ of class $\cS$ with regular punctures.

The locus of opers $\mathbf{L}$ may be interpreted as a quantization of the Coulomb moduli space $\mathbf{B}$ (or equivalently, of the spectral curves sitting above them). Indeed, the internal Coulomb parameters $u$ as well as external mass parameters $m$ carry mass dimensions. It is natural to introduce an additional parameter $\epsilon$ with mass dimension one such that all terms in the oper equation have the same mass dimension. In the semi-classical limit $\epsilon \to 0$ the family of opers then limits to the Coulomb moduli space $\mathbf{B}$. 

\subsection{$SL(2)$ opers}\label{sec:sl2opers}

An $SL(2)$ oper is locally described by a scalar differential equation of the form
\begin{equation}\label{eqn:oper2}
 \mathbf{D}  y = y''(z)+ t_2(z) y(z)= 0,
\end{equation}
where $y(z)$ is a $(-\half)$-differential on $C$. To determine the transformation properties of the coefficient $t_2(z)$, we consider what happens to the differential equation~(\ref{eqn:oper2}) under a holomorphic change of coordinates.

Under the holomorphic coordinate change $z \mapsto z(w)$ the $(-\half)$-differential $y(z)$ transforms into
\begin{align}
\widetilde{y}(w) = y(z(w)) \left( \frac{dz}{dw} \right)^{-\half}. 
\end{align}
This implies that
\begin{align}
\widetilde{y}''(w) 
&= (z'(w))^{\frac{3}{2}} \left( y''(z(w))- \half \{ w,z\} \, y(z(w)) \right).
\end{align}
where the brackets $\{ \cdot, \cdot \}$ denote the Schwarzian derivative
\begin{align}
\{ w , z \} = \frac{w'''(z)}{w'(z)} - 
\frac{3}{2} \left( \frac{w''(z)}{w'(z)} \right)^2 = -\{ z,w\} / z'(w)^2
\end{align}

Under a holomorphic coordinate change $z \mapsto z(w)$ the differential equation~(\ref{eqn:oper2}) thus transforms into
\begin{align}
0 &=  \widetilde{y}''(w) + \widetilde{t}_2(w) \, \widetilde{y}(w) \\
& = (z'(w))^{\frac{3}{2}} \left( y''(z(w))+ \half \{ w,z\} \, y(z(w)) + (z'(w))^{-2} \, \widetilde{t}_2(w) \, y(z(w))  \right), \notag
\end{align}
where we have not specified yet how the coefficient $t_2(z)$ transforms. 

Now, demanding the differential equation~(\ref{eqn:oper2}) be invariant under the holomorphic coordinate change $z \mapsto z(w)$, we find that
\begin{align}\label{eqn:t2transformation}
\widetilde{t}_2(w) &= (z'(w))^2 \left( t_2(z(w)) - \half \{w,z\} \right) \\
& = (z'(w))^2 \, t_2(z(w)) + \half \{z,w\} \notag
\end{align} 
In other words, the coefficient $t_2$ should transform as a so-called projective connection on $C$. 

Observe that the transformation properties~(\ref{eqn:t2transformation}) of the coefficient $t_2$ show that the difference between any two $SL(2)$ opers is a quadratic differential on $C$. Thus the space of  $SL(2)$ opers $\mathbf{L}$ is an affine space modelled on the quadratic differentials.

\subsubsection*{$SL(2)$ flat connection}\label{sec:sl2flat}

The differential equation
\begin{equation}\label{eqn:sl2operlocal}
 \mathbf{D} y = y''(z)+ t_2(z) y(z)= 0,
\end{equation}
can be put in the form of an $SL(2)$ flat connection
\begin{align}\label{eqn:sl2flatconnlocal}
\nabla^\mathrm{oper} Y =  d Y + A Y = \frac{d Y(z)}{dz} dz + A_z \, dz \, Y(z) = 0
\end{align}
where 
\begin{align}
Y(z) = \left( \begin{array}{c} -y'(z) \\ y(z) \end{array} \right) \quad \mathrm{and} \quad A_z = \left( \begin{array}{cc} 0 & -t_2(z) \\ 1 & 0  \end{array} \right).
\end{align}

While under a change of variables $z \to z(w)$ we have that 
\begin{align}
\widetilde{y}(w) = y(z(w)) \left( \frac{dz}{dw} \right)^{-\half} \equiv y(z(w)) s(w),
\end{align}
the section $Y$ transforms as
\begin{align}\label{eq:jet}
\widetilde{Y}(w) = \left( \begin{array}{cc} s(w)^{-1} & - s'(w) \\ 0 & s(w) \end{array} \right) Y(z(w)) \equiv U^{-1}(w) Y(z(w)),
\end{align}
and hence obeys
\begin{align}
d \widetilde{Y} + \widetilde{A} \widetilde{Y} = \frac{d \widetilde{Y}(w)}{dw} dw + \widetilde{A}_w \, dw \, \widetilde{Y}(w) = 0,
\end{align}
with 
\begin{align}
\widetilde{A} = U^{-1} d U + U^{-1} A U = \widetilde{A}_w\, dw.
\end{align}

Since the new connection form is indeed 
\begin{align}
 \widetilde{A}_w = \left( \begin{array}{cc} 0 & - \frac{t_2(z(w))}{s(w)^4} + \frac{s''(w)}{s(w)}  \\ 1 & 0 \end{array} \right) = \left( \begin{array}{cc} 0 & - \widetilde{t}_2(w)  \\ 1 & 0 \end{array} \right),
\end{align}
we find that the $SL(2)$ oper $\mathbf{D}$ defined locally by equation (\ref{eqn:sl2operlocal}) is equivalent to the $SL(2)$ flat connection $\nabla^\mathrm{oper}$ defined locally by equation~(\ref{eqn:sl2flatconnlocal}).

More invariantly, the transformation property~(\ref{eq:jet}) says that $Y$ transforms as  a 1-jet, and implies that we have converted the oper $\mathbf{D}$ into the flat connection $\nabla^\mathrm{oper}$ in the rank~2 vector bundle of 1-jets of $(-\frac{1}{2})$-differentials on $C$. 

\subsubsection*{Fuchsian $SL(2)$ opers}

The space $\mathbf{L}$ of $SL(2)$ opers on any surface $C$ with regular punctures consists of Fuchsian $SL(2)$ opers on $C$, which are locally defined by a Fuchsian differential equation of degree 2. We require that the local exponents of these $SL(2)$ opers at each puncture $z_l$ are given by\footnote{It is a simple exercise to check that the exponents of an $SL(2)$ oper add up to $\frac{3}{2}$ at each puncture, hence (\ref{eqn:localexpSU(2)}) is the most symmetric choice.}
\begin{align}\label{eqn:localexpSU(2)}
\half \pm \frac{m_l}{2},
\end{align}
in terms of the mass parameters $m_l$.
This implies that the Fuchsian $SL(2)$ opers are $SL(2)$ flat connections with a fixed semi-simple conjugacy classes 
\begin{align}
\mathcal{C}_l = \mathrm{diag} \left( - e^{- \pi i m_l }, - e^{ \pi i m_l } \right).
\end{align}
at each puncture $z_l$.
For reference, recall that we fixed the residues of the differentials $\sqrt{\varphi_2}$ in $\mathbf{B}$ at each puncture $z_l$ to be $\pm \frac{m_l}{2}$.

\begin{example} Locus of opers for $T_2[\mathbb{P}^1_{0,1,\infty}]$. 

Recall that for a fixed choice of residues $\pm \frac{m_l}{2}$, the three-punctured sphere $\mathbb{P}^1_{0,1,\infty}$ admits the unique quadratic differential 
\begin{align}
\varphi_2(z) =  - \frac{m_0^2}{4 z^2} - \frac{m_1^2}{4(z-1)^2} - \frac{m^2_{\infty}-m_0^2 - m_1^2}{4 z(z-1)} 
\end{align}
with at most second-order poles at all punctures. The corresponding $SL(2)$ oper is given by
\begin{align}\label{eqn:hypergeom}
\mathbf{D} \, y(z) =  y''(z)+t_2(z) y(z) = 0,
\end{align}
with
\begin{align}
t_2(z) = \frac{\Delta_0}{z^2} + \frac{\Delta_1}{(z-1)^2} + \frac{\Delta_{\infty}-\Delta_0 - \Delta_1}{z(z-1)},
\end{align}
and
\begin{align}
\Delta_l = \frac{1-m_l^2}{4}. 
\end{align}
This is equivalent (after a simple and standard transformation) to the classical Gauss' hypergeometric differential equation. Note that the local exponents of the $SL(2)$ oper~(\ref{eqn:hypergeom})  are indeed given by  $\frac{1}{2} \pm \frac{m_l}{2}$, and that $\eps^2 \,  t_2(z)$ reduces to $\varphi_2(z)$ in the semi-classical limit $\eps \to 0$ discussed in \S\ref{sec:semiclassical}. 

\end{example}

The hypergeometric oper~(\ref{eqn:hypergeom}) in the limit $m_l \to 0$ corresponds to a distinguished projective structure, namely the one induced by the Fuchsian uniformization of $\mathbb{P}^1_{0,1,\infty}$. Indeed, the three-punctured sphere $\mathbb{P}^1_{0,1,\infty}$ is uniformized by the modular lambda function
\begin{align}
\lambda:  \mathbb{H} &\to \mathbb{P}^1_{0,1,\infty} \\
w &\mapsto z=\lambda(w) \notag
\end{align}
which is invariant under the discrete group $\Gamma(2) \subset PSL_2(\mathbb{R})$.

The uniformization oper $\mathbf{D}_{\mathrm{unif}}$ on the three-punctured sphere is thus represented by the differential operator
\begin{align}
\mathbf{D}_{\mathrm{unif}} &= \partial^2_w \, = \partial^2_z + \half \{ w,z \}  = \partial^2_z - \frac{ \{z,w\} }{\, 2 z'(w)^2} \\
&= \partial^2_z + \frac{ 1-z+z^2}{4 z^2 (z-1)^2 }= \partial^2_z +  \frac{1}{4 z^{2}}+\frac{1}{4(z-1)^{2}} - \frac{1}{4 z (z-1)},\label{eqn:unif3punctured} \notag
\end{align}
where in the first line we have used the transformation law for projective connections. Both its local exponents are equal to $\frac{1}{2}$. 

Note that the hypergeometric oper~(\ref{eqn:hypergeom}) itself is of the form  
\begin{align}
\mathbf{D}_{\mathrm{unif}} + \varphi_2.
\end{align} 

\begin{example} Locus of opers for $T_2[\mathbb{P}^1_{0,q,1,\infty}]$. 

The four-punctured sphere $\mathbb{P}^1_{0,q,1,\infty}$ admits the 1-dimensional space of quadratic differentials 
\begin{align}
\varphi_2(z) =  - \frac{m_0^2}{4z^2}- \frac{m^2}{4(z-q)^2}  - \frac{m_1^2}{4(z-1)^2} - \frac{m^2_{\infty}-m_0^2 -m^2 - m_1^2}{4z(z-1)} + \frac{ u}{z(z-q)(z-1)} 
\end{align}
with regular singularities at all punctures. 

The corresponding 1-dimensional family of $SL(2)$ opers are defined by the differential equation
\begin{align}\label{eqn:oper4-punctured}
\mathbf{D}\, y(z) = y''(z)+t_2(z) y(z) = 0,
\end{align}
with
\begin{align}
t_2(z) = \frac{\Delta_0}{z^2} + \frac{\Delta}{(z-q)^2} + \frac{\Delta_1}{(z-1)^2} + \frac{\Delta_{\infty}-\Delta_0 -\Delta - \Delta_1}{z(z-1)} + \frac{ H}{z(z-q)(z-1)},
\end{align}
where $H$ is a free complex parameter, the so-called accessory parameter, 
and
\begin{align}
\Delta_l = \frac{1-m_l^2}{4}. 
\end{align}
The differential equation~(\ref{eqn:oper4-punctured}) is known as Heun's differential equation. It is the most general Fuchsian equation of order 2 with four singularities. 

As before we may write the Heun's opers in the form
\begin{align}
\mathbf{D} = \mathbf{D}_0 + \varphi_2
\end{align}
with respect to the base oper
\begin{equation}\label{eqn:baseop2}
\mathbf{D}_0 = \partial^2_z   + \frac{1}{4 z^2} + \frac{1}{4(z-q)^2} + \frac{1}{4(z-1)^2} - \frac{1}{2z(z-1)} + \frac{\mathrm{const}}{z(z-q)(z-1)},
\end{equation}
but unlike before we are not forced to fix the arbitrary constant. 

\end{example}

In the limit $q \to 0$ the four-punctured sphere $\mathbb{P}^1_{0,q,1,\infty}$ can be thought of as degenerating into two three-punctured spheres. In the same limit, 
the family of Heun's opers degenerates into a pair of hypergeometric opers. 

More precisely, if we define $\ell$ through 
\begin{align}\label{eqn:ell}
H &= \Delta_\ell - \Delta_0 - \Delta + \mathcal{O}(q),
\end{align}
with $\Delta_{l}=\frac{1-\ell^{2}}{4}$ in the limit $q \to 0$, the family of Heun's opers~(\ref{eqn:oper4-punctured}) has two interesting limits:
\begin{enumerate}
\item In the limit $q\to 0$ the family reduces to the hypergeometric oper~(\ref{eqn:hypergeom}) with parameters $(\ell, m_{1}, m_{\infty})$.
\item If we first map $z\mapsto q t$ and then take the limit $q \to 0$, the family reduces to the hypergeometric oper~(\ref{eqn:hypergeom}) with parameters $(m_{0}, m, \ell)$. 
\end{enumerate}
The definition of $\ell$ through equation~(\ref{eqn:ell}) will be justified in \S\ref{sec:monopers}, where it will be the eigenvalue of the monodromy around the pants curve enclosing $0$ and $q$.

Using the AGT correspondence, the effective twisted superpotential for the superconformal $SU(2)$ theory can be found as a series expansion of the accessory parameter $H$ in $q$, through a generalized Matone relation \cite{Alday:2009fs,Drukker:2009id,Ashok:2015gfa}. This is however \emph{not} the route that we take in this paper. Instead, we aim to find the effective twisted superpotential directly from the oper monodromies. (Our strategy might be useful though for establishing similar generalized Matone relations beyond $SU(2)$ theories.)

\subsection{$SL(3)$ opers}

An $SL(3)$ oper is locally described by a differential equation of the form
\begin{equation}\label{eqn:oper3}
 \mathbf{D}  y =  y'''(z) + t_2(z)  y'(z) + t_3(z) y(z) = 0,
\end{equation}
where $y(z)$ is now a section of $K_C^{-1}$, i.e.~a $(-1)$-differential on $C$. To determine the transformation properties of the coefficients $t_2(z)$ and $t_3(z)$, let us consider again what happens to the differential equation~(\ref{eqn:oper3}) under a holomorphic change of coordinates.

Under a holomorphic coordinate change $z \mapsto z(w)$ the $(-1)$-differential $y(z)$ transforms as
\begin{align}
\widetilde{y}(w) = y(z(w)) \left( \frac{dz}{dw} \right)^{-1} \equiv s(w)\, y(z(w)) .
\end{align}
This implies that
\begin{align}
\widetilde{y}'(w) &=   y'(z(w)) + s'(w)  \, y(z(w)) .
\end{align}
and
\begin{align}
\widetilde{y}'''(w) 
&= \frac{y'''(z(w))}{s(w)^2} +  \left( \frac{2 s''(w)}{s(w)} - \frac{s'(w)^2}{s(w)^2} \right)  y'(z(w)) + s'''(w) y(z(w)).
\end{align}

Under the holomorphic coordinate change $z \mapsto z(w)$ the differential equation~(\ref{eqn:oper3}) thus transforms into
\begin{align}
0 &=  \widetilde{y}'''(w) + \widetilde{t}_2(w) \, \widetilde{y}'(w) + \widetilde{t}_3(w) \, \widetilde{y}(w)  \\
& = \frac{1}{s(w)^{2}} \Big( y'''(z(w))+ \left( s(w)^2 \, \widetilde{t}_2(w) + 2 s(w) s''(w) - s'(w)^2 \right) y'(z(w)) +  \\
& \qquad +  \left( s(w)^{3} \, \widetilde{t}_3(w) + s(w)^2 s'(w) \, \widetilde{t}_2(w) + s(w)^2 s'''(w) \right) y(z(w))  \Big),
\end{align}
where we have not specified yet how the coefficients $t_2(z)$ and $t_3(z)$ transform. 

Now, since the differential equation~(\ref{eqn:oper3}) must be invariant under the holomorphic coordinate change $z \mapsto z(w)$, we find that 
\begin{align}\label{eqn:t2transformationsl3}
\widetilde{t}_2(w) &= (s(w))^{-2} \left( t_2(z(w)) - 2 s(w) s''(w) + s'(w)^2  \right) \\
& = (z'(w))^2 \, t_2(z(w)) + 2 \{z,w\}.
\end{align} 
and 
\begin{align}\label{eqn:t3transformationsl3p}
\widetilde{t}_3(w) = \frac{t_3(z(w))}{s(w)^3} -  \frac{s'(w)}{s(w)} \, \widetilde{t}_2(w) - \frac{s'''(w)}{s(w)}. 
\end{align}
Equation~(\ref{eqn:t2transformationsl3}) says that the coefficient $t_2(z)/4$ transforms as a projective connection. 

To find how $t_3(z)$ transforms, we read off from equation~(\ref{eqn:t2transformationsl3}) that
\begin{align}
\half \partial_w \widetilde{t}_2(w) - \half \frac{\partial_z t_2(z(w))}{s(w)^{3} } 
& =  -  \frac{s'(w)}{s(w)} \, \widetilde{t}_2(w) - \frac{s'''(w)}{s(w)}.
\end{align}
Substituting this into equation~(\ref{eqn:t3transformationsl3p}) yields 
\begin{align}\label{eqn:t3transformationsl3}
\widetilde{t}_3(w) - \half \partial_w \widetilde{t}_2(w) & = (z'(w))^3 \left( t_3(z(w)) - \half \partial_z t_2(z(w)) \right). 
\end{align}
In other words, the combination $t_3(z) - \half t'_2(z)$ transforms as a 3-differential. 

\subsubsection*{$SL(3)$ flat connection}

Note that the differential equation
\begin{equation}\label{eqn:sl3operlocal}
 \mathbf{D} y = y'''(z)+ t_2(z) y'(z) + t_3(z) y(z) = 0,
\end{equation}
can be put in the form of an $SL(3)$ flat connection
\begin{align}\label{eqn:sl3flatconnlocal}
\nabla^\mathrm{oper} Y = d Y + A Y = \frac{d Y(z)}{dz} dz + A_z \, dz \, Y(z) = 0
\end{align}
where 
\begin{align}
Y(z) = \left( \begin{array}{c} y''(z) \\ -y'(z) \\ y(z) \end{array} \right) \quad \mathrm{and} \quad A_z = \left( \begin{array}{ccc} 0 & -t_2(z) & t_3(z) \\ 1 & 0 & 0 \\ 0 & 1 & 0 \end{array} \right).
\end{align}

While under a change of variables $z \to z(w)$ we have that 
\begin{align}
\widetilde{y}(w) = y(z(w)) \left( \frac{dz}{dw} \right)^{-1} \equiv y(z(w)) s(w),
\end{align}
the section $Y$ transforms as
\begin{align}\label{eqn:2jet}
\widetilde{Y}(w) = \left( \begin{array}{ccc} s(w)^{-1} &  -\frac{s'(w)}{s(w)} & s''(w) \\ 0 & 1 &  -s'(w) \\ 0 & 0 & s(w) \end{array} \right) Y(z(w)) \equiv U^{-1}(w) Y(z(w)),
\end{align}
and hence obeys
\begin{align}
d \widetilde{Y} + \widetilde{A} \widetilde{Y} = \frac{d \widetilde{Y}(w)}{dw} dw + \widetilde{A}_w \, dw \, \widetilde{Y}(w) = 0,
\end{align}
with 
\begin{align}
\widetilde{A} = U^{-1} d U + U^{-1} A U = \widetilde{A}_w\, dw.
\end{align}

Since the new connection form is indeed 
\begin{align}
 \widetilde{A}_w =  \left( \begin{array}{ccc} 0 & - \widetilde{t}_2(w) & \widetilde{t}_3(w) \\ 1 & 0 & 0 \\ 0 & 1 & 0 \end{array} \right),
\end{align}
with 
\begin{align}
\widetilde{t}_2(w) &=  \frac{t_2(z(w))}{s(w)^2} + \frac{s'(w)^2}{s(w)^2} -\frac{2 s''(w)}{s(w)} \\
\widetilde{t}_3(w) &= \frac{t_3(z(w))}{s(w)^3} - \frac{s'(w) t_2(z(w))}{s(w)^3} - \frac{s'(w)^3}{s(w)^3}+\frac{2 s'(w) s''(w)}{s(w)^2} -\frac{s'''(w)}{s(w)} 
\end{align}
in agreement with equations~(\ref{eqn:t2transformationsl3}) and (\ref{eqn:t3transformationsl3}), we find that the $SL(3)$ oper defined locally by equation (\ref{eqn:sl3operlocal}) is equivalent to the $SL(3)$ flat connection defined locally equation~(\ref{eqn:sl3flatconnlocal}).

More invariantly, the transformation property~(\ref{eqn:2jet}) says that $Y$ transforms as a 2-jet, and implies that we have converted the oper $\mathbf{D}$ into the flat connection $\nabla^\mathrm{oper}$ in the rank 3 vector bundle $E$ of 2-jets of sections of $K_C^{-1}$. 

\subsubsection*{Fuchsian $SL(3)$ opers}

The space $\mathbf{L}$ of $SL(3)$ opers on a surface $C$ with regular punctures consists of all Fuchsian $SL(3)$ opers on $C$, which are locally given by a Fuchsian differential equation of degree 3, with various possible restrictions at each puncture depending on the type of the puncture.

\subsubsection*{Maximal punctures}

Suppose for a moment that the surface $C$ has only maximal punctures. The space of $SL(3)$ opers on $C$ consists of all Fuchsian $SL(3)$ opers on $C$, where we require that the local exponents at each such puncture $z_l$ are given by\footnote{It is a simple exercise to check that the exponents of an $SL(3)$ oper add up to $3$ at each puncture, hence (\ref{eqn:localexpSU(3)}) is the most symmetric choice.} 
\begin{align}\label{eqn:localexpSU(3)}
 1+\frac{m_{l,1}}{2}, \,  1+\frac{m_{l,2}}{2}, \,  1-\frac{m_{l,1}}{2}-\frac{m_{l,2}}{2},
\end{align}
in terms of the mass parameters $m_{l,1}$ and $m_{l,2}$. This implies that the resulting $SL(3)$ opers have fixed semi-simple conjugacy class
\begin{align}\label{eqn:conjclassSU3}
\mathcal{C}_l = \diag \left(e^{ \pi i m_{l,1}} ,e^{ \pi i m_{l,2}}, e^{ - \pi i (m_{l,1}+ m_{l,2}) } \right),
\end{align}
at each maximal puncture $z_l$.

\begin{example} Locus of opers for $T_3[\mathbb{P}^1_{0,1,\infty}]$. 

As summarized in equations (\ref{eqn:K=3phi2}) and (\ref{eqn:K=3phi3}), the three-punctured sphere $\mathbb{P}^1_{0,1,\infty}$ with three maximal punctures admits a 1-dimensional family of differentials
\begin{align}
\varphi_2 &= \frac{c_\infty z^2 - (c_0-c_1+c_\infty) z + c_0}{z^2(z-1)^2} (dz)^2 \\\
\varphi_3 &= \frac{d_\infty z^3 + u z^2 + (d_0+d_1-d_\infty-u) z - d_0}{z^3(z-1)^3} (dz)^3.
\end{align}

The 1-dimensional family of Fuchsian $SL(3)$ opers on $\mathbb{P}^1_{0,1,\infty}$ with three maximal punctures may be parametrized as 
\begin{align}\label{eqn:FuchsianSL3}
y'''(z) + t_2(z) y'(z) + t_3(z) y(z),
\end{align}
with coefficients
\begin{align}
t_2 &=  \frac{(1+ c_\infty) z^2 - (1+ c_0-c_1+c_\infty) z + (1+c_0)}{z^2(z-1)^2} \label{eqn:SL(3)operC301}\\
t_3 &= \frac{d_\infty z^3 + u z^2 + (d_0+d_1-d_\infty-u) z - d_0}{z^3(z-1)^3} + \frac{1}{2} t_2' \label{eqn:SL(3)operC302}.
\end{align}
Its local exponents are indeed given by
\begin{align} 
z=0: &\quad 1+\frac{m_{0,1}}{2},1+\frac{m_{0,2}}{2},1-\frac{m_{0,1}}{2}-\frac{m_{0,2}}{2} \notag \\
z=1: &\quad 1+\frac{m_{1,1}}{2},1+\frac{m_{1,2}}{2},1-\frac{m_{1,1}}{2}-\frac{m_{1,2}}{2}  \label{exponentsSU3}\\
z=\infty: &\quad 1+\frac{m_{\infty,1}}{2},1 +\frac{m_{\infty,2}}{2},1-\frac{m_{\infty,1}}{2}-\frac{m_{\infty,2}}{2}. \notag
\end{align}

The Fuchsian $SL(3)$ oper~(\ref{eqn:FuchsianSL3}) may be written in the form
\begin{align}
y'''(z) +  \left( 4 t^\mathrm{unif}_2 + \varphi_2(z) \right) y'(z) + \left( 2 \partial_z t^\mathrm{unif}_2(z) +  \frac{1}{2} \partial_z \varphi_2(z)  + \varphi_3 \right) y(z), \label{eqn:SL3operphi}
\end{align}
where 
\begin{align}
t^\mathrm{unif}_2(z) = \frac{ 1-z+z^2}{z^2 (z-1)^2 } 
\end{align}
is the coefficient of the Fuchsian uniformization oper~(\ref{eqn:unif3punctured}). The Fuchsian $SL(3)$ oper~(\ref{eqn:FuchsianSL3}) thus again has an interpretation in terms of Fuchsian uniformization. In fact, in the limit $m_{l,i} \to 0$ it is equal to a lift of the Fuchsian uniformatization oper. 

This is most easily seen by rewriting  the $SL(3)$ flat connection defined locally by equation~(\ref{eqn:sl3flatconnlocal}) in the form\footnote{This is the ``canonical form" as in e.g. \cite{2005math......1398B,frenkelloop}, which makes the $\varphi$-action more obvious. This is also the form in which it plays a role in \cite{2016arXiv160702172D}, in the scaling limit of Hitchin section.}
\begin{align}
\frac{d \widetilde{Y}(z)}{dz} dz + \left( \begin{array}{ccc} 0 & - \frac{1}{2} t_2(z) & t_3(z) - \frac{1}{2} \partial_z t_2(z)  \\ 1 & 0 & - \frac{1}{2} t_2(z) \\ 0 & 1 & 0 \end{array} \right) dz \, \widetilde{Y}(z) = 0.
\end{align}
Indeed, in this form it is clear that the $SL(3)$ oper defined locally by
\begin{align}
 y'''(z)+ t_2(z) \, y'(z) + \frac{1}{2} \partial_z t_2(z) \, y(z) = 0,
\end{align}
is the lift of the $SL(2)$ oper defined locally by
\begin{align}
 y'(z) + \frac{1}{4} t_2(z) \, y(z) = 0,
\end{align}
using the homomorphism $\rho: sl(2) \to sl(3)$ given by the spin 1 representation of $sl(2)$. 

The lift of the Fuchsian uniformization oper 
\begin{align}\label{eqn:3unif3punctured}
\mathbf{D}^{(3)}_{\mathrm{unif}}  =& \, \partial_z^3 + 4 t_2^\mathrm{unif} \, \partial_z + 2 \left( \partial_z t_2^\mathrm{unif}  \right)
\end{align}
has all three exponents equal to $1$. 
\end{example} 

More generally, if the underlying surface $C$ has complex structure moduli, such as for instance for $\mathbb{P}^1_{0,q,1,\infty}$, the $SL(3)$ base oper $\mathbf{D}_0^{(3)}$ may be described as the lift (using the homomorphism $\rho: sl(2) \to sl(3)$ given by the spin 1 representation) of the $SL(2)$ base oper $\mathbf{D}_0$, which in the example of $\mathbb{P}^1_{0,q,1,\infty}$ is written down in~(\ref{eqn:baseop2}).

\subsubsection*{Minimal punctures}

Suppose that $C$ has minimal punctures as well. We may obtain the space of Fuchsian $SL(3)$ opers on $C$ with minimal punctures from the locus of Fuchsian $SL(3)$ opers on $C$ with only maximal punctures, by simply enforcing the monodromy around the minimal punctures to be diagonal with two equal eigenvalues, i.e. a multiple of a reflection matrix. 

This requires tuning two of the local exponents at each minimal puncture as well as tuning one internal parameter for each minimal puncture. These constraints can be expressed in terms of the differential equation~(\ref{eqn:oper3}) as follows:
\begin{enumerate}[(i)]
\item We set the mass parameters at the minimal puncture $z_l$ equal to $ m_l \pm 1$. 
\item We require that if we multiply the differential equation 
\begin{align}
\widetilde{\mathbf{D}} y(z) = \mathbf{D}  (z-z_l)^{\frac{1-m_l}{2}} y(z) = 0
\end{align} 
by a factor $(z-z_l)^\frac{1+m_l}{2}$, such that the leading coefficient has an order 1 zero at $z=z_l$, the resulting differential equation has analytic coefficients at $z=z_l$. 
\end{enumerate}

This second condition implies that two of the solutions of the diffential equation $\widetilde{\mathbf{D}} y(z) =0$ are holomorphic at $z=z_l$ (see for instance \cite{Beukers1989}). In return, that implies that the local monodromy of the $SL(3)$ oper defined by the differential operator $\mathbf{D}$ around the puncture $z_l$ is a multiple of a reflection matrix. 

\begin{example} Locus of opers for $T_3[\mathbb{P}^1_{0,\underline{1},\infty}]$. 

Suppose that $z=1$ is a minimal puncture. This imposes the constraints
\begin{align} \label{eq:exponentsminimal}
m_{1,1}^{\mathrm{bif}} &= m_1 - 1 \\
m_{1,2}^{\mathrm{bif}} &= m_1 + 1, \nonumber
\end{align}
as well as 
\begin{align}\label{eqn:uconstraintquantum}
u^{\mathrm{bif}} = \frac{\left( m_1-1 \right) m_1 \left( m_1+1 \right)}{8} - d_0 - 2 d_{\infty} + \frac{m_1}{2} (c_0 - c_{\infty}).
\end{align}
on the family of Fuchsian $SL(3)$ opers defined by the coefficients~(\ref{eqn:SL(3)operC301}) and (\ref{eqn:SL(3)operC302}). This hence fixes the oper uniquely. 

The resulting differential equation 
\begin{align} \label{eqn:genhyperoper}
\mathbf{D}^{\mathrm{bif}} y(z) = 0
\end{align}
can be written in the form of the generalized hypergeometric differential equation
\begin{align}\label{eqn:SL(3)operC21}
 \Big[ z (\theta + \alpha_1)(\theta + \alpha_2)(\theta + \alpha_3) - (\theta + \beta_1-1)(\theta + \beta_2-1)(\theta + \beta_3-1) \Big]\widetilde{y}(z)=0
\end{align}
where $\theta = z \partial_z$, with coefficients
\begin{align}
\alpha_1 &= \tfrac{1}{2}(-m_{\infty,1}+m_1+m_{0,3}-1 + 2 \beta_3), \notag \\
\alpha_2 &= \tfrac{1}{2}(-m_{\infty, 2}+m_1+m_{0,3}-1 + 2 \beta_3), \notag \\
\alpha_3 &=\tfrac{1}{2}( -m_{\infty, 3}+m_1+m_{0,3}-1 + 2 \beta_3),\\
\beta_1&=  \tfrac{1}{2}(- m_{0,1} + m_{0,3} + 2 \beta_3), \notag \\
\beta_2 &= \tfrac{1}{2}( - m_{0,2} + m_{0,3} + 2 \beta_3), \notag
\end{align}
where $m_{0,3} = -m_{0,1} -m_{0,2}$, $m_{\infty,3} = -m_{\infty,1} -m_{\infty,2}$ and 
\begin{align}
\widetilde{y}(z)= z^{-\frac{\beta_1+\beta_2+\beta_3}{3}} (z-1)^{-\frac{\alpha_1+\alpha_2+\alpha_3-\beta_1-\beta_2-\beta_3+3}{3}} y(z).
\end{align}

Indeed, it is known that the monodromy group for the generalized hypergeometric differential equation~(\ref{eqn:genhyperoper}) is rigid and characterized by a reflection matrix at $z=1$ \cite{Beukers1989}.

Finally, comparing the constraint~(\ref{eqn:uconstraintquantum}) on the opers with the constraint~(\ref{eqn:uconstraint}) on the differentials, we notice a (quantum) difference. This implies that the generalized hypergeometric oper cannot be written in the form~(\ref{eqn:SL3operphi}) for any choice of the coefficient $t_2^{\mathrm{unif}}$.

\end{example}

\begin{example} 
Locus of opers for $T_3[\mathbb{P}^1_{0,\underline{q},\underline{1},\infty}]$. 

The space of $SL(3)$ opers on the four-punctured sphere $\mathbb{P}^1_{0,\underline{q},\underline{1},\infty}$ with two maximal punctures at $z=0$ and $z=\infty$ and two minimal punctures at $z=q$ and $z=1$ is 2-dimensional. It may be obtained from the 4-dimensional family of Fuchsian $SL(3)$ opers on the four-punctured sphere $\mathbb{P}^1_{0,q,1,\infty}$ with four maximal punctures by imposing the conditions for a minimal puncture at $z=q$ and $z=1$. 

The resulting family of opers may be written down as the differential equations
\begin{align}\label{eqn:genHeun}
\mathbf{D}\, y (z) = y'''(z) + t_2(z) y'(z) + t_3(z) y(z)=0,
\end{align}
with coefficients
\begin{align}
t_2 &= \frac{1+c_0}{z^2} + \frac{1+c}{(z-q)^2} + \frac{1+c_1}{(z-1)^2} + \frac{c_\infty-c_0-c-c_1-2}{z(z-1)} + \frac{H_1}{z(z-q)(z-1)} \label{eqn:genHeun2} \\
t_3 &=\frac{d_0}{z^3} + \frac{d}{(z-q)^3}  + \frac{d_1}{(z-1)^3} + \frac{d_\infty- d_0 - d-d_1}{z(z-q)(z-1)} + \label{eqn:genHeun3}
\\ & \quad + \frac{ (1-q) (4 c_0 - 3 m^2  - 3 m_1^2 - 4 c_\infty  +6) m_1 }{8 z (z-1)^2 (z-q)}  + \frac{H_2}{z^2(z-q)(z-1)} \notag \\
& \quad - \frac{H_1 }{z(z-1)^{2}(z-q)^{2}} \left(\frac{m_1}{2}(z-q)+\frac{m}{2}(z-1) \right) + \frac{1}{2} t_2'.\notag
\end{align}

We call this the family of \emph{generalized Heun's opers}. For any member of this family the  monodromy around either minimal puncture is semi-simple with two equal eigenvalues. 

Note that the coefficients $t_2$ and $t_3 - \frac{1}{2} t_2'$ from equations~(\ref{eqn:genHeun2}) and (\ref{eqn:genHeun3}) only differ with the differentials~(\ref{eqn:K=3phi24punctures2}) and (\ref{eqn:K=3phi24punctures3}) in terms that have a smaller mass dimension. This difference goes to zero in the semi-classical limit $\eps \to 0$ discussed in section \ref{sec:semiclassical}. 
\end{example}

In the limit $q \to 0$ the four-punctured sphere $\mathbb{P}^1_{0,\underline{q},\underline{1},\infty}$ degenerates into two three-punctured spheres. In the same limit, 
the family of generalized Heun's opers degenerates into a pair of generalized hypergeometric opers. 

If we assume that  
\begin{align}\label{eqn:H10H20}
H_1 &= 1+c_0+c-c_\ell + \mathcal{O}(q)\\
H_2 &= d_0 -  d_\ell + \frac{m}{2} (c_0 - c_\ell) - \frac{(m-1)m(m+1)}{8}+ \mathcal{O}(q)
\end{align}
in the limit $q \to 0$, the family of generalized Heun's opers~(\ref{eqn:genHeun}) has two interesting limits: 
\begin{enumerate}
\item In the limit $q\to 0$ the family reduces to the generalized hypergeometric oper~(\ref{eqn:SL(3)operC21}) with coefficients $(m_{0,i} \mapsto \ell_i)$, $m_{1,i}$ and $m_{\infty,i}$.
\item If we first map $z\mapsto q t$ and then take the limit $q \to 0$, the family reduces to the generalized hypergeometric oper~(\ref{eqn:SL(3)operC21}) with coefficients $m_{0,i}$, $(m_{1,i} \mapsto m_{i})$ and $(m_{\infty,i} \to \ell_i)$. 
\end{enumerate}
The assumptions~(\ref{eqn:H10H20}) will be justified in \S\ref{sec:monopers}.

\subsection{Semi-classical limit}\label{sec:semiclassical}

It is natural to introduce an additional parameter $\epsilon$ with mass dimension 1 such that all terms in the Fuchsian differential equations have the same mass dimension. The corresponding locus of opers $\mathbf{L}_\eps$ is a complex Lagrangian subspace of the moduli space of flat $\eps$-connections. In the semi-classical limit $\epsilon \to 0$ the locus $\mathbf{L}_\eps$ limits to the space of quadratic (and higher if $K>2$) differentials $\mathbf{B}$. 

In the following we often leave out the $\eps$ for uncluttered notation, but at any stage it is a simple matter to reproduce the $\eps$-dependence.

\section{Monodromy of opers}\label{sec:monopers}

In this section we study the monodromy representation of the opers in the locus $\mathbf{L}$ in our main examples. For the superconformal $SU(2)$ theory with four flavors this is Heun's differential equation~(\ref{eqn:oper4-punctured}), while for the superconformal $SU(3)$ theory with six flavors this is its generalization (\ref{eqn:genHeun}) to $K=3$. Since both are families of opers on a punctured sphere, there are no complications due to tricky coordinate transformations, and the monodromy representation is simply found as the fundamental system of solutions to the respective differential equations.  

The relevant differential equations are too complicated for one to write down the monodromy representation explicitly in $q$. We use the fact that the underlying Riemann surface is the four-punctured sphere $\mathbb{P}^1_{0,q,1,\infty}$ and write down the expressions in a series expansion in $q$, following and expanding arguments of \cite{Lay4347,Kazakov2001,Menotti:2014kra}.\footnote{While finishing this paper we noticed that similar strategy has been used in~\cite{2016arXiv160403082I} to calculate the Painlev\'e VI tau-function for small $q$. } We are helped by the fact that the leading contribution when $q \to 0$ is described by the (generalized) hypergeometric differential equation, whose monodromy has been explicitly computed \cite{Beukers1989}.

The same method may be applied to compute the monodromy representation of any family of opers of class $\cS$ in a perturbation series in the complex structure parameters $q$, when analytic expressions are known for the oper monodromies in the limit $q \to 0$.

\subsection{Heun's differential equation}\label{sec:monHeun}

In this subsection we compute the monodromy representation of Heun's differential equation~(\ref{eqn:oper4-punctured}) in a perturbation series in $q$. 

To compare to the monodromy representation~(\ref{eq:mon4psphereK=2avLT}) of any flat $SL(2)$ connection on the four-punctured sphere $\mathbb{P}_{0,q,1,\infty}$ in terms of average length-twist coordinates, we fix the monodromy $\mathbf{M}^\mathrm{oper}_\alpha$ around the punctures $z=1$ and $z=\infty$ such that it has trace
\begin{align}
\Tr \mathbf{M}^\mathrm{oper}_\alpha = -2 \cos (\pi \ell).
\end{align}
The parameter $\ell$ will later play the role of the Coulomb parameter $a$.

As computed in \cite{Menotti:2014kra} , fixing the monodromy $\mathbf{M}^\mathrm{oper}_\alpha$ as before determines a series expansion of the accessory parameter $H$ in $q$  
\begin{align}\label{eqn:Hinell}
H = \sum_{k=0}^\infty q^k \, H_k, 
\end{align}
with for instance
\begin{align}
H_0 = \Delta_\ell - \Delta_0 - \Delta
\end{align}
and 
\begin{align}\label{eq:H1}
H_1 = \frac{(\Delta_\ell-\Delta_0+\Delta)(\Delta_\ell-\Delta_\infty+\Delta_1)}{2 \Delta_\ell} - H_0.
\end{align}

What is left to do is to compute the monodromy $\mathbf{M}^\mathrm{oper}_\beta$ of Heun's differential equation~(\ref{eqn:oper4-punctured}) around the punctures at $z=0$ and $z=\infty$ in a perturbation series in $q$. Our strategy for this is as follows:

\begin{enumerate}

\item We define the rescaled Heun's differential equation by substituting $z=q t$ in Heun's differential equation itself. We construct solutions $v_1(t)$ and $v_2(t)$ of the rescaled Heun's differential equation in a neigbourhood of $t=0$, in a perturbation series in $q$. 

\item We analytically continue the solutions $v_1(t)$ and $v_2(t)$ to $t=\infty$ while keeping $z=q t$ finite, but very small. We re-organize the functions $w_1(z) = v_1 (t/q)$ and $w_2(z) = v_2(t/q)$, which are solutions of Heun's differential equation itself, around $z=0$ in a perturbation series in $q$. 

\item We analytically continue the solutions $w_1(z)$ and $w_2(z)$ to $z=\infty$. 

\end{enumerate}

These three steps together determine the connection matrix $S_\mathrm{total}(q)$ that relates the local solutions of Heun's differential equation near the puncture at $z=0$ to the local solutions near the puncture at $z=\infty$. Say that $M_0$ and $M_\infty$ are the local monodromies around $z=0$ and $z=\infty$, respectively. Then the monodromy matrix of Heun's differential equation around the punctures $z=0$ and $z=\infty$ is found as 
\begin{align}\label{eqn:monoperpertq}
\mathbf{M}^\mathrm{oper}_\beta = M_0  \, S_\mathrm{total}(q) \, M_\infty \, \left(S_\mathrm{total}(q)\right)^{-1}. 
\end{align}
The computation is illustrated in Figure~\ref{fig:4pspherestretchedpaths} and the result is summarized in equation~(\ref{eqn:MonBOperHeunq0+1}). 

\insfigscaled{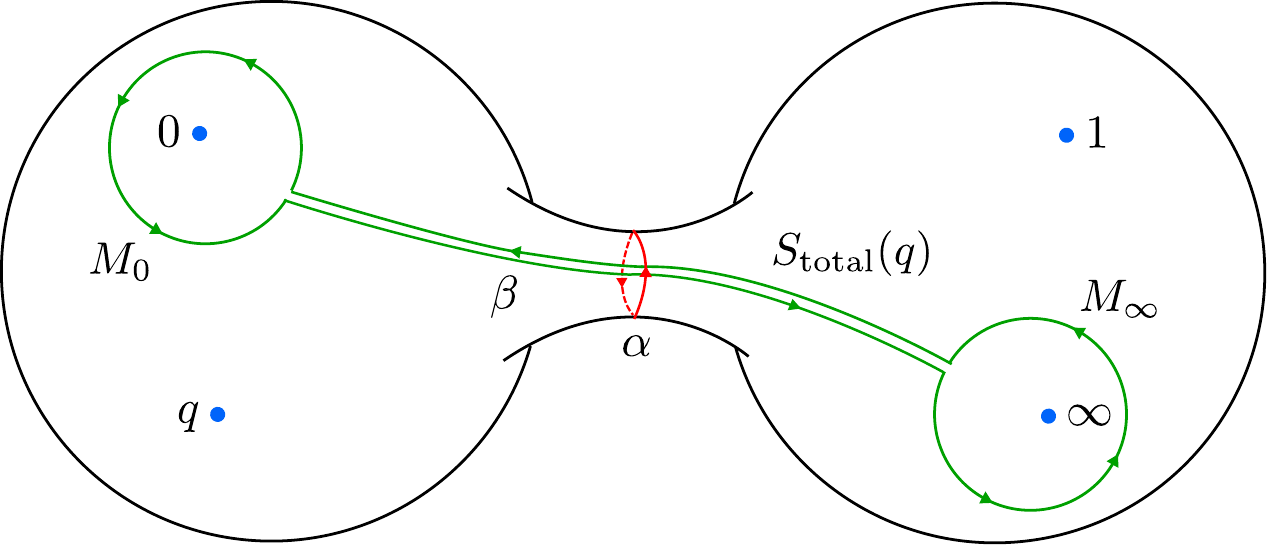}{0.85}{Decomposition of the cycle $\beta$ on $\mathbb{P}^1_{0,q,1,\infty}$ into four paths corresponding to the computation of the monodromy matrix $M_\beta$ on $\mathbb{P}^1_{0,q,1,\infty}$ as in equation~(\ref{eqn:monoperpertq}).}

\subsubsection*{Step 1: Perturbation of the rescaled Heun's differential equation}

We first define the rescaled Heun's differential equation by substituting $z=qt$ in equation~(\ref{eqn:oper4-punctured}) and expand it in a perturbation series in $q$. We also expand its solutions $v(t,\ell)$ in $q$ as
\begin{align}
v(t,\ell) = \sum_{k=0}^\infty q^k \, v^{(k)}(t,\ell).
\end{align}

The leading contribution $v^{(0)}(t,\ell)$ is determined by the hypergeometric differential equation 
\begin{align}
\partial_t^2 \, v^{(0)}(t,\ell) + Q_0(t,\ell) \, v^{(0)}(t,\ell) = 0,
\end{align} 
with 
\begin{align}
Q_0(t,\ell) = \frac{\Delta_0 - (\Delta_\ell - \Delta + \Delta_0)t + \Delta_\ell t^2}{t^2 (1-t)^2}.
\end{align}
We find the two independent solutions
\begin{align}
v^{(0)}_1(t,\ell) &= t^{ \frac{1-m_0}{2}} (1-t)^{\frac{1+m}{2}} 
{_2}F_1 \left(\frac{1-\ell+m-m_0}{2},\frac{1+\ell+m-m_0}{2},1-m_0,t \right) \\
v^{(0)}_2(t,\ell) &= t^{\frac{1+m_0}{2}} (1-t)^{\frac{1+m}{2}} 
{_2}F_1 \left(\frac{1-\ell+m+m_0}{2},\frac{1+\ell+m+m_0}{2},1+m_0,t \right), 
\end{align}
where ${_2}F_1 \left( a,b,c,t \right)$ is the Gauss hypergeometric function
\begin{align}
{_2}F_1 \left( a,b,c,t \right) = \sum_{k=0}^\infty \frac{(a)_k (b)_k }{(c)_k} \frac{t^k}{k!},
\end{align}
with $(x)_k:=x(x+1)\ldots(x+k-1)$ the Pochhammer symbol.

The next-to-leading contribution $v^{(1)}(t,\ell)$ is determined by the equation
\begin{align}
\partial_t^2 \, v^{(1)}(t,\ell) + Q_0(t,\ell) \, v^{(1)}(t,\ell) + Q_1(t,\ell) \, v^{(0)}(t,\ell)= 0,
\end{align}
with 
\begin{align}
Q_1(t,\ell) = \frac{(\Delta_\ell + \Delta_1-\Delta_\infty)(\Delta_\ell+\Delta_1-\Delta_\infty-2 \Delta_\ell t)}{2 \, \Delta_\ell \, t\, (1-t)}.
\end{align}
Its solutions can be found in two ways. 

First, we may use general perturbation theory to write the solution in the form 
\begin{align}\label{eqn:pertv}
v^{(1)}_r(t,\ell) &= S_{r1} v^{(0)}_1(t,\ell) + S_{r2} v^{(0)}_2(t,\ell)  
\end{align}
with 
\begin{align}
S_{r1} &= \frac{1}{m_0} \int_0^t v^{(0)}_2(s,\ell) Q_1(s,\ell) v^{(0)}_r(s,\ell) ds \\
S_{r2} &= -\frac{1}{m_0} \int_0^t v^{(0)}_1(s,\ell) Q_1(s,\ell) v^{(0)}_r(s,\ell) ds.
\end{align}
In a perturbation series in $t$ we find 
\begin{align}
S_{11} =  - t \left( \sigma + \mathcal{O}(t) \right) & \qquad  S_{12} =  - t^{1-m_0} \left( \frac{\sigma}{m_0-1} + \mathcal{O}(t) \right) \\
S_{21} =  - t^{1+m_0} \left( \frac{\sigma}{m_0+1} + \mathcal{O}(t) \right) & \qquad S_{22} =  - t \left( \sigma + \mathcal{O}(t) \right) 
\end{align}
with 
\begin{align}
\sigma = \frac{(\ell^2 +m_0^2 - m^2 -1)(\ell^2 +m_1^2 - m_\infty^2 -1)}{8 m_0 (\ell^2-1)}. 
\end{align}

Note that there is the freedom of adding any multiple of $v^{(0)}_r(t,\ell)$ to $v^{(1)}_r(t,\ell)$. This only changes the boundary conditions of $v_r(t,\ell)$ at $t=0$. The choice made in equation~(\ref{eqn:pertv}) fixes 
\begin{align}
v^{(1)}_r(t=0,\ell)=0.
\end{align}
This choice implies that the $t$-expansion of $v^{(1)}_1(t,\ell)$ starts off with a term proportional to $t^{\frac{3-m_0}{2}}$, and that the $t$-expansion of $v^{(1)}_2(t,\ell)$ starts off with a term proportional to $t^{\frac{3+m_0}{2}}$:
\begin{align}
v^{(1)}_1(t,\ell) &= - t^{\frac{3-m_0}{2}} \left( \frac{(\ell^2+m_0^2 - m^2 -1)(\ell^2 + m_1^2 - m_\infty^2 -1)}{8 (\ell^2-1)(m_0-1)} + \mathcal{O}(t) \right)\\
v^{(1)}_2(t,\ell) &= t^{\frac{3+m_0}{2}} \left( \frac{(\ell^2+m_0^2 - m^2 -1)(\ell^2 + m_1^2 - m_\infty^2 -1)}{8 (\ell^2-1)(m_0+1)} + \mathcal{O}(t) \right)
\end{align}

Alternatively, we could make an ansatz of the form 
\begin{align}\label{eqn:pertLSv}
\widetilde{v}^{(1)}_r(t,\ell) &= A^{(1)}_{r,-1} v^{(0)}_r(t,\ell+2) + A^{(1)}_{r,0} v^{(0)}_r(t,\ell)  + A^{(1)}_{r,1} v^{(0)}_r(t,\ell-2).
\end{align}
This ansatz confirms the value~(\ref{eq:H1}) for $H_1$ and fixes 
\begin{align}
A^{(1)}_{1,-1} &= -\frac{(\ell\pm m - m_0+1)(\ell^2 + m_1^2 - m_\infty^2 -1)}{16 \ell (\ell+1)^2} \\
A^{(1)}_{1,1} &= -\frac{(\ell\pm m + m_0-1)(\ell^2 + m_1^2 - m_\infty^2 -1)}{16 \ell (\ell-1)^2} \\
A^{(1)}_{2,-1} &= -\frac{(\ell\pm m + m_0+1)(\ell^2 + m_1^2 - m_\infty^2 -1)}{16 \ell (\ell+1)^2} \\
A^{(1)}_{2,1} &= +\frac{(\ell\pm m - m_0-1)(\ell^2 + m_1^2 - m_\infty^2 -1)}{16 \ell (\ell-1)^2},
\end{align}
where $(\ell \pm b) = (\ell+b)(\ell-b)$.

The coefficient $A_{r,0}^{(1)}$ is left underdetermined, corresponding to the freedom of adding a multiple of $v^{(0)}_r(t,\ell)$ to $v^{(1)}_r(t,\ell)$. Comparing to equation~(\ref{eqn:pertv}) we choose
\begin{align}
A^{(1)}_{r,0} = - A^{(1)}_{r,1}  - A^{(1)}_{r,-1},
\end{align}
to fix the boundary condition $v^{(1)}_r(t=0,\ell)=0$. Indeed, we then find that 
\begin{equation}
\widetilde{v}^{(1)}_r(t,\ell)  = v^{(1)}_r(t,\ell).
\end{equation}

The expansion of $v^{(1)}(t,\ell)$ in terms of hypergeometric functions as in equation~(\ref{eqn:pertLSv}) will be useful to analytically continue to $t=\infty$. 

We can continue this perturbation to any order in $q$ by expanding \cite{Lay4347}
\begin{align}
v^{(k)}_r(t,\ell) &= \sum_{j=-k}^k A^{(k)}_{r,j} \, v^{(0)}_r(t,\ell-2j).
\end{align}
and find for instance that 
\begin{align}
A^{(2)}_{1,-2} &=\frac{(\ell\pm m - m_0+3)(\ell\pm m - m_0+1)(\ell\pm m_1 \pm m_\infty+1)}{512 \ell (\ell+1)^2(\ell+2)^2(\ell+3)} \\
& \quad +  \frac{(\ell\pm m - m_0+3)(\ell\pm m - m_0+1)(m_1^2-1)}{128 \ell (\ell+1)(\ell+2)(\ell+3)} \notag \\
A^{(2)}_{1,2} &= \frac{(\ell\pm m + m_0-3)(\ell\pm m + m_0-1)(\ell\pm m_1 \pm m_\infty-1)}{512 \ell (\ell-1)^2(\ell-2)^2(\ell-3)} \\
& \quad +  \frac{(\ell\pm m + m_0-3)(\ell\pm m + m_0-1)(m_1^2-1)}{128 \ell (\ell-1)(\ell-2)(\ell-3)} \notag .
\end{align}

\subsubsection*{Step 2: Solutions for $|z|< 1$}

We have just seen that the solutions $v_r(t,\ell)$ of the rescaled Heun equation may be expanded in $q$ as
\begin{align}\label{eqn:v0attinfty}
v_r(t,\ell) =\,  & v_r^{(0)}(t,\ell) \\
& + q \left( A^{(1)}_{r,-1} \, v_r^{(0)}(t,\ell+2) + A^{(1)}_{r,0} \, v_r^{(0)}(t,\ell)  + A^{(1)}_{r,1} \, v_r^{(0)}(t,\ell-2) \right) \notag \\
&  \quad + q^2 \sum_{k=-2}^2 A^{(2)}_{r,k} \, v_r^{(0)}(t,\ell-2k) \notag \\
&  \quad \quad+ \cO(q^3)  \notag
\end{align}
Instead of considering the solutions $v_r(t,\ell)$ around $t=0$, we now want to analytically continue them to $|t|\gg1$. 

For $|t|\gg1$ the hypergeometric functions $v_1^{(0)}(t)$ and $v_2^{(0)}(t)$ may be expanded as 
\begin{align}\label{eqn:approxt>>1}
v^{(0)}_r(t,\ell-2j) & = (-1)^{\frac{1-2m+m_0}{2}}(-t)^{\frac{1-\ell}{2}+j} B_t[\ell-2j]_{r1} \\
& \qquad \qquad \left( 1 + \frac{(\ell- 2j)^2+ m^2 - m_0^2-1}{4(\ell-2j+1) }\, t^{-1} +  \cO\left( t^{-2} \right) \right) \notag \\
& + (-1)^{\frac{1-2m+m_0}{2}}(-t)^{\frac{1+\ell}{2}-j}  B_t[\ell-2j]_{r2} \\
& \qquad \qquad \left( 1 -  \frac{(\ell-2j)^2 + m^2 - m_0^2-1}{4(\ell-2j-1) }\, t^{-1} +  \cO\left(t^{-2} \right) \right) \notag
\end{align}
with
\begin{align}
B_t[\ell] & = \left( \begin{array}{cc} 
\dfrac{\Gamma[-\ell]\Gamma[1- m_0]}{\Gamma[\frac{1- \ell + m - m_0}{2}]\Gamma[\frac{1-\ell- m- m_0}{2}]}  & \dfrac{\Gamma[\ell]\Gamma[1- m_0]}{\Gamma[\frac{1+ \ell+ m - m_0}{2}]\Gamma[\frac{1+\ell- m- m_0}{2}]}  \\
\dfrac{\Gamma[-\ell]\Gamma[1+ m_0]}{\Gamma[\frac{1- \ell+ m + m_0}{2}]\Gamma[\frac{1-\ell- m+ m_0}{2}]} & \dfrac{\Gamma[\ell]\Gamma[1+ m_0]}{\Gamma[\frac{1+ \ell+ m +m_0}{2}]\Gamma[\frac{1+\ell- m + m_0}{2}]} \end{array} \right).
\end{align}
This implies that the solutions $v_r(t,\ell)$ have the expansion
\begin{align}\label{eqn:vpertq}
v_r(t,\ell) =\, (-1)^{\frac{1-2m+m_0}{2}}(-t)^{\frac{1-\ell}{2}} \Big(& \sum_{l=0}^{\infty} B_t[\ell- 2l]_{r1} \, A^{(l)}_{r,l} \, (-z)^l  + \cO\left(t^{-1} \right) \Big)  \\
 \, + (-1)^{\frac{1-2m+m_0}{2}}(-t)^{\frac{1+\ell}{2}}  & \Big( \sum_{l=0}^{\infty} B_t[\ell+2l]_{r2} \, A^{(l)}_{r, -l} \, (-z)^l  + \cO\left( t^{-1} \right) \Big).   \notag
\end{align}
for $|t|\gg 1$, yet $|z| = |qt| < 1$.

\subsubsection*{Leading order in $q$ }

Write the solutions $w_r(z,\ell)= v_r\left(\frac{z}{q},\ell\right)$
to the unrescaled Heun equation~(\ref{eqn:oper4-punctured}) in a $q$-expansion as 
\begin{align}
w_r(z,\ell)= \sum_k q^k \, w^{(k)}_r(z,\ell).
\end{align}
Equation~(\ref{eqn:vpertq}) implies that the leading contribution $w^{(0)}_r(z,\ell)$ is given by 
\begin{align}
w^{(0)}_r(z,\ell) &=  (-1)^{\frac{1-2m+m_0}{2}} \left(-\frac{z}{q}\right)^{\frac{1-\ell}{2}} B_t[\ell]_{r1}  \left( \sum_{l=0}^{\infty} \frac{B_t[\ell- 2l]_{r1} \, A^{(l)}_{r,l}}{B_t[\ell]_{r1} } \, (-z)^l  \right)   \\
&\quad + (-1)^{\frac{1-2m+m_0}{2}} \left(-\frac{z}{q}\right)^{\frac{1+\ell}{2}}  B_t[\ell]_{r2}  \left( \sum_{l=0}^{\infty} \frac{B_t[\ell+2l]_{r2} \, A^{(l)}_{r, -l}}{B_t[\ell]_{r2} } \, (-z)^l  \right).  
\end{align}

The coefficients in front of $(-z)$ are
\begin{align}
\frac{B_t[\ell-2]_{r1} \,  A^{(1)}_{r,1}}{B_t[\ell]_{r1} } & = \frac{(\ell^2+m_1^2 - m_\infty^2-1)}{4(\ell-1)}\\
\frac{B_t[\ell+2]_{r2} \,  A^{(1)}_{r,-1}}{B_t[\ell]_{r2}} & = - \frac{(\ell^2+m_1^2 - m_\infty^2-1)}{4(\ell+1)},
\end{align}
whereas the coefficients in front of $(-z)^2$ are
\begin{align}
\frac{B_t[\ell-4]_{r1} \,  A^{(2)}_{r,2}}{B_t[\ell]_{r1} } & =  
\frac{(\ell-m_1 \pm m_\infty-3)(\ell-m_1 \pm m_\infty-1)}{32(\ell-1)(\ell-2)} \\
& \quad + \frac{(m_1+1)(\ell-m_1 \pm m_\infty-1)}{8 (\ell-1)} + \frac{(m_1^2-1)}{8} \notag \\
\frac{B_t[\ell+4]_{r2} \,  A^{(2)}_{r,-2}}{B_t[\ell]_{r2}} & =  
\frac{(\ell+m_1 \pm m_\infty+3)(\ell+m_1 \pm m_\infty+1)}{32(\ell+1)(\ell+2)} \\
& \quad - \frac{(m_1+1)(\ell+m_1 \pm m_\infty+1)}{8 (\ell+1)} + \frac{(m_1^2-1)}{8}. \notag
\end{align}

This suggests that $w^{(0)}_r(z,\ell)$ can be rewritten in the form
\begin{align}\label{eqn:wrescaledHeun-Heun1}
\left( \begin{array}{c} w^{(0)}_1(z,\ell) \\ w^{(0)}_2(z,\ell) \end{array} \right) 
&= 
B_t[\ell] \, T \left( \begin{array}{c}  y^{(0)}_{1}(z,\ell)  \\   y^{(0)}_{2}(z,\ell) \end{array} \right) 
\end{align}
with 
\begin{align}
T & = \left( \begin{array}{cc}  q^{\frac{\ell-1}{2}} & 0 \\  
0 & q^{\frac{-\ell-1}{2}}  \end{array} \right),
\end{align}
and
\begin{align}
y^{(0)}_{1}(z,\ell) &= (-1)^{\frac{1-2m+m_0}{2}}\,(1-z)^{\frac{1+m_1}{2}} (-z)^{\frac{1-\ell}{2}} \\
& \qquad {_2F_1}\left(\frac{1-\ell+m_1-m_\infty}{2},\frac{1-\ell+m_1+m_\infty}{2},1-\ell,z\right) \notag \\
y^{(0)}_{2}(z,\ell) &= (-1)^{\frac{1-2m+m_0}{2}}\,(1-z)^{\frac{1+m_1}{2}} (-z)^{\frac{1+\ell}{2}}\\
& \qquad{_2F_1} \left(\frac{1+\ell+m_1-m_\infty}{2},\frac{1+\ell+m_1+m_\infty}{2},1+\ell,z\right), \notag
\end{align}
which is a basis of solutions to the unrescaled Heun equation~(\ref{eqn:oper4-punctured})  at $q=0$.

Indeed, since $w^{(0)}_r(z,\ell)$ is a solution to the unrescaled Heun equation at $q=0$, and since we have verified equation~(\ref{eqn:wrescaledHeun-Heun1}) up to order $z^2$, equation~(\ref{eqn:wrescaledHeun-Heun1}) must hold to any order.

\subsubsection*{Next-to-leading order in $q$}

To find the next-to-leading contribution $w^{(1)}_r$ in $q$ we substitute 
the $t^{-1}$-expansion~(\ref{eqn:approxt>>1}) of $v_r^{(0)}$ into the $q$-expansion (\ref{eqn:v0attinfty}) of $v_r$. The resulting expansion is
\begin{align}
w^{(1)}_r(z,\ell) =\, & (-1)^{\frac{1-2m+m_0}{2}}  \left(-\frac{z}{q}\right)^{\frac{1-\ell}{2}}   \sum_{m=-1}^{\infty} W^{(1)}_{1,m} \, z^{m} \\
 &+ \, (-1)^{\frac{1-2m+m_0}{2}}  \left(-\frac{z}{q}\right)^{\frac{1+\ell}{2}} \sum_{m=-1}^{\infty} W^{(1)}_{2,m} \, z^{m} 
\end{align}
with 
\begin{align}
W^{(1)}_{1,-1} &= B_t[\ell]_{r1} \, \frac{\ell^2+m^2-m_0^2-1}{4(\ell+1)}\\
W^{(1)}_{1,0} &=- \, B_t[\ell-2]_{r1} \, A^{(1)}_{r,1}  \frac{(\ell-2)^2+m^2- m_0^2-1}{4(\ell-1)} + B_t[\ell]_{r1} \, A^{(1)}_{r,0}  \\
W^{(1)}_{2,-1} &= - B_t[\ell]_{r2} \,  \frac{\ell^2+m^2-m_0^2-1}{4(\ell-1)} \\
W^{(1)}_{2,0} &=  \, B_t[\ell+2]_{r2} \, A^{(1)}_{r,-1} \frac{(\ell+2)^2+m^2- m_0^2-1}{4(\ell+1)} + \, B_t[\ell]_{r2} \, A^{(1)}_{r,0}
\end{align}
and so forth. 

This expansion is consistent with the closed form
\begin{align}\label{eqn:wrescaledHeun-Heun}
\left( \begin{array}{c} w^{(1)}_1(z,\ell) \\ w^{(1)}_2(z,\ell) \end{array} \right) 
&= 
B_t[\ell] \, T \left( \begin{array}{c}  y^{(1)}_{1}(z,\ell)  \\   y^{(1)}_{2}(z,\ell) \end{array} \right), 
\end{align}
where 
\begin{align}\label{eqn:pertHeunz0}
y^{(1)}_r(z,\ell) &= C^{(1)}_{r,-1} \, y^{(0)}_r(z,\ell+2) + C^{(1)}_{r,0} y^{(0)}_r(z,\ell) + C^{(1)}_{r,1}  y^{(0)}_r(z,\ell-2), 
\end{align}
with coefficients 
\begin{align}
 C^{(1)}_{1,-1} &=  \frac{1-\ell^2+ m_0^2  - m^2 }{4 (\ell+1)}  \\
  C^{(1)}_{1,0} &= \frac{\ell}{4} \left( 1+ \frac{(m_1^2-m_\infty^2)(m_0^2-m^2)}{(\ell+1)^2(\ell-1)^2}  \right) \\
  &\quad -\frac{(m_1^2-m_\infty^2)(m_0^2-m^2)}{4 (\ell+1)^2(\ell-1)^2} +\frac{(m_1^2-m_\infty^2)(1-m_0)}{4 (\ell+1)(\ell-1)} - \frac{m_0}{4} \notag \\ 
   C^{(1)}_{1,1} &= \frac{(\ell^2 + m^2 - m_0^2-1)(\ell \pm m_1 \pm m_\infty-1) }{64 \, \ell(\ell-1)^3 (\ell-2)} 
   \end{align}
   and $C^{(1)}_{2,k}(\ell) = C^{(1)}_{1,k}(-\ell)$. 

Indeed, since $y^{(0)}_{1}(z,\ell) + q y^{(1)}_{1}(z,\ell) + \mathcal{O}(q^2)$ is a solution of the unrescaled Heun equation in a perturbation series in $q$, and since we can verify equation~(\ref{eqn:wrescaledHeun-Heun}) up to second order in $z$, it must hold to any order in $z$.

\subsubsection*{Solutions at $z=\infty$}

Analytically continuing to $z=\infty$ gives 
\begin{align}
 \left( \begin{array}{c} y^{(0)}_{1}(z,\ell) \\ y^{(0)}_{2}(z,\ell) \end{array} \right) 
 \approx B_z[\ell]  \left( \begin{array}{c} 
(-1)^{\frac{1-2m + m_0}{2}} (-z)^{\frac{1-m_\infty}{2}} \\
(-1)^{\frac{1-2m + m_0}{2}} (-z)^{\frac{1+m_\infty}{2}} 
 \end{array} \right)
\end{align}
with
\begin{align}
B_z[\ell] &=  \left( \begin{array}{cc} \dfrac{\Gamma[1-\ell]\Gamma[- m_\infty]}{\Gamma[\frac{1- \ell-m_1 - m_\infty}{2}]\Gamma[\frac{1-\ell+ m_1- m_\infty}{2}]} & 
\dfrac{\Gamma[1-\ell]\Gamma[ m_\infty]}{\Gamma[\frac{1- \ell - m_1 + m_\infty}{2}]\Gamma[\frac{1-\ell+ m_1+ m_\infty}{2}]} \\
\dfrac{\Gamma[1+\ell]\Gamma[- m_\infty]}{\Gamma[\frac{1+ \ell-m_1 - m_\infty}{2}]\Gamma[\frac{1+\ell+ m_1- m_\infty}{2}]}  &
\dfrac{\Gamma[1+\ell]\Gamma[ m_\infty]}{\Gamma[\frac{1+ \ell-m_1 + m_\infty}{2}]\Gamma[\frac{1+\ell+ m_1+ m_\infty}{2}]}  \end{array} \right).
\end{align}

This implies that 
\begin{align}
 \left( \begin{array}{c} y^{(1)}_{1}(z,\ell) \\ y^{(1)}_{2}(z,\ell) \end{array} \right) 
 \approx  \, S_z[\ell] 
 \left( \begin{array}{c} 
(-1)^{\frac{1-2m + m_0}{2}} (-z)^{\frac{1-m_\infty}{2}} \\
(-1)^{\frac{1-2m + m_0}{2}} (-z)^{\frac{1+m_\infty}{2}} 
 \end{array} \right)
\end{align}
where 
\begin{align}
S_z[\ell]_{rs} = 
C^{(1)}_{r,-1} \, B_z[\ell+2]_{rs} + C^{(1)}_{r,0} \, B_z[\ell]_{rs} + C^{(1)}_{r,1} \, B_z[\ell-2]_{rs}  
\end{align}

Hence 
\begin{align}
\left( \begin{array}{c} w_1(z,\ell) \\ w_2(z,\ell) \end{array} \right) 
& \approx S_\mathrm{total}(q)  \left( \begin{array}{c} 
(-1)^{\frac{1-2m + m_0}{2}} (-z)^{\frac{1-m_\infty}{2}} \\
(-1)^{\frac{1-2m + m_0}{2}} (-z)^{\frac{1+m_\infty}{2}} 
 \end{array} \right)
\end{align}
with
\begin{align}
S_\mathrm{total}[q] = B_t[\ell] \, T \, B_z[\ell] \left(1   + q \,  B_z[\ell]^{-1} S_z[\ell] + \mathcal{O}(q^2) \right).
\end{align}

\subsubsection*{Step 4: Monodromy}

Say that 
\begin{align}
M_0 =  \left( \begin{array}{cc} e^{ \pi i (1-m_0) } & 0 \\ 0 & e^{ \pi i (1+m_0) } \end{array} \right) 
\end{align}
and 
\begin{align}
M_\infty =  \left( \begin{array}{cc} e^{ \pi i (1+m_\infty) } & 0 \\ 0 & e^{ \pi i (1-m_\infty) } \end{array} \right) 
\end{align}
are the local monodromy matrices at zero and infinity, respectively. Then the monodromy matrix of Heun's differential equation around the punctures $z=0$ and $z=\infty$ is given by 
\begin{align}
\mathbf{M}^\mathrm{oper}_\beta = M_0  \, S_\mathrm{total}[q] \, M_\infty \, S_\mathrm{total}[q]^{-1}. 
\end{align}

We compute the inverse of $\mathbf{M}_\beta$ using that
\begin{align}\label{eq:inverseBz}
B_z[\ell]^{-1} = B_t[-\ell]_{\{m \to m_1,\, m_0 \to - m_\infty\}}
\end{align}
and that 
\begin{align}
S_\mathrm{total}[q]^{-1} = (1-q \, B_z[\ell]^{-1} \, S_z[\ell]) \, (B_t[\ell] \, T \,  B_z[\ell])^{-1} + \mathcal{O}(q^2).
\end{align}
We then find
\begin{align}
\mathbf{M}^\mathrm{oper}_\beta = \mathbf{M}^{\mathrm{oper},(0)}_\beta  + q \, \mathbf{M}^{\mathrm{oper},(1)}_\beta  + \mathcal{O}(q^2)
\end{align}
with 
\begin{align}\label{eqn:TrMonB0}
\mathbf{M}^{\mathrm{oper},(0)}_\beta = M_0  \, B_t[\ell] \, T \, B_z[\ell] \, M_\infty \, B_z[\ell]^{-1} \, T^{-1} \, B_t[\ell]^{-1}.
\end{align}
and 
\begin{align}
\mathbf{M}^{\mathrm{oper},(1)}_\beta &= M_0  \, B_t[\ell] \, T \, B_z[\ell] \, \left( B_z[\ell]^{-1} \, S_z[\ell] \, M_\infty - M_\infty \, B_z[\ell]^{-1} \, S_z[\ell] \right)  \\
& \qquad \qquad \, B_z[\ell]^{-1} \, T^{-1} \, B_t[\ell]^{-1}.
\end{align}

\subsubsection*{Leading order monodromy}

In the limit $q \to 0$ the four-punctured sphere $\mathbb{P}^1_{0,q,1,\infty}$ may be approximated by gluing two three-punctured spheres $\mathbb{P}^1_{0,1,\infty}$ using the plumbing fixture method. In the same limit Heun's differential equation~(\ref{eqn:oper4-punctured}) may be approximated by the two hypergeometric differential equations~(\ref{eqn:hypergeom}), one on each three-punctured sphere. It is well-known that $B_t[\ell]$ and $B_z[\ell]^{-1}$ are the connection matrices for these hypergeometric differential equations, respectively. Equation~(\ref{eqn:TrMonB0}) shows that the leading order contribution in $q$ to the monodromies of Heun's differential equation may simply be found from the monodromies of the hypergeometric differential equation by splicing in the gluing matrix $T$. (See \cite{Lay4347,Kazakov2001} for an alternative proof.)

To leading order in $q$ we calculate that
\begin{align}
\Tr \mathbf{M}^{\mathrm{oper},0}_\beta & = D_- \, q^{-\ell} + D_\circ +  D_+ \, q^{\ell}, \label{eqn:MonBOperHeunq0}
\end{align}
where
\begin{align}
D_- & = -4 \pi^2  \frac{\Gamma[ 1 + \ell ]^2 \Gamma[\ell ]^2 }{ \Gamma\Big[ \frac{1}{2} + \frac{ \ell \pm m_0 \pm m }{2}\Big] \Gamma\Big[ \frac{1}{2} + \frac{\ell  \pm m_1 \pm m_\infty }{2}\Big]  } \\
D_+ & = -4 \pi^2  \frac{\Gamma[ 1 - \ell ]^2 \Gamma[- \ell ]^2 }{ \Gamma\Big[ \frac{1}{2} - \frac{ \ell \pm m_0 \pm m }{2}\Big] \Gamma\Big[ \frac{1}{2} - \frac{\ell  \pm m_1 \pm m_\infty }{2}\Big]  } 
\end{align}
and
\begin{align}
D_\circ & =  \frac{ \cos (\pi \ell, \pi m_0, \pi m_1) + \cos(\pi \ell, \pi m, \pi m_\infty) + \cos (\pi m_0, \pi m_\infty) + \cos (\pi m, \pi m_1 )}{ \frac{1}{2} \sin^2 (\pi \ell)}, \label{eqn:D0rank2}
\end{align}
where we defined 
\begin{align}
\cos(x_1,\ldots,x_n) = \cos(x_1) \cdots \cos(x_n).
\end{align}

\subsubsection*{Next-to-leading order monodromy}

At next-to-leading order in $q$ we find  
\begin{align}
 B_z[\ell]^{-1} \, S_z[\ell] \, M_\infty - M_\infty \, B_z[\ell]^{-1} \, S_z[\ell] = \left( \begin{array}{cc} 0 & \delta M_+ \\ \delta M_- & 0 \end{array} \right)
\end{align}
with
\begin{align}
\delta M_{+} =  \frac{2 i \pi^2 \left(C_{1,0}^{(1)}-C_{2,0}^{(1)}\right)}{\sin[\pi \ell]} \frac{\Gamma[1 + m_\infty]}{\Gamma[1 - m_\infty]} \frac{1}{ \Gamma\left[ \frac{1 \pm \ell \pm m_1 + m_\infty}{2}\right]}
\end{align}
and 
\begin{align}
\delta M_{-} =  \frac{2 i \pi^2 \left(C_{2,0}^{(1)}- C_{1,0}^{(1)}\right)}{\sin[\pi \ell]} \frac{\Gamma[1 - m_\infty]}{\Gamma[1 + m_\infty]} \frac{1}{ \Gamma\left[ \frac{1 \pm \ell \pm m_1 + m_\infty}{2}\right]}
\end{align}
This leads to
\begin{align}
\Tr \mathbf{M}^{\mathrm{oper},1}_\beta & = D_- \left(C_{2,0}^{(1)}-C_{1,0}^{(1)}\right) \, q^{-\ell}  +  D_+ \left(C_{1,0}^{(1)}-C_{2,0}^{(1)}\right) \, q^{\ell} \label{eqn:MonBOperHeunq1}.
\end{align}
with 
\begin{align}
C_{1,0}^{(1)}-C_{2,0}^{(1)} \, = \, \frac{\ell}{2} \left( 1+ \frac{(m_1^2-m_\infty^2)(m_0^2-m^2)}{(\ell+1)^2(\ell-1)^2}  \right).
\end{align}

\subsubsection*{The result}

Up to order $q$ we thus find that
\begin{align}
\Tr \mathbf{M}^\mathrm{oper}_\beta & = D_- \, q^{-\ell} \left(1- c_1 q  + \mathcal{O}(q^2)\right) + D_\circ +  D_+ \, q^{\ell} \left(1+  c_1 q  + \mathcal{O}(q^2)\right) , \label{eqn:MonBOperHeunq0+1}
\end{align}
with 
\begin{align}
D_- & = -4 \pi^2  \frac{\Gamma[ 1 + \ell ]^2 \Gamma[\ell ]^2 }{ \Gamma\Big[ \frac{1}{2} + \frac{ \ell \pm m_0 \pm m }{2}\Big] \Gamma\Big[ \frac{1}{2} + \frac{\ell  \pm m_1 \pm m_\infty }{2}\Big]  } \\
D_+ & = -4 \pi^2  \frac{\Gamma[ 1 - \ell ]^2 \Gamma[- \ell ]^2 }{ \Gamma\Big[ \frac{1}{2} - \frac{ \ell \pm m_0 \pm m }{2}\Big] \Gamma\Big[ \frac{1}{2} - \frac{\ell  \pm m_1 \pm m_\infty }{2}\Big]  }, 
\end{align}
whereas
\begin{align}
D_\circ & =  \frac{ \cos (\pi m_0, \pi m_\infty) + \cos (\pi m, \pi m_1 ) + \cos (\pi \ell, \pi m_0, \pi m_1) + \cos(\pi \ell, \pi m, \pi m_\infty)}{\frac{1}{2} \sin^2 (\pi \ell)}, \label{eqn:D0rank2}
\end{align}
and
\begin{align}
c_1= \frac{\ell}{2} \left( 1+ \frac{(m_1^2-m_\infty^2)(m_0^2-m^2)}{(\ell+1)^2(\ell-1)^2}  \right).
\end{align}
Using the same techniques one can in principle compute the oper monodromies to any order in $q$.

We rewrite this result in terms of perturbative and instanton corrections to the effective twisted superpotential of the superconformal $SU(2)$ theory coupled to four hypers in \S\ref{sec:SU2superpotential}.

\subsection{Generalized Heun's differential equation}

The monodromies of the generalized Heun's differential equation~(\ref{eqn:genHeun}) may be computed perturbatively in an expansion in $q$ in the same way,. Here we content ourselves with the leading contribution in $q$. 

Again, we start with fixing the coefficients in the expansion
\begin{align}
H_1 &= \sum_{k=0}^\infty q^k H_{1,k} \\
H_2 &= \sum_{k=0}^\infty q^k H_{2,k} 
\end{align}
of the accessory parameters $H_1$ and $H_2$ in equation~(\ref{eqn:genHeun}), by requiring that the monodromy $\mathbf{M}_\alpha^\mathrm{oper}$ around the punctures $z=1$ and $z=\infty$ has traces 
\begin{align}\label{eq:monMAoper}
\mathrm{Tr}\,  \mathbf{M}_\alpha^\mathrm{oper} &= e^{2 \pi i \left(1+\frac{\ell_1}{2}\right)} + e^{2 \pi i \left(1+\frac{\ell_2}{2}\right)} + e^{2 \pi i \left(1-\frac{\ell_1}{2}-\frac{\ell_2}{2}\right)},\\
\mathrm{Tr}  \left( \mathbf{M}_\alpha^\mathrm{oper} \right)^{-1} &= e^{2 \pi i \left(1-\frac{\ell_1}{2}\right)} + e^{2 \pi i \left(1-\frac{\ell_2}{2}\right)} + e^{2 \pi i \left(1+\frac{\ell_1}{2}+\frac{\ell_2}{2}\right)}, \notag
\end{align}
for some fixed complex numbers $\ell_1$ and $\ell_2$.
This determines the leading coefficients~(\ref{eqn:H10H20}), i.e.
\begin{align}
H_{1,0} &= 1+c_0+c-c_\ell, \\
H_{2,0} &= d_0 -  d_\ell + \frac{m}{2} (c_0 - c_\ell) - \frac{(m-1)m(m+1)}{8}.
\end{align}

Recall that, on the one hand, the generalized Heun's equation~(\ref{eqn:genHeun}) limits to the generalized hypergeometric oper~(\ref{eqn:SL(3)operC21}) with coefficients $m_{0,i}$, $(m_{1,i} \mapsto m_{i})$ and $(m_{\infty,i} \to \ell_i)$ if we first replace $z \mapsto q t$ and then take the limit $q \to 0$. A basis of three independent solutions of the limiting generalized hypergeometric differential equation at $t=0$ is given by the generalized hypergeometric functions
\begin{align}
v^{(0)}_r(t) &= t^{1+m_{0,r}} {_3 F_{2}}(\alpha_{r,1},\alpha_{r,2},\alpha_{r,3},\beta_{r,j},\beta_{r,k},t) \\
&= t^{1+m_{0,r}}  \sum_{n=0}^\infty \dfrac{(\alpha_{r,1})_n (\alpha_{r,2})_n (\alpha_{r,3})_n}{(\beta_{r,j})_n (\beta_{r,k})_n} \dfrac{t^n}{n!}, \notag
\end{align}
with $j \neq r$ and $k \neq r$, and with coefficients 
\begin{align}
\alpha_{r,j} &= \frac{1+m+m_{0,r}-\ell_j}{2}, \quad
\beta_{r,j} = 1+\frac{m_{0,r}-m_{0,j}}{2}, 
\end{align}
where $\ell_3 = -\ell_1-\ell_2$ and $m_{0,3} = - m_{0,1}-m_{0,2}$.

The analytic continuation of the solutions $v^{(0)}_r(t)$ from $t=0$ to $t=\infty$ is described by the connection matrix with coefficients 
\begin{align}
B_t[\ell]_{ij} & =  \prod_{k \neq i} \prod_{l \neq j}\,
\dfrac{\Gamma[\frac{\ell_j - \ell_l}{2}] \Gamma[1+ \frac{m_{0,i}-m_{0,k}}{2}] }{\Gamma[ \frac{1-\ell_l+m+m_{0,i}}{2}] \Gamma[ \frac{1 + \ell_j-m- m_{0,k} }{2}]}.
\end{align}

On the other hand, the generalized Heun's equation~(\ref{eqn:genHeun}) limits to the generalized hypergeometric oper~(\ref{eqn:SL(3)operC21}) with coefficients $(m_{0,i} \mapsto \ell_i)$, $m_{1,i}$ and $m_{\infty,i}$ if we just take the limit $q \to 0$. 

The analytic continuation of its solutions $y^{(0)}_i(z)$ from $z=0$ to $z=\infty$ is thus determined by the connection matrix
\begin{align}
B_z[\ell]_{ij} & =   \prod_{k \neq i} \prod_{l \neq j}\,
\dfrac{\Gamma[\frac{m_{\infty,j} - m_{\infty,l}}{2}] \Gamma[1+ \frac{\ell_{i}-\ell_{k}}{2}] }{\Gamma[ \frac{1-m_{\infty,l}+m_1+\ell_{i}}{2}] \Gamma[ \frac{1 + m_{\infty,j}-m_1- \ell_{k} }{2}]}.
\end{align}
Similar to equation~(\ref{eq:inverseBz}) the connection matrices $B_t[\ell]$ and $B_z[\ell]$ are related by
\begin{align}
B_z[l]^{-1} = B_t[-l]_{\{m \mapsto m_1,m_0 \mapsto -m_\infty\}}.
\end{align}

Going through the same steps as for the Heun's differential equation in the previous subsection, we find that the leading contribution to the monodromy matrix of the generalized Heun's equation around the punctures $z=0$ and $z=\infty$ is computed by the expression
\begin{align}
\mathbf{M}_\beta^{\mathrm{oper},0} = M_0 \, B_t[\ell] \, T \, B_z[\ell] \, M_\infty \, B_z[\ell]^{-1} \, T^{-1} \, B_t[\ell]^{-1}, 
\end{align}
where now
\begin{align}
T &= \diag\left(q^{\frac{-\ell_1}{2}}, q^{\frac{-\ell_2}{2}} , q^{-\frac{\ell_3}{2}} \right),\\
M_0 &= \diag \left( e^{2 \pi i \left(1-\frac{m_{0,1}}{2}\right) }, e^{2 \pi i \left(1-\frac{m_{0,2}}{2}\right) }, e^{2 \pi i \left(1-\frac{m_{0,3}}{2} \right) } \right), \\
M_\infty &= \diag \left( e^{2 \pi i \left(1+\frac{m_{\infty,1}}{2}\right) }, e^{2 \pi i \left(1+\frac{m_{\infty,2}}{2}\right) }, e^{2 \pi i \left(1+\frac{m_{\infty,3}}{2} \right) } \right).
\end{align}

We break the computation of $\mathbf{M}_\beta^{\mathrm{oper},0}$ up in smaller pieces. We find that 
\begin{align} 
& \left( B_z[\ell] \, M_\infty \, B_z[\ell]^{-1} \right)_{ij}  = e^{\frac{\pi i}{2} (2m_1-2+\ell_i+\ell_j)} \Big( \delta_{i,j} + \\
& 2i e^{-\frac{3 \pi i}{2} (m_1-1)} \, \dfrac{ \prod_{k=1}^3 \cos\left( \frac{\pi(\ell_j+m_1-m_{\infty,k})}{2} \right) }
{ \prod_{m \neq j} \sin\left( \frac{\pi(\ell_j-\ell_m)}{2} \right) } \,
\frac{ \prod_{k=1}^3 \Gamma \Big[ \frac{1 + \ell_j + m_1  - m_{\infty,k}}{2} \Big] \prod_{l \neq i} \Gamma \Big[ 1 + \frac{\ell_i - \ell_l}{2} \Big] }{ \prod_{k=1}^3 \Gamma \Big[ \frac{1  + \ell_i + m_1- m_{\infty,k}}{2} \Big] \prod_{m \neq j} \Gamma \Big[ 1 + \frac{\ell_j - \ell_m}{2} \Big]   } \Big) \notag,
\end{align}
whereas
\begin{align}
\left( B_t[\ell]^{-1} \, M_0 \, B_t[\ell]  \right)_{ij} = \left( B_z[-\ell] \, M_\infty \, B_z[-\ell]^{-1} \right)_{\{m_1 \mapsto m,m_\infty \mapsto -m_0\},ij}.
\end{align}

Substituting these expressions into $\mathbf{M}_\beta^{\mathrm{oper},0}$,
we find that the leading order contribution to the traces is given by\footnote{Expressions for $\underline{D}_\circ$ and $\overline{D}_\circ$ are available upon request.} 
\begin{align}
\Tr \mathbf{M}^{\mathrm{oper},0}_\beta & = \underline{D}_\circ + \underline{D}(\ell_1,\ell_3) \, q^{ \frac{\ell_3-\ell_1}{2}} + \underline{D}(\ell_2,\ell_3) \,  
q^{\frac{\ell_3-\ell_2}{2}} + \underline{D}(\ell_1,\ell_2) \, q^{\frac{\ell_2-\ell_1}{2}}   \label{eq:monMBoper} \\
& \quad + \underline{D}(\ell_2,\ell_1) \, q^{\frac{\ell_1-\ell_2}{2}} + \underline{D}(\ell_3,\ell_2)\, q^{ \frac{\ell_2-\ell_3}{2}} + \underline{D}(\ell_3,\ell_1) \, q^{\frac{\ell_1-\ell_3}{2}} \notag \\
\Tr  \left( \mathbf{M}^{(\mathrm{oper},0}_\beta \right)^{-1} & = \overline{D}_\circ + \overline{D}(\ell_1,\ell_3) \, q^{ \frac{\ell_3-\ell_1}{2}} + \overline{D}(\ell_2,\ell_3) \,  
q^{\frac{\ell_3-\ell_2}{2}} + \overline{D}(\ell_1,\ell_2) \, q^{\frac{\ell_2-\ell_1}{2}}    \label{eq:monMBoperinv} \\
& \quad + \overline{D}(\ell_2,\ell_1) \, q^{\frac{\ell_1-\ell_2}{2}} + \overline{D}(\ell_3,\ell_2)\, q^{ \frac{\ell_2-\ell_3}{2}} + \overline{D}(\ell_3,\ell_1) \, q^{\frac{\ell_1-\ell_3}{2}} \notag
\end{align}
where
\begin{align}
\underline{D}(\ell_k,\ell_l) &=  -4 \pi^2 e^{- \pi i \left(\frac{m+m_1-2}{2} \right)} \dfrac{D_{\star}(\ell_k,\ell_l)}{D_{\uparrow}(\ell_k) D_{\downarrow}(\ell_l)}, \\ \overline{D}(\ell_k,\ell_l) &=  -4 \pi^2 e^{ \pi i \left(\frac{m+m_1-2}{2}\right)} \dfrac{D_{\star}(\ell_k,\ell_l)}{D_{\uparrow}(\ell_k) D_{\downarrow}(\ell_l)},
\end{align}
and
\begin{align}
D_{\star}(\ell_k,\ell_l) &=   \Gamma \Big[1 + \frac{\ell_k-\ell_l}{2}\Big]^2 \Gamma \Big[ \frac{\ell_k-\ell_l}{2} \Big]^2  \Gamma \Big[1- \ell_l - \frac{\ell_k}{2} \Big]  \\
& \qquad  \times \Gamma \Big[ 1+ \ell_k + \frac{\ell_l}{2}\Big]  \Gamma \Big[\ell_k + \frac{\ell_l}{2}  \Big]  \Gamma \Big[- \frac{\ell_k}{2} - \ell_l\Big] \notag, 
\end{align}
whereas
 \begin{align}
D_{\uparrow}(\ell_k) &=   \prod_{j=1}^3 \Gamma \Big[ \frac{1-m - m_{0,j} + \ell_k}{2} \Big]  \Gamma \Big[\frac{1+m_1 - m_{\infty,j} + \ell_k}{2} \Big]
\end{align}
and
\begin{align}
D_{\downarrow}(\ell_k) &=   \prod_{j=1}^3 \Gamma \Big[ \frac{1+m + m_{0,j} - \ell_k}{2} \Big]  \Gamma \Big[ \frac{1- m_1 + m_{\infty,j} - \ell_k}{2} \Big]. 
 \end{align}

We rewrite this result in terms of perturbative corrections to the effective twisted superpotential of the superconformal $SU(3)$ theory coupled to six hypers in \S\ref{sec:SU3superpotential}.

\section{Generating function of opers}\label{sec:genfunopers}

The locus of (framed) opers forms a complex Lagrangian subspace inside the moduli space of (framed) flat connections $\cM_{\mathrm{flat}}^{\cC}(C,SL(K))$. Given any set of Darboux coordinates $\{x_i,y_i \}$ on $\cM_{\mathrm{flat}}^{\cC}(C,SL(K))$ we can thus define a generating function $W^\mathrm{oper} (x,\eps)$ of the space of opers by the coupled set of equations 
\begin{align}
y_i =  \frac{\partial W^\mathrm{oper} (x, \eps)}{\partial x_i}.
\end{align} 

In this section we find the generating function of opers $W^\mathrm{oper}(x,\eps)$ in our two main examples, the superconformal $SU(2)$ theory with four flavors and the superconformal $SU(3)$ theory with six flavors, with respect to the length-twist coordinates $L^i$ and $T_i$ defined in the first part of this paper. We do this by comparing the formulae for the oper monodromies in \S\ref{sec:monopers} to the formulae for the monodromies in terms of the length-twist coordinates $L^i$ and $T_i$ in \S\ref{section:spectral-monodromy}. 

Since the spectral twist coordinates $T_i$ are only determined up multiplication by a simple monomial in the (exponentiated) mass parameters, due to the ambiguity in the choice of a Fenchel-Nielsen spectral network, the generating function $W^\mathrm{oper}(x,\eps)$ that we find in this section is determined up to a linear factor of the form $m x$.   

\subsection{Superconformal $SU(2)$ theory with $N_f=4$}\label{sec:SU2superpotential}

Comparing the monodromy traces $\mathbf{M}^\mathrm{oper}_\alpha$ of the opers around the pants curve $\alpha$ to the monodromy traces $\mathbf{M}_\alpha$ in terms of the length-twist coordinates $L$ and $T$, gives the identifications 
\begin{align}
M_l &= - e^{\pi i  m_l},\\
L &= - e^{\pi i  \ell}.
\end{align}
These identifications in particular imply that the constant term $N_\circ$ in equation~(\ref{eq:K=2monMBav}) agrees with the constant term $D_\circ$ in equation~(\ref{eqn:MonBOperHeunq0+1}).

Next, we want to find the twist $T$ as a function of the length $L$ on the locus of opers.

\subsubsection*{Leading order contribution in $q$}

Comparing the leading order contribution in $q$ to the oper monodromy $\mathbf{M}^\mathrm{oper}_\beta$, as computed in  equation~(\ref{eqn:MonBOperHeunq0}), to the monodromy $\mathbf{M}_\beta$ in terms of the length-twist coordinates $L$ and $T$, as computed in equation~(\ref{eq:K=2monMBav}), shows that  up to leading order in $q$
\begin{align}
T+ \frac{1}{T} = \frac{D_+}{\sqrt{N}}\, q^\ell + \frac{D_-}{\sqrt{N}} \, q^{-\ell}, \label{eqn:locusopersK2}
\end{align}
where
\begin{align}
N(\ell) = \frac{16\, \cos \left( \frac{\pi \ell \pm \pi m_0 \pm \pi m}{2} \right) \cos\left(\frac{ \pi \ell \pm \pi m_1 \pm \pi m_\infty}{2}\right)}{ \sin(\pi \ell)^4}.
\end{align}
and $D_- = D_+|_{\ell = -\ell} $ with 
\begin{align}
D_+(\ell)& = -4 \pi^2  \frac{\Gamma[ 1-\ell ]^2 \Gamma[-\ell ]^2 }{ \Gamma\Big[ \frac{1}{2} - \frac{ \ell \pm m_0 \pm m }{2}\Big] \Gamma\Big[ \frac{1}{2} - \frac{\ell  \pm m_1 \pm m_\infty }{2} \Big]  }.
\end{align}

Repeatedly using the identity 
\begin{equation}\label{eq:GammaId}
\cos \left( \frac{\pi x}{2} \right) \Gamma \big[ \frac{1}{2} + \frac{x}{2} \Big] \Gamma \Big[\frac{1}{2}- \frac{x}{2} \Big] = \pi,
\end{equation}
we find that
\begin{align}
\frac{D_+}{\sqrt{N}} = \sqrt{\frac{\Gamma[\frac{1}{2} + \frac{\ell \pm m_0 \pm m}{2}] \Gamma[\frac{1}{2} + \frac{\ell \pm m_1 \pm m_\infty}{2}] }{\Gamma[\frac{1}{2} - \frac{\ell \pm m_0 \pm m}{2}] \Gamma[\frac{1}{2} - \frac{\ell \pm m_1 \pm m_\infty}{2}]} } \frac{\Gamma[ 1 - \ell] \Gamma[ -\ell]}{\Gamma[ 1 + \ell]\Gamma[ \ell] },
\end{align}
and hence that 
\begin{align}
\frac{D_-}{\sqrt{N}} = \frac{\sqrt{N}}{D_+}
\end{align}

This implies that equation~(\ref{eqn:locusopersK2}) is solved by
\begin{align}
T  &= \sqrt{\frac{\Gamma[\frac{1}{2} + \frac{\ell \pm m_0 \pm m}{2}] \Gamma[\frac{1}{2} + \frac{\ell \pm m_1 \pm m_\infty}{2}] }{\Gamma[\frac{1}{2} - \frac{\ell \pm m_0 \pm m}{2}] \Gamma[\frac{1}{2} - \frac{\ell \pm m_1 \pm m_\infty}{2}]} } \frac{\Gamma[ 1 - \ell] \Gamma[- \ell]}{\Gamma[ 1 + \ell]\Gamma[ \ell] } q^{\ell}
\end{align}
up to leading order in $q$.

\subsubsection*{Classical and 1-loop contribution}

Since the generating function of opers $W^\mathrm{oper}(\ell,q)$ is defined by 
\begin{align}
\frac{1}{2} \log T =   \frac{\partial W^\mathrm{oper}(\ell,q)}{\partial \ell} 
\end{align}
on the locus of opers, we find that 
\begin{align}
\frac{\partial W^\mathrm{oper}(\ell,q)}{\partial \ell} 
&= \frac{\ell}{2} \log q + \frac{1}{4} \log \frac{\Gamma[\frac{1}{2} + \frac{\ell \pm m_0 \pm m}{2}] }{\Gamma[\frac{1}{2} - \frac{\ell \pm m_0 \pm m}{2}] } + \frac{1}{4} \log \frac{\Gamma[\frac{1}{2} + \frac{\ell \pm m_1 \pm m_\infty}{2}] }{\Gamma[\frac{1}{2} - \frac{\ell \pm m_1 \pm m_\infty}{2}] } \\
& \quad +  \frac{1}{2} \log \frac{\Gamma[ 1 - \ell] }{\Gamma[ \ell] }+  \frac{1}{2} \log \frac{\Gamma[- \ell] }{\Gamma[1+ \ell] } + \mathcal{O}(q). \notag
\end{align}

To make contact with known formulae, we write the last equation in terms of the special function
\begin{equation}\label{eq:defUpsilon}
\Upsilon(x) = \int_{\frac{1}{2}}^x du \log \frac{\Gamma(u)}{\Gamma(1-u)},
\end{equation}
which has the property that
\begin{align}
\frac{\partial}{\partial x} \Upsilon(\beta + \gamma x) = \gamma \log \frac{\Gamma[\beta + \gamma x]}{\Gamma[1-\beta - \gamma x]}
\end{align}
as well as 
\begin{align}
\Upsilon[1-x] = \Upsilon[x].
\end{align}

We thus find
\begin{align}
W^\mathrm{oper}(\ell,q) &= W^\mathrm{oper}_\mathrm{clas}(\ell,\tau) + W^\mathrm{oper}_{\mathrm{1-loop}}(\ell) + \mathcal{O}(q) \label{eqn:leadingW}
\end{align}
with 
\begin{align}\label{eqn:genoperclass}
W^\mathrm{oper}_\mathrm{clas}(\ell,\tau) &= \frac{\ell^2}{4} \log q,
\end{align}
and 
\begin{align}\label{eqn:genoperq0}
 W^\mathrm{oper}_{\mathrm{1-loop}}(\ell)&= W^\mathrm{oper}_{\mathrm{anti-hyp}}(\ell,m_0,m) 
+ W^\mathrm{oper}_{\mathrm{vector}}(\ell)+ W^\mathrm{oper}_{\mathrm{hyp}}(\ell,m_1,m_\infty) 
\end{align}
with 
\begin{align}
W^\mathrm{oper}_{\mathrm{vector}}(\ell)&=  - \frac{1}{2} \, \Upsilon [- \ell ]- \frac{1}{2} \, \Upsilon[\ell]\\
W^\mathrm{oper}_{\mathrm{anti-hyp}}(\ell,m_0,m)& =  \frac{1}{2} \, \Upsilon\big[\frac{1}{2} + \frac{\ell \pm m_0 \pm m}{2} \big]\\
W^\mathrm{oper}_{\mathrm{hyp}}(\ell,m_1,m_\infty)& =  \frac{1}{2}\,  \Upsilon\big[ \frac{1}{2} + \frac{\ell \pm m_1 \pm m_\infty}{2} \big],
\end{align}
up to an integration constant that is independent of $\ell$. 

If we identify the length coordinate $\ell$ with the Coulomb parameter $a$, and compare to the expression for the Nekrasov-Shatashvili effective twisted superpotential for the $SU(2)$ gauge theory coupled to four hypermultiplets, given in equations~(\ref{eq:Wtree}), (\ref{eq:W1-loop}) and (\ref{eq:1-inst}), we find that 
\begin{align}
W^\mathrm{oper}_\mathrm{clas}(a,\tau) &= \widetilde{W}^\mathrm{eff}_\mathrm{clas}(a,\tau) \\
W^\mathrm{oper}_\mathrm{1-loop}(a) &= \widetilde{W}^\mathrm{eff}_\mathrm{1-loop}(a).
\end{align}
In particular, $ W^\mathrm{oper}_{\mathrm{1-loop}}(\ell) $ is equal to half the classical Liouville action on the nodal four-punctured sphere. 

This computation is similar to and agrees with that in \cite{Vartanov:2013ima}.

\subsubsection*{1-instanton correction}\label{sec:1-instanton}

The 1-instanton correction $W^\mathrm{oper}_1(\ell,q)$ in the generating function of opers, 
\begin{align}
W^\mathrm{oper}(\ell,q) &= W^\mathrm{oper}_\mathrm{clas}(\ell) \log q  + W^\mathrm{oper}_{\mathrm{1-loop}}(\ell) + \, W^\mathrm{oper}_1(\ell)\,q + \mathcal{O}(q^2),
\end{align}
is computed by the next-to-leading order correction in $q$ in equation~(\ref{eqn:MonBOperHeunq0+1}) as 
\begin{align}
W^\mathrm{oper}_1(\ell) &= \frac{\ell^2}{8} + \frac{ (m_0^2-m^2)(m_\infty^2-m_1^2)}{8 (\ell+1)(\ell-1)} \label{eqn:1instanton}\\
&=  \frac{(\ell \pm m_0 + m+1)(\ell + m_1 \pm m_\infty+1)}{16 \ell(\ell+1)} \\ 
& \qquad + \frac{(\ell \pm m_0 - m -1)(\ell - m_1 \pm m_\infty-1)}{16 \ell(\ell-1)} \notag \\
& \qquad \quad - \frac{1}{8} \, (m^2-m_0^2+m_1^2-m_\infty^2 -1)- \frac{1}{2}(1+m)(1+m_1), \notag 
\end{align}
up to an integration constant that is independent of $\ell$.

Comparing this to the 1-instanton contribution to the Nekrasov-Shatashvili effective twisted superpotential for the $SU(2)$ theory with four hypermultiplets, given in equation~(\ref{eq:1-inst}), we conclude that 
\begin{align}
W^\mathrm{oper}_1(a) &= \, \widetilde{W}_1^\mathrm{eff}(a),
\end{align}
after setting the integration constant. That is, $W^\mathrm{oper}_1(a)$ computes the 1-instanton correction to the  Nekrasov-Shatashvili effective twisted superpotential, up to a ``spurious'' factor that does not depend on the Coulomb parameter $a$.

So far we have hidden the dependence on $\eps$, but let us now reintroduce this by scaling all parameters $a$ and $m_k$ as $a \mapsto \frac{a}{\eps}$ and $m_l \mapsto \frac{m_l}{\eps}$, respectively. It follows that the $\eps$-expansion of $W^\mathrm{oper}_1(a)$ is simply
\begin{align}\label{eqn:epsexpansionW1}
W^\mathrm{oper}_1(a,\eps) & = \frac{1}{\eps^2} \frac{a^4 + (m_0^2-m^2)(m_\infty^2-m_1^2)}{4 a^2 } \\
& \qquad + \sum_{k=0}^{\infty} \eps^{2k} \frac{  (m_0^2-m^2)(m_\infty^2-m_1^2)}{4 a^{2k+4} } . \notag
\end{align}
In particular, it does not have any odd powers in $\eps$.

\subsection{Superconformal $SU(3)$ theory with $N_f=6$}\label{sec:SU3superpotential}

Comparing the monodromy traces $\mathbf{M}^{\mathrm{oper}}_\alpha$ of the generalized Heun's opers around the pants curve $\alpha$, given in equation~(\ref{eq:monMAoper}), to the monodromy traces $\mathbf{M}_\alpha$ in terms of the higher length-twist coordinates $L_1,L_2,T_1,T_2$, given in equation~(\ref{eq:K=3monMAav}), yields the identifications 
\begin{align}\label{eq:identLSU3}
L_1 = e^{ \pi i \ell_1 }, \quad L_2 = e^{ \pi i \ell_2 }.
\end{align}
Equating the eigenvalues of the local monodromies at the punctures, by comparing equations~(\ref{eqn:nonab-masses-23}) to (\ref{eqn:conjclassSU3}) for maximal punctures and (\ref{eqn:nonab-masses-14}) to (\ref{eq:exponentsminimal}) for minimal punctures, yields the identifications
\begin{align}\label{eq:identMSU3}
M_{0,i} = e^{\pi i m_{0,i} }, \quad M = -e^{\pi i m  }, \quad M_1 = -e^{\pi i m_1}, \quad  M_{\infty,i} = e^{\pi i m_{\infty,i}}.
\end{align}  

Next, we want to find the twists $T_1$ and $T_2$ as a function of the lengths $L_1$ and $L_2$ on the locus of generalized Heun's opers.

\subsubsection*{Leading order contribution}

To leading order in $q$ we need to equate equations~(\ref{eq:K=3monMBav}) and~(\ref{eq:K=3monMBavinv}), which capture the monodromy along the 1-cycle $\beta$ on $C$ in terms of the twist coordinates $T_1$ and $T_2$ as 
\begin{align}
\Tr \mathbf{M}_\beta  &= \underline{N}_{\circ} + \underline{N}(L_1,L_3) T_1 + \underline{N}(L_2,L_3) T_2 + \underline{N}(L_1,L_2) \frac{T_1}{T_2} \\
& \quad + \underline{N}(L_2,L_1) \frac{T_2}{T_1}  + \frac{\underline{N}(L_3,L_2)}{T_2}+ \frac{\underline{N}(L_3,L_1)}{T_1},  \notag \\
\Tr \mathbf{M}_\beta^{-1} &= \overline{N}_{\circ} + \overline{N}(L_1,L_3) T_1 + \overline{N}(L_2,L_3) T_2 + \overline{N}(L_1,L_2) \frac{T_1}{T_2} \\
& \quad + \overline{N}(L_2,L_1) \frac{T_2}{T_1} + \frac{\overline{N}(L_3,L_2)}{T_2}+ \frac{\overline{N}(L_3,L_1)}{T_1}, \notag
\end{align}
to equations~(\ref{eq:monMBoper}) and~(\ref{eq:monMBoperinv}), respectively, which capture the monodromy of the generalized Heun's equation to the leading order in $q$ as
\begin{align}
\Tr \mathbf{M}^{\mathrm{oper},0}_\beta & = \underline{D}_\circ + \underline{D}(\ell_1,\ell_3) \, q^{ \frac{\ell_3-\ell_1}{2}} + \underline{D}(\ell_2,\ell_3) \,  
q^{\frac{\ell_3-\ell_2}{2}} + \underline{D}(\ell_1,\ell_2) \, q^{\frac{\ell_2-\ell_1}{2}}   \\
& \quad + \underline{D}(\ell_2,\ell_1) \, q^{\frac{\ell_1-\ell_2}{2}} + \underline{D}(\ell_3,\ell_2)\, q^{ \frac{\ell_2-\ell_3}{2}} + \underline{D}(\ell_3,\ell_1) \, q^{\frac{\ell_1-\ell_3}{2}} \notag \\
\Tr  \left( \mathbf{M}^{(\mathrm{oper},0}_\beta \right)^{-1} & = \overline{D}_\circ + \overline{D}(\ell_1,\ell_3) \, q^{ \frac{\ell_3-\ell_1}{2}} + \overline{D}(\ell_2,\ell_3) \,  
q^{\frac{\ell_3-\ell_2}{2}} + \overline{D}(\ell_1,\ell_2) \, q^{\frac{\ell_2-\ell_1}{2}}     \\
& \quad + \overline{D}(\ell_2,\ell_1) \, q^{\frac{\ell_1-\ell_2}{2}} + \overline{D}(\ell_3,\ell_2)\, q^{ \frac{\ell_2-\ell_3}{2}} + \overline{D}(\ell_3,\ell_1) \, q^{\frac{\ell_1-\ell_3}{2}} \notag
\end{align}
 and solve $T_1$ and $T_2$ as a function of $L_1$ and $L_2$.

 With the identifications~(\ref{eq:identLSU3}) and (\ref{eq:identMSU3}) we can check that $\underline{N}_\circ$ equals $\underline{D}_\circ$ and that $\overline{N}_\circ$ equals $\overline{D}_\circ$.

Furthermore, since
\begin{align}
\frac{\overline{N}(L_k,L_l)}{ \underline{N}(L_k,L_l)} = 
\frac{\overline{D}(\ell_k,\ell_l) }{ \underline{D}(\ell_k,\ell_l) } = M_1 M_4,
\end{align}
it is sufficient to solve the equation
\begin{align} \label{eqn:comparetracesSU(3)}
\Tr \mathbf{M}_\beta = \Tr \mathbf{M}^{\mathrm{oper},0}_\beta 
\end{align}
for $T_1$ and $T_2$.
 
By repeatedly using the identity~(\ref{eq:GammaId}) we can simplify the quotient
\begin{align}
\frac{\underline{D}(\ell_k,\ell_l)}{\underline{N}(\ell_k,\ell_l)} &= \widetilde{D}_{\uparrow}(\ell_k)\widetilde{D}_{\downarrow}(\ell_l) \widetilde{D}_{*}(\ell_k,\ell_l), 
\end{align}
to a product of Gamma-functions. 
 
Here, 
\begin{align}
\widetilde{D}_{\uparrow}(\ell_k) &:= \frac{8 \pi^3 i }{D_{\uparrow}(\ell_k) N(\ell_k)}  = 
\prod_{j=1}^3 \sqrt{ \frac{\Gamma\Big[ \frac{1+m + m_{0,j} - \ell_k}{2} \Big] \Gamma \Big[ \frac{1 - m_1 + m_{\infty,j} - \ell_k)}{2} \Big]} {\Gamma\Big[ \frac{1-m - m_{0,j} + \ell_k}{2} \Big] \Gamma \Big[ \frac{1 + m_1 - m_{\infty,j} + \ell_k)}{2} \Big]} }, 
\end{align}
whereas
\begin{align}
\widetilde{D}_{\downarrow}(\ell_k) &= \frac{8 \pi^3 i }{D_{\downarrow}(\ell_k)N_\diamond(\ell_k)}= \widetilde{D}_{\uparrow}(\ell_k)^{-1}. 
\end{align}

Furthermore,
\begin{align}
& \widetilde{D}_{\star}(\ell_k,\ell_l) := \frac{D_{\star}(\ell_k,\ell_l) N_{\star}(\ell_k,\ell_l)}{16 \pi^4}  \\
& = \quad \frac{  \Gamma \Big[1+\frac{\ell_k-\ell_l}{2} \Big]  \Gamma \Big[\frac{ \ell_k-\ell_l}{2} \Big]   } { \Gamma \Big[ \frac{\ell_l-\ell_k}{2} \Big] \Gamma \Big[1+ \frac{\ell_l-\ell_k}{2} \Big]}   \sqrt{ \frac{ \Gamma \Big[ 1+ \ell_k+ \frac{\ell_l}{2} \Big]  
\Gamma \Big[ \ell_k+ \frac{\ell_l}{2} \Big] 
\Gamma \Big[1- \frac{\ell_k}{2} - \ell_l) \Big]  \Gamma \Big[-  \frac{\ell_k}{2}-\ell_l \Big] }{ \Gamma \Big[ - \ell_k- \frac{\ell_l}{2} \Big]  
\Gamma \Big[ 1 - \ell_k- \frac{\ell_l}{2} \Big] 
\Gamma \Big[ \frac{\ell_k}{2} + \ell_l) \Big]  \Gamma \Big[1 + \frac{\ell_k}{2} + \ell_l \Big] } } \notag.
\end{align}

It follows from the last four equations that 
\begin{align}
\frac{\underline{D}(\ell_k,\ell_l)}{\underline{N}(\ell_k,\ell_l) }  &= \frac{\underline{N}(\ell_l,\ell_k)}{\underline{D}(\ell_l,\ell_k) }
\end{align}
and also that
\begin{align}
\frac{\underline{D}(\ell_1,\ell_3)}{\underline{N}(\ell_1,\ell_3) } \frac{\underline{N}(\ell_2,\ell_3)}{\underline{D}(\ell_2,\ell_3) } &= \frac{\underline{D}(\ell_1,\ell_2)}{\underline{N}(\ell_1,\ell_2) }. 
\end{align}

This implies that equation~(\ref{eqn:comparetracesSU(3)}) is solved by the coupled system of equations
\begin{align}\label{eq:systemforTSU3}
T_1  &= \frac{\underline{D}(\ell_3,\ell_1)}{\underline{N}(\ell_3,\ell_1) } q^{\frac{\ell_1-\ell_3}{2}}  \\
T_2 &= \frac{\underline{D}(\ell_3,\ell_2)}{\underline{N}(\ell_3,\ell_2)} q^{\frac{\ell_2-\ell_3}{2}}. \notag
\end{align}

The generating function $W^\mathrm{oper}(\ell_1,\ell_2)$ of the locus of generalized Heun's opers is defined as
\begin{eqnarray}
\frac{1}{2} \log T_1  =   \partial_{\ell_1} W^{\mathrm{oper}}(\ell_1,\ell_2)   \\
\frac{1}{2} \log T_2  = \partial_{\ell_2} W^{\mathrm{oper}} (\ell_1,\ell_2), \notag
\end{eqnarray}
so that for instance
\begin{align}
  &\partial_{\ell_1} W^{\mathrm{oper}}(\ell_1,\ell_2)     \, = \notag \\
  & \frac{1}{4} \log \frac{ \Gamma \Big[1+\frac{\ell_3-\ell_1}{2} \Big]^2  \Gamma \Big[\frac{ \ell_3-\ell_1}{2} \Big]^2   } { \Gamma \Big[ \frac{\ell_1-\ell_3}{2} \Big]^2 \Gamma \Big[1+ \frac{\ell_1-\ell_3}{2} \Big]^2}   \frac{ \Gamma \Big[ 1+ \frac{\ell_3-\ell_2}{2} \Big]  
\Gamma \Big[ \frac{\ell_3-\ell_2}{2} \Big] 
\Gamma \Big[1+ \frac{\ell_2-\ell_1}{2}  \Big]  \Gamma \Big[ \frac{\ell_2-\ell_1}{2} \Big] }{ \Gamma \Big[  \frac{\ell_2-\ell_3}{2} \Big]  
\Gamma \Big[ 1 + \frac{\ell_2-\ell_3}{2} \Big] 
\Gamma \Big[ \frac{\ell_1-\ell_2}{2} \Big]  \Gamma \Big[1 + \frac{\ell_1-\ell_2}{2}  \Big]  } \\
  & \frac{1}{4} \sum_{j=1}^3 \log \frac{\Gamma\Big[ \frac{1-m - m_{0,j} + \ell_1}{2} \Big] \Gamma\Big[ \frac{1+m + m_{0,j} - \ell_3}{2} \Big] \Gamma \Big[ \frac{1 + m_1 - m_{\infty,j} + \ell_1)}{2} \Big]  \Gamma \Big[ \frac{1 - m_1 + m_{\infty,j} - \ell_3)}{2} \Big]} {\Gamma\Big[ \frac{1+m + m_{0,j} - \ell_1}{2} \Big] \Gamma\Big[ \frac{1-m - m_{0,j} + \ell_3}{2} \Big]\Gamma \Big[ \frac{1 - m_1 + m_{\infty,j} - \ell_1)}{2} \Big]  \Gamma \Big[ \frac{1 + m_1 - m_{\infty,j} + \ell_3)}{2} \Big]} .  \notag
\end{align}

To leading order in $q$, in terms of the function $\Upsilon(x)$ defined in equation~(\ref{eq:defUpsilon}), we thus find that
\begin{align}\label{eqn:leadingW3}
W^{\mathrm{oper}}(\ell_1,\ell_2) &=   \frac{\ell_1^2 + \ell_2^2 + \ell_1 \ell_2}{2} \log q + W^{\mathrm{oper}}_{\mathrm{anti-hyp}}(\ell_1,\ell_2, m,m_{0,1},m_{0,2}) \\& \qquad 
+ W^{\mathrm{oper}}_{\mathrm{vector}}(\ell_1,\ell_2)+ W^{\mathrm{oper}}_{\mathrm{hyp}}(\ell_1,\ell_2,m_1,m_{\infty,1},m_{\infty,2}) +O(q)\notag
\end{align}
where 
\begin{align}
W^{\mathrm{oper}}_{\mathrm{vector}}(\ell_1,\ell_2)&= -
 \frac{1}{2} \sum_{j=1}^3 \Upsilon \big[ \frac{\ell_j -\ell_{j+1}}{2}  \big] - \frac{1}{2}  \sum_{j=1}^3  \Upsilon \big[ \frac{\ell_{j+1} -\ell_{j}}{2 }  \big]\\
W^{\mathrm{oper}}_{\mathrm{anti-hyp}}(\ell_1,\ell_2,m,m_{0,1},m_{0,2})& = \frac{1}{2} \sum_{j,k=1}^3 \Upsilon\big[\frac{1 - m - m_{0,k} +\ell_j}{2}\big]\\
W^{\mathrm{oper}}_{\mathrm{hyp}}(\ell_1,\ell_2,m_1,m_{\infty,1},m_{\infty,2})& = \frac{1}{2} \sum_{j,k=1}^3 \Upsilon\big[ \frac{1+m_1-m_{\infty,k}+\ell_j}{2} \big],
\end{align}
up to an integration constant that is independent in $\ell_1$ and $\ell_2$. 

If we identify the length coordinates $\ell_i$ with the Coulomb parameters $a_i$, the above expressions agree with the classical and 1-loop contributions to the Nekrasov-Shatashvili effective twisted superpotential $\widetilde{W}^{\mathrm{eff}}$ for the $SU(3)$ gauge theory coupled to six hypermultiplets. 

More precisely, the 1-loop contribution to $\exp \widetilde{W}^{\mathrm{eff}}$ may be computed as a product of determinants of differential operators. There is a certain freedom in its definition due to the regularization of divergencies, which implies that it is only determined up to a phase \cite{Pestun:2007rz,Vartanov:2013ima}. For a distinguished choice of phase, the 1-loop contribution to $\exp \widetilde{W}^{\mathrm{eff}}$ may be identified with the square-root of the product of two Toda three-point functions with one semi-degenerate primary field, first computed in \cite{Fateev:2005gs},  in the Nekrasov-Shatashvili  (or $c \to \infty$) limit. The 1-loop contribution to $W^{\mathrm{oper}}$, as found in equation~(\ref{eqn:leadingW3}), agrees with the 1-loop contribution to $\widetilde{W}^{\mathrm{eff}}$ in this ``Toda scheme''.

Instanton contributions to $W^{\mathrm{oper}}$ may be obtained by computing the monodromies of the generalized Heun's differential equation up to a higher order in $q$, following the strategy of \S\ref{sec:monHeun}. We leave this for future work.

\section{WKB asymptotics}\label{sec:WKB}

Given an $\eps$-oper $\nabla^\mathrm{oper}_\eps$ there is yet another method to compute its monodromy representation. This is sometimes called the ``exact WKB method'' \cite{Voros83,exactWKB,2014arXiv1401.7094I}. We will review this approach in \S\ref{sec:exactWKB}, following \cite{exactWKB}. 

In \S\ref{sec:WKBab} we compare the monodromy representation for the oper $\nabla^\mathrm{oper}_\eps$ obtained from the exact WKB method to that obtained from the abelianization mapping. We conclude $\nabla^\mathrm{oper}_\eps$ is abelianized by the Borel sums (in the direction $\vartheta = \arg \eps$) of its WKB solutions. 

As a consequence, this implies that the spectral coordinate $\cX_\gamma(\nabla^\mathrm{oper}_\eps)$ has an asymptotic expansion in the limit $\eps \to 0$ given by  
\begin{align}\label{eq:quantumperiod}
\cX_\gamma(\nabla^\mathrm{oper}_\eps) \sim \exp \left( \oint S_\mathrm{odd}(\eps)dz \right),
\end{align}
where $S_\mathrm{odd}(\eps)$ is a solution to the Ricatti equation~(\ref{eqn:Ricatti}). These WKB-asymptotics relate the Nekrasov-Rosly-Shatashvili correspondence to the approach of computing the $\eps$-asymptotics of the effective twisted superpotential $\widetilde{W}^\mathrm{eff}(a,q,\eps)$ using quantum periods (pioneered in \cite{Mironov:2009uv} for the pure $SU(2)$ gauge theory).

In \S\ref{sec:quantumperiod} we conclude that while the $\eps$-asymptotics of the effective twisted superpotential may be found by computing quantum periods, the analytic result is found by computing the Borel sums of the quantum periods in a critical direction $\vartheta_0$ corresponding to a Fenchel-Nielsen network.

Whereas we restrict ourselves to $K=2$ in this section, a similar discussion holds for higher rank.

\subsection{Monodromy representation from exact WKB}\label{sec:exactWKB}

We start off with a brief review of the exact WKB method, following \cite{exactWKB,2014arXiv1401.7094I}. 

Let $\eps$ be a small complex parameter with phase $\vartheta$. Fix an $SL(2)$ $\eps$-oper $\nabla^\mathrm{oper}_\eps$ on $C$ locally given by the differential operator 
\begin{align}\label{eq:sl2opereps}
\mathbf{D}(\eps) = \eps^2 \partial_z^2 - Q(z,\eps),
\end{align}
where $Q(z,\eps) = \sum_{j=0}^N Q_j(z) \eps^j$ is a polynomial in $\eps$ with coefficients $Q_j(z)$ that are meromorphic on $C$, satisfying conditions outlined in  \cite{2014arXiv1401.7094I}. The principal part $Q_0(z)$ of $Q(z,\eps)$ defines a meromorphic quadratic differential $\varphi_2 = Q_0(z) (dz)^2$ on $C$. 

The zeroes and poles of $\varphi_2$ on $C$ are called turning points and singular points, respectively. Stokes curves are paths on $C$ emanating from the turning points such that 
\begin{align}
e^{-i \vartheta} \sqrt{\varphi_2} (v) \in \R
\end{align}
for every nonzero tangent vector $v$ to the path. We orient the Stokes curves such that the real part of $e^{-i \vartheta} \int^z \sqrt{\varphi_2}$ increases along the trajectory in the positive direction. We assign signs $+$ and $-$ to the singular poles so that the trajectories with positive directions flow from $-$ to $+$. 

Stokes curves oriented away from turning points are called dominant, while those oriented towards turning points are called recessive. The Stokes curves, the turning and the singular points form a graph on $C$ which is called the Stokes graph $\cG_\vartheta(\varphi_2)$. 

The WKB ansatz for the solutions to the differential equation
\begin{align}\label{eqn:wkbdiffeqn}
\left( \eps^2 \partial_z^2 - Q(z,\eps) \right) \psi(z) = 0
\end{align}
can be written in the form
\begin{align}\label{eqn:WKBansatz}
\psi_{\pm}(z) = \frac{1}{\sqrt{S_\mathrm{odd}}} \exp \left( \pm \int^z_{z_0} S_\mathrm{odd} \, dz \right),  
\end{align}
with base-point $z_0$. Here
\begin{align}
S_\mathrm{odd} = \sum_{n=0}^\infty \eps^{2n-1} S_{2n-1}
\end{align}
is the odd part to the formal solution $S = \sum_{k=-1}^\infty \eps^{k} S_k $ of the Riccati equation
\begin{align}\label{eqn:Ricatti}
S'(z) + S^2 = \eps^{-2} Q_0(z).
\end{align}
Note that $S_{-1} = \sqrt{Q_0}$.

Suppose that the differential $\varphi_2$ is generic, such that there are no saddle trajectories. Then the WKB solutions $\psi_{\pm}$ are Borel resummable (in the direction $\vartheta$) in each connected region of $C \backslash \cG_\vartheta(\varphi_2)$ \cite{koikeschafke}. The Borel sums of $\psi_{\pm}$ give analytic solutions to the differential equation~(\ref{eqn:wkbdiffeqn}).

Any solution $\psi_{\pm}^l$ obtained upon Borel resummation in a region $l$ can be analytically continued into a neighbouring region $l'$. It is related to the solution $\psi_{\pm}^{l'}$ obtained upon Borel resummation in the region $l'$ by a so-called connection formula. 

Say that we cross a dominant Stokes line clockwise with regard to the turning point $b$ that it emanates from. Then the Borel sums $\psi_{b,\pm}^l$ and $\psi_{b,\pm}^{l'}$ of 
\begin{align}
\psi_{b,\pm} = \frac{1}{\sqrt{S_\mathrm{odd}}} \exp \left( \pm \int^z_{b} S_\mathrm{odd} \, dz \right),  
\end{align}
on either side of the Stokes line are related by the transformation
\begin{align}\label{eqn:VdomStokes}
\psi_{b,-}^l &= \psi_{b,-}^{l'}  \\
\psi_{b,+}^l &= \psi_{b,+}^{l'} + i \psi_{b,-}^{l'}, \notag
\end{align}
Indeed, $\psi^l_{b,+}$ is dominant in this region, and hence is allowed to pick up a recessive contribution without changing the WKB asymptotics.

Say that we cross a recessive Stokes line clockwise with regard to the turning point $b$ that it emanates from. Then the Borel sums $\psi_{b,\pm}^l$ and $\psi_{b,\pm}^{l'}$ on either side of the Stokes line are related by the transformation
\begin{align}\label{eqn:VrecStokes}
\psi_{b,+}^l &= \psi_{b,+}^{l'}  \\
\psi_{b,-}^l &= \psi_{b,-}^{l'} + i \psi_{b,+}^{l'}, \notag
\end{align}
Indeed, in this situation $\psi^l_{b,-}$ is dominant, and hence is allowed to pick up a recessive contribution without changing the WKB asymptotics.

With the above data we can construct a monodromy representation for the $SL(2)$ oper $\nabla^\mathrm{oper}_\eps$. Suppose that we want to compute the monodromy matrix along a path $C_0$ with begin and end-point at $z_0$ with respect to the Borel sums of the WKB solutions $\psi_\pm$. Label the Stokes regions that the path $C_0$ crosses as $U_l$ (with $l$ increasing) and say that $\psi_\pm^l$ is the Borel sum in Stokes region $U_l$. Then we can determine the monodromy matrix along $C_0$ by computing the basis transformation that relates $\psi_\pm^{l+1}$ to $\psi_\pm^l$. 

Let $b$ be the turning point that the Stokes line emanates from. The connection formulae tell us how to relate the (local) Borel sums $\psi^l_{b, \pm}$ in the neighbouring regions $U_l$ across Stokes lines. The transformation is of the form  
\begin{align}\label{eqn:connformula}
(\psi_{b,+}^l,\psi_{b,-}^l) = (\psi_{b,+}^{l+1},\psi_{b,-}^{l+1}) \, V^{l,l+1},
\end{align}
where the matrix $V^{l,l+1}$  is determined by equation~(\ref{eqn:VdomStokes}) or~(\ref{eqn:VrecStokes}). It depends on the type of the Stokes line and the direction of crossing.

We would like to know the transformation in terms of the Borel sums $\psi^l_{\pm}$, which are defined with respect to the base-point $z_0$.  Now $\psi_\pm^l$ differs from $\psi^l_{b, \pm}$ by the transformation
\begin{align}
(\psi_{+}^l,\psi_{-}^l) = (\psi_{b,+}^l,\psi_{b,-}^l) \, D_{z_0}^b,
\end{align}
where $D_{z_0}^b$ is the Borel sum of the matrix
\begin{align}
\left( \begin{array}{cc} \exp \left(+ \int^b_{z_0} S_\mathrm{odd} \, dz \right) & 0 \\ 0 & \exp \left(- \int^b_{z_0} S_\mathrm{odd} \, dz \right) \end{array} \right).
 \end{align}
(Note that since the integrals do not depend on the position $z$, the Borel summed $D_{z_0}^b$ does not depend on the Stokes region.)

Hence we find that $\psi_\pm^{l+1}$ and $\psi_\pm^l$ are related by the transformation
\begin{align}
(\psi_{+}^l,\psi_{-}^l) =  (\psi_{+}^{l+1},\psi_{-}^{l+1}) \, \widetilde{V}_{z_0}^{l,l+1},
\end{align}
with connection matrix
\begin{align}\label{eqn:Vz0l}
\widetilde{V}_{z_0}^{l,l+1} = (D_{z_0}^b)^{-1} \, V^{l,l+1} \, D_{z_0}^b.
\end{align}
The monodromy matrix along the path $C_0$ is then found by multiplying all connection matrices $\widetilde{V}_{z_0}^{l,l+1}$ along it. 

If the differential equation~(\ref{eqn:wkbdiffeqn}) is Fuchsian, the resulting monodromy representation may be expressed in terms of the characteristic exponents at the regular singular points and the Borel sums (in the direction $\vartheta$) of the contour integrals
\begin{align}
\exp  \left( V_\gamma \right) := \exp \left( \oint_\gamma S_\mathrm{odd} \, dz \right)
\end{align}
along 1-cycles $\gamma$ on the covering $\Sigma$. The exponent $V_\gamma$ is called the \emph{Voros symbol} for the cycle $\gamma$.

So far we have kept the phase $\vartheta = \arg \epsilon$ fixed. The connection formulae~(\ref{eqn:connformula}) describe the analytic continuation of the Borel sums $\psi_{b,\pm}$ of the WKB solutions in the $z$-plane. Let us now consider what happens if we vary the phase $\vartheta$. We make the dependence on $\vartheta$ explicit in the notation by writing $\psi_{b,\pm}^\vartheta$. 

Suppose a Stokes line crosses a point $z\in C$ at a critical phase $\vartheta_0$. Then the Borel sums $\psi_{b,\pm}^{\vartheta_0-\delta}$ and $\psi_{b,\pm}^{\vartheta_0+\delta}$ are not equal, but related by connection formulae similar to~(\ref{eqn:connformula}) in a neighbourhood of the point $z$, for small enough $\delta$. This is called the Stokes phenomenon. The Borel sums $\psi_{b,\pm}^\vartheta$ do have the same asymptotic expansion, given by $\psi_{b,\pm}$, in the whole sector $\{ \eps \in \C\,|\, |\theta - \vartheta_0| < \pi/2, |\eps| \ll 1 \}$. 

The Borel sums of the Voros symbols $V_\gamma$ are affected by the Stokes phenomenon as well. The Voros symbol $V_\gamma$ is Borel summable (in the direction $\vartheta$) if the cycle $\gamma$ does not intersect with a saddle trajectory of the Stokes graph $\cG_\vartheta(\varphi_2)$. This Borel summability is broken if a saddle trajectory appears, say at the phase $\vartheta_0$. The Borel sums of the Voros symbol $V_\gamma$ in the directions $\vartheta_0 \pm \delta$ are related by "jump formulae", for sufficiently small $\delta$ (see for instance \cite{2014arXiv1401.7094I} for explicit expressions). Of course, both Borel sums do have the same asymptotic expansion in the limit $|\epsilon| \to 0$, given by the Voros symbol $V_\gamma$ itself.

\subsection{Relating exact WKB to abelianization}\label{sec:WKBab}

The above procedure of finding the monodromy representation using the exact WKB method is very similar to finding the monodromy representation of a flat $SL(2)$ connection $\nabla$ using the abelianization mapping. In fact, in this section we show that the resulting monodromy representations are equivalent on the locus of opers.\footnote{While finalizing this paper we heard about an alternative argument from Andrew Neitzke.} 

Fix an $\epsilon$-oper $\nabla^\mathrm{oper}_\eps$ locally given by~(\ref{eq:sl2opereps}), with fixed phase $\vartheta = \arg \eps$. It is easy to see that the corresponding Stokes graph $\cG_\vartheta(\varphi_2)$ and spectral network $\cW_\vartheta(\varphi_2)$ are equivalent notions. Indeed, the Stokes curves of the Stokes graph $\cG_\vartheta(\varphi_2)$ have the same definition as the walls of the spectral network $\cW_\vartheta(\varphi_2)$. Furthermore, the labels of the walls determine the orientations of the Stokes curves and vice versa. In particular, notice that the labels of the walls in the spectral network are chosen in such a way that the recessive (or small) section $s_i$ of the flat connection $\nabla^\mathrm{oper}$ stays invariant across a wall. 

Let us remind ourselves how we compute the monodromy along the path $C_0$ for any flat $SL(2)$ connection $\nabla$ using the abelianization method with respect to the spectral network $\cW_\vartheta(\varphi_2)$ (see \S\ref{sec:ab-l-t}, or for some more detail \S6 and \S7 of \cite{Hollands:2013qza}). Suppose the flat $SL(2)$ connection $\nabla$ is abelianized with respect to $\cW_\vartheta(\varphi_2)$ by the equivariant connection $\nabla^\mathrm{ab}$. To find the monodromy of $\nabla$ along $C_0$ we cut the path $C_0$ into a collection of smaller paths $\wp$ that do not cross any walls nor branch-cuts of $\cW_\vartheta(\varphi_2)$. The monodromy along $C_0$ is then given by the product of abelian parallel transport matrices $D_\wp$ over all paths $\wp$, where we splice in a branch-cut matrix when crossing a branch-cut and a unipotent matrix $\cS_w$ when crossing a wall. 

As is shown in \cite{Hollands:2013qza}, and reviewed in \S\ref{sec:ab-l-t}, the abelianization mapping is unique for any $K=2$ Fock-Goncharov or (resolution of a) $K=2$ Fenchel-Nielsen spectral network. Furthermore, the unipotent matrices $\cS_w$ are of a rather special form. They have 1's on the diagonal, and the only nonzero off-diagonal component of $\cS_w$ can be written as the abelian parallel transport of $\nabla^\mathrm{ab}$ along an auxiliary (or detour) path that starts at a lift of the basepoint $w$, follows the wall in the opposite orientation, circles around the branch-point $b$, and returns to the other lift of the basepoint $w$.  

More precisely, the previous description is valid if we choose the branch-cut matrix 
\begin{align}
\left( \begin{array}{cc} 0 & 1 \\ -1 & 0 \end{array} \right),
\end{align}
as is conventional. If instead we would choose the branch-cut matrix to be
\begin{align}\label{eqn:branchcutdiag}
\left( \begin{array}{cc} 0 & i \\ i & 0 \end{array} \right),
\end{align}
the nonzero off-diagonal component of $\cS_w$ is multiplied by an additional factor $\pm i$.

Let us now decompose the connection matrix $\widetilde{V}^{l,l+1}_{z_0}$ from equation~(\ref{eqn:Vz0l}) as 
\begin{align}
\widetilde{V}^{l,l+1}_{z_0} = (D_{z_0}^w)^{-1} \, \cS_w \, D_{z_0}^w.
\end{align}
Then the matrix
\begin{align}\label{eqn:Sl}
\cS_w = (D_{w}^b)^{-1} \, V^{l,l+1} \,  D_{w}^b.
\end{align}
has as only nonzero off-diagonal component the Borel sum of 
\begin{align}\label{eqn:auxWKB}
\exp \left( \pm  \int^b_{w} S_\mathrm{odd} \, dz \right) \left(\pm i \right) \exp \left( \mp  \int^w_{b} S_\mathrm{odd} \, dz \right),
\end{align}
where two signs in the exponentials are opposite and depend on the orientation of the Stokes curve, whereas the sign in front of the factor $i$ depends on the direction of crossing the Stokes curve. 

It follows that the monodromy representation for the oper $\nabla^\mathrm{oper}_\eps$ obtained using the exact WKB method can be brought in the form of a monodromy representation obtained using the abelianization mapping. In fact, it shows that $\nabla^\mathrm{oper}_\eps$ is abelianized in each region $l$ by the Borel sums $\psi^l_\pm$ (in the direction $\vartheta$) of the WKB solutions $\psi_\pm$ of the differential equation~(\ref{eqn:wkbdiffeqn}).

Indeed, with respect to this basis the abelian parallel transport matrix is the Borel sum of the matrix 
\begin{align}
D_\wp = \left( \begin{array}{cc} \exp \left(+ \int_\wp S_\mathrm{odd} \, dz \right) & 0 \\ 0 & \exp \left(- \int_\wp S_\mathrm{odd} \, dz \right) \end{array} \right),
 \end{align}
while the branch-cut matrix is of the non-conventional form~(\ref{eqn:branchcutdiag}), due to the square-root $\sqrt{S_\mathrm{odd}}$ in the denominator of the definition of $\psi_\pm$, and the nonzero off-diagonal component of the unipotent matrix $\cS_w$ is the Borel sum of the expression~(\ref{eqn:auxWKB}).

Thus we conclude the monodromies obtained from the exact WKB method are equivalent to the monodromies obtained through the abelianization mapping on the locus of $\eps$-opers. 

In particular, this relation shows that the spectral coordinates $\log \cX_\gamma (\nabla^\mathrm{oper}_\eps)$ are equal to the Borel sums (in the direction $\vartheta$) of the Voros periods $V_\gamma$. As an immediate consequence it follows that the spectral coordinates $\cX_\gamma(\nabla^\mathrm{oper}_{\eps'})$ have the WKB asymptotics
\begin{align}\label{eqn:WKBasymp}
\cX_\gamma (\nabla^\mathrm{oper}_{\eps'}) \sim \exp \left( \oint_\gamma S_\mathrm{odd}(\eps') \, dz \right)
\end{align}
in the limit $\eps' \to 0$ with $|\arg \eps' - \vartheta| < \pi/2$. (This was already shown by a different argument in \cite{GAIOTTO2013239}.) We emphasize that while the spectral coordinates $\cX_\gamma$ are thus rather sensitive to the choice of the phase $\vartheta$, their WKB asymptotics are not. For example, the spectral coordinates for the two resolutions of a Fenchel-Nielsen network generated by a Strebel differential differ, but their WKB asymptotics in the limit $\eps \to 0$ agree.

The right-hand side of equation~(\ref{eqn:WKBasymp}) is also known as a quantum period
\begin{align}
\Pi_\gamma(\eps) =\exp \left( \oint_\gamma S_\mathrm{odd}(\eps) \, dz \right).
\end{align}

\subsection{Quantum periods and non-perturbative corrections}\label{sec:quantumperiod}

In the main part of this paper we computed the generating function of the space of $\eps$-opers on the four-punctured sphere $\mathbb{P}^1_{0,q,1,\infty}$, with respect to the complexified length-twist coordinates $(\ell,\tau)$, by comparing monodromy representations. The complexified length-twist coordinates were realized as spectral coordinates 
\begin{align}
L & = - \exp( \pi i \ell) \\
T &=  - \exp (2 \tau) \notag
\end{align}
by abelianizing with respect to a Fenchel-Nielsen network $\cW$. The discussion in \S\ref{sec:WKBab} indicates an alternative way of computing this generating function. 

Fix the phase $\vartheta_0$ and the mass parameters $m_l$ such that $e^{- 2 i \vartheta_0} \varphi_2$ is a Strebel differential, generating a Fenchel-Nielsen network isotopic to $\mathcal{W}$. This is certainly possible in the weakly coupling limit $q \to 0$ where
\begin{align}
u = \frac{a_0^2}{2} + \mathcal{O}(q),
\end{align}
with $a_0 = \oint_A \sqrt{\varphi_2}$. 

According to the discussion in \S\ref{sec:WKBab}, the length-twist coordinates $(\ell,\tau^\pm)$ restricted to the space of $\eps$-opers may be computed as the Borel sums of the Voros symbols $V_A$ and $V_B$, respectively, in the direction $\arg \eps=\vartheta_0 \pm \delta$ for sufficiently small $\delta$. Let us denote these Borel sums as $V_A^\pm$ and $V_B^\pm$, respectively.

The generating function of $\eps$-opers is found by inverting the relation
\begin{align}
\frac{\ell}{2} &=V^+_A(H,q,\eps)=V^-_A(H,q,\eps), 
\end{align}
where $H$ denotes the accessory parameter, and substituting the result into the expression for $V_B^\pm(H,q,\eps)$, to find
\begin{align}
\frac{\partial W^\mathrm{oper}(\ell,q,\eps)}{\partial \ell} & = \frac{V^+_B(\ell,q,\eps)}{4} +\frac{V^-_B(\ell,q,\eps)}{4}.   
 \end{align}
By construction, this generating function agrees analytically with the generating function of opers as computed in \S\ref{sec:genfunopers} (after reintroducing the $\eps$-dependence in the latter). 

The $\eps$-asymptotics of the generating function $W^\mathrm{oper}(\ell,q,\eps)$ are simply obtained by computing the Voros symbols $V_A$ and $V_B$ in an $\eps$-expansion as quantum periods. This relates the Nekrasov-Rosly-Shatashvili correspondence to the approach of computing the $\eps$-asymptotics of the NS superpotential $\widetilde{W}^\mathrm{eff}(a,q,\eps)$ using quantum periods \cite{Mironov:2009uv}. 

The exact NS superpotential $\widetilde{W}^\mathrm{eff}(a,q,\eps)$ is found by Borel resumming its asymptotic expansion in a critical direction $\vartheta_0$ corresponding to a Fenchel-Nielsen network. In particular, we find that in this critical direction the NS superpotential does \emph{not} acquire any non-perturbative corrections.

\bibliographystyle{utphys}
\bibliography{NRS-SU3bifund}

\providecommand{\href}[2]{#2}\begingroup\raggedright\begin{thebibliography}{10}

\bibitem{donagi1996supersymmetric}
R.~Donagi and E.~Witten, ``{Supersymmetric Yang-Mills theory and integrable
  systems},'' {\em Nuclear Physics B} {\bf 460} (1996), no.~2, 299--334.

\bibitem{GAIOTTO2013239}
D.~Gaiotto, G.~W. Moore, and A.~Neitzke, ``{Wall-crossing, Hitchin systems, and
  the WKB approximation},'' {\em Advances in Mathematics} {\bf 234} (2013) 239
  -- 403.

\bibitem{Gaiotto:2009we}
D.~Gaiotto, ``{N=2 dualities},''
\href{http://www.arXiv.org/abs/0904.2715}{{\tt 0904.2715}}.

\bibitem{Neitzke:2014cja}
A.~Neitzke, ``{Hitchin Systems in $\mathcal{N} =$ 2 Field Theory},'' in {\em
  New Dualities of Supersymmetric Gauge Theories}, J.~Teschner, ed.,
  pp.~53--77.
\newblock 2016.
\newblock
\href{http://www.arXiv.org/abs/1412.7120}{{\tt 1412.7120}}.
\newblock

\bibitem{Nekrasov:2002qd}
N.~A. Nekrasov, ``{Seiberg-Witten prepotential from instanton counting},'' {\em
  Adv. Theor. Math. Phys.} {\bf 7} (2003), no.~5, 831--864,
\href{http://www.arXiv.org/abs/hep-th/0206161}{{\tt hep-th/0206161}}.

\bibitem{Nekrasov:2003rj}
N.~Nekrasov and A.~Okounkov, ``{Seiberg-Witten theory and random partitions},''
  {\em Prog. Math.} {\bf 244} (2006) 525--596,
\href{http://www.arXiv.org/abs/hep-th/0306238}{{\tt hep-th/0306238}}.

\bibitem{Nakajima:2003uh}
H.~Nakajima and K.~Yoshioka, ``{Lectures on instanton counting},'' in {\em {CRM
  Workshop on Algebraic Structures and Moduli Spaces Montreal}}.
\newblock 2003.
\newblock
\href{http://www.arXiv.org/abs/math/0311058}{{\tt math/0311058}}.
\newblock

\bibitem{Nekrasov:2009rc}
N.~A. Nekrasov and S.~L. Shatashvili, ``{Quantization of Integrable Systems and
  Four Dimensional Gauge Theories},'' in {\em {Proceedings, 16th International
  Congress on Mathematical Physics (ICMP09): Prague, Czech Republic, August
  3-8, 2009}}, pp.~265--289.
\newblock 2009.
\newblock
\href{http://www.arXiv.org/abs/0908.4052}{{\tt 0908.4052}}.
\newblock

\bibitem{Pestun:2007rz}
V.~Pestun, ``{Localization of gauge theory on a four-sphere and supersymmetric
  Wilson loops},'' {\em Commun. Math. Phys.} {\bf 313} (2012) 71--129,
\href{http://www.arXiv.org/abs/0712.2824}{{\tt 0712.2824}}.

\bibitem{Vartanov:2013ima}
J.~Teschner and G.~S. Vartanov, ``{Supersymmetric gauge theories, quantization
  of $\mathcal{M}_{\mathrm{flat}}$, and conformal field theory},'' {\em Adv.
  Theor. Math. Phys.} {\bf 19} (2015) 1--135,
\href{http://www.arXiv.org/abs/1302.3778}{{\tt 1302.3778}}.

\bibitem{Nekrasov:2011bc}
N.~Nekrasov, A.~Rosly, and S.~Shatashvili, ``{Darboux coordinates, Yang-Yang
  functional, and gauge theory},'' {\em Nucl. Phys. Proc. Suppl.} {\bf 216}
  (2011) 69--93,
\href{http://www.arXiv.org/abs/1103.3919}{{\tt 1103.3919}}.

\bibitem{teschner2011}
J.~Teschner, ``{Quantization of the Hitchin moduli spaces, Liouville theory and
  the geometric Langlands correspondence I},'' {\em Adv. Theor. Math. Phys.}
  {\bf 15} (04, 2011) 471--564.

\bibitem{Alday:2009aq}
L.~F. Alday, D.~Gaiotto, and Y.~Tachikawa, ``{Liouville Correlation Functions
  from Four-dimensional Gauge Theories},'' {\em Lett. Math. Phys.} {\bf 91}
  (2010) 167--197,
\href{http://www.arXiv.org/abs/0906.3219}{{\tt 0906.3219}}.

\bibitem{beilinson1991quantization}
A.~Beilinson and V.~Drinfeld, ``{Quantization of Hitchin’s integrable system
  and Hecke eigensheaves},'' 1991.

\bibitem{2005math......1398B}
A.~{Beilinson} and V.~{Drinfeld}, ``{Opers},''
  \href{http://www.arXiv.org/abs/math/0501398}{{\tt math/0501398}}.

\bibitem{BenZvi:1999kx}
D.~Ben-Zvi and E.~Frenkel, ``{Spectral curves, opers and integrable systems},''
\href{http://www.arXiv.org/abs/math/9902068}{{\tt math/9902068}}.

\bibitem{Gaiotto:2014bza}
D.~Gaiotto, ``{Opers and TBA},''
\href{http://www.arXiv.org/abs/1403.6137}{{\tt 1403.6137}}.

\bibitem{2016arXiv160702172D}
O.~{Dumitrescu}, L.~{Fredrickson}, G.~{Kydonakis}, R.~{Mazzeo}, M.~{Mulase},
  and A.~{Neitzke}, ``{Opers versus nonabelian Hodge},''
  \href{http://www.arXiv.org/abs/1607.02172}{{\tt 1607.02172}}.

\bibitem{Nekrasov:2010ka}
N.~Nekrasov and E.~Witten, ``{The Omega Deformation, Branes, Integrability, and
  Liouville Theory},'' {\em JHEP} {\bf 09} (2010) 092,
\href{http://www.arXiv.org/abs/1002.0888}{{\tt 1002.0888}}.

\bibitem{Gaiotto:2012rg}
D.~Gaiotto, G.~W. Moore, and A.~Neitzke, ``{Spectral networks},'' {\em Annales
  Henri Poincare} {\bf 14} (2013) 1643--1731,
\href{http://www.arXiv.org/abs/1204.4824}{{\tt 1204.4824}}.

\bibitem{Hollands:2013qza}
L.~Hollands and A.~Neitzke, ``{Spectral Networks and Fenchel–Nielsen
  Coordinates},'' {\em Lett. Math. Phys.} {\bf 106} (2016), no.~6, 811--877,
\href{http://www.arXiv.org/abs/1312.2979}{{\tt 1312.2979}}.

\bibitem{Lay4347}
W.~Lay and S.~Y. Slavyanov, ``Heun{\textquoteright}s equation with nearby
  singularities,'' {\em Proceedings of the Royal Society of London A:
  Mathematical, Physical and Engineering Sciences} {\bf 455} (1999), no.~1992,
  4347--4361.

\bibitem{Kazakov2001}
A.~Y. Kazakov, ``Coalescence of two regular singularities into one regular
  singularity for the linear ordinary differential equation,'' {\em Journal of
  Dynamical and Control Systems} {\bf 7} (2001), no.~1, 127--149.

\bibitem{Menotti:2014kra}
P.~Menotti, ``{On the monodromy problem for the four-punctured sphere},'' {\em
  J. Phys.} {\bf A47} (2014), no.~41, 415201,
\href{http://www.arXiv.org/abs/1401.2409}{{\tt 1401.2409}}.

\bibitem{exactWKB}
T.~Kawai and Y.~Takei, {\em Algebraic Analysis of Singular Perturbation
  Theory}.
\newblock No.~227 in Translations of Mathematical Monographs. American
  Mathematical Society, 2005.

\bibitem{Mironov:2009uv}
A.~Mironov and A.~Morozov, ``{Nekrasov Functions and Exact Bohr-Zommerfeld
  Integrals},'' {\em JHEP} {\bf 04} (2010) 040,
\href{http://www.arXiv.org/abs/0910.5670}{{\tt 0910.5670}}.

\bibitem{Chacaltana:2010ks}
O.~Chacaltana and J.~Distler, ``{Tinkertoys for Gaiotto Duality},'' {\em JHEP}
  {\bf 11} (2010) 099,
\href{http://www.arXiv.org/abs/1008.5203}{{\tt 1008.5203}}.

\bibitem{Williams2016}
H.~Williams, ``Toda systems, cluster characters, and spectral networks,'' {\em
  Communications in Mathematical Physics} {\bf 348} (2016), no.~1, 145--184.

\bibitem{liu2008jenkins}
J.~Liu, ``Jenkins--strebel differentials with poles,'' {\em Commentarii
  Mathematici Helvetici} {\bf 83} (2008), no.~1, 211--240.

\bibitem{Gabella:2017hpz}
M.~Gabella, P.~Longhi, C.~Y. Park, and M.~Yamazaki, ``{BPS Graphs: From
  Spectral Networks to BPS Quivers},'' {\em JHEP} {\bf 07} (2017) 032,
\href{http://www.arXiv.org/abs/1704.04204}{{\tt 1704.04204}}.

\bibitem{Longhi:2016wtv}
P.~Longhi, ``{Wall-Crossing Invariants from Spectral Networks},''
\href{http://www.arXiv.org/abs/1611.00150}{{\tt 1611.00150}}.

\bibitem{neitzkeswn}
A.~Neitzke, ``{swn-plotter}.''
  \url{http://www.ma.utexas.edu/users/neitzke/mathematica/swn-plotter.nb}.

\bibitem{Hollands:2016kgm}
L.~Hollands and A.~Neitzke, ``{BPS states in the Minahan-Nemeschansky ${E_6}$
  theory},'' {\em Commun. Math. Phys.} {\bf 353} (2017), no.~1, 317--351,
\href{http://www.arXiv.org/abs/1607.01743}{{\tt 1607.01743}}.

\bibitem{Goldman1986}
{Goldman, William M.}, ``{Invariant functions on Lie groups and Hamiltonian
  flows of surface group representations},'' {\em {Inventiones mathematicae}}
  {\bf 85} (1986), no.~2, 263--302.

\bibitem{10.2307/2007011}
S.~Wolpert, ``{The Fenchel-Nielsen Deformation},'' {\em Annals of Mathematics}
  {\bf 115} (1982), no.~3, 501--528.

\bibitem{10.2307/2007075}
S.~Wolpert, ``{On the Symplectic Geometry of Deformations of a Hyperbolic
  Surface},'' {\em Annals of Mathematics} {\bf 117} (1983), no.~2, 207--234.

\bibitem{GOLDMAN1984200}
W.~M. Goldman, ``The symplectic nature of fundamental groups of surfaces,''
  {\em Advances in Mathematics} {\bf 54} (1984), no.~2, 200 -- 225.

\bibitem{tan_1994}
S.~Tan, ``{Complex Fenchel-Nielsen coordinates for quasi-Fuchsian
  structures},'' {\em Internat. J. Math.} {\bf 5} (1994) 239–251.

\bibitem{kourouniotis_1994}
C.~Kourouniotis, ``{Complex length coordinates for quasi-fuchsian groups},''
  {\em Mathematika} {\bf 41} (1994), no.~1, 173–188.

\bibitem{goldman2009trace}
W.~M. Goldman, ``Trace coordinates on fricke spaces of some simple hyperbolic
  surfaces,'' \href{http://www.arXiv.org/abs/0901.1404}{{\tt 0901.1404}}.

\bibitem{Dimofte:2011jd}
T.~Dimofte and S.~Gukov, ``{Chern-Simons Theory and S-duality},'' {\em JHEP}
  {\bf 05} (2013) 109,
\href{http://www.arXiv.org/abs/1106.4550}{{\tt 1106.4550}}.

\bibitem{Gaiotto:2012db}
D.~Gaiotto, G.~W. Moore, and A.~Neitzke, ``{Spectral Networks and Snakes},''
  {\em Annales Henri Poincare} {\bf 15} (2014) 61--141,
\href{http://www.arXiv.org/abs/1209.0866}{{\tt 1209.0866}}.

\bibitem{Hollands:toappear}
L.~Hollands and A.~Neitzke {\em to appear}.

\bibitem{2013arXiv1306.0876F}
E.~{Frenkel} and C.~{Teleman}, ``{Geometric Langlands Correspondence Near
  Opers},'' \href{http://www.arXiv.org/abs/1306.0876}{{\tt 1306.0876}}.

\bibitem{Okuda:2010ke}
T.~Okuda and V.~Pestun, ``{On the instantons and the hypermultiplet mass of
  N=2* super Yang-Mills on $S^{4}$},'' {\em JHEP} {\bf 03} (2012) 017,
\href{http://www.arXiv.org/abs/1004.1222}{{\tt 1004.1222}}.

\bibitem{Alday:2009fs}
L.~F. Alday, D.~Gaiotto, S.~Gukov, Y.~Tachikawa, and H.~Verlinde, ``{Loop and
  surface operators in N=2 gauge theory and Liouville modular geometry},'' {\em
  JHEP} {\bf 01} (2010) 113,
\href{http://www.arXiv.org/abs/0909.0945}{{\tt 0909.0945}}.

\bibitem{Drukker:2009id}
N.~Drukker, J.~Gomis, T.~Okuda, and J.~Teschner, ``{Gauge Theory Loop Operators
  and Liouville Theory},'' {\em JHEP} {\bf 02} (2010) 057,
\href{http://www.arXiv.org/abs/0909.1105}{{\tt 0909.1105}}.

\bibitem{Ashok:2015gfa}
S.~K. Ashok, M.~Billó, E.~Dell'Aquila, M.~Frau, R.~R. John, and A.~Lerda,
  ``{Non-perturbative studies of N=2 conformal quiver gauge theories},'' {\em
  Fortsch. Phys.} {\bf 63} (2015) 259--293,
\href{http://www.arXiv.org/abs/1502.05581}{{\tt 1502.05581}}.

\bibitem{frenkelloop}
E.~Frenkel, {\em Langlands Correspondence for Loop Groups}.
\newblock Cambridge University Press, 2007.

\bibitem{Beukers1989}
F.~Beukers and G.~Heckman, ``Monodromy for the hypergeometric function
  nfn-1.,'' {\em Inventiones mathematicae} {\bf 95} (1989), no.~2, 325--354.

\bibitem{2016arXiv160403082I}
A.~{Its}, O.~{Lisovyy}, and A.~{Prokhorov}, ``{Monodromy dependence and
  connection formulae for isomonodromic tau functions},''
  \href{http://www.arXiv.org/abs/1604.03082}{{\tt 1604.03082}}.

\bibitem{Fateev:2005gs}
V.~A. Fateev and A.~V. Litvinov, ``{On differential equation on four-point
  correlation function in the Conformal Toda Field Theory},'' {\em JETP Lett.}
  {\bf 81} (2005) 594--598,
\href{http://www.arXiv.org/abs/hep-th/0505120}{{\tt hep-th/0505120}}.

\bibitem{Voros83}
A.~Voros, ``The return of the quartic oscillator,'' {\em Annales de l'Institut
  Henri Poincare, Section A, Physique Theorique} (1983), no.~39(3), 211--338.

\bibitem{2014arXiv1401.7094I}
K.~{Iwaki} and T.~{Nakanishi}, ``{Exact WKB analysis and cluster algebras},''.

\bibitem{koikeschafke}
T.~Koike and R.~Sch\"afke, ``{On the Borel summability of WKB solutions of
  Sch\"odinger equations with polynomial potentials and its application}.'' in
  preparation.

\end{thebibliography}\endgroup

\end{document}